\documentclass[
preprint,
  amsmath,amssymb,
  aps,
]{revtex4-2}

\usepackage{graphicx}
\usepackage{amsmath}
\usepackage{slashed}
\usepackage{bm}
\usepackage{lineno}
\usepackage{hyperref}
\usepackage{lipsum}
\usepackage[table]{xcolor} 
\usepackage{cleveref}

\makeatletter
\def\l@subsubsection#1#2{}
\makeatother

\begin{document}
\title[Breit-Wheeler Process and Vacuum Birefringence]{Report on Progress in Physics: Observation of the Breit-Wheeler Process and Vacuum Birefringence in Heavy-Ion Collisions}
\preprint{Report on Progress}

\author{James Daniel Brandenburg$^{1,2}$, 
Janet Seger$^3$,
Zhangbu Xu$^1$,
Wangmei Zha$^4$}

\address{$^1$ Brookhaven National Laboratory, Upton NY 11973-5000}
\address{$^2$ Center for Frontiers in Nuclear Science affiliate at Stony Brook University, Stony Brook NY, 11794-38000}
\address{$^3$ Creighton University, Omaha NE, 68178}
\address{$^4$ University of Science and Technology of China, Hefei, China}

\date{Received:  / Accepted: }

\begin{abstract}
    This Report reviews the effort over several decades to observe the linear Breit-Wheeler process ($\gamma\gamma \rightarrow e^+e^-$) and vacuum birefringence in high-energy particle and heavy-ion collider experiment.  This Report, motivated by STAR's recent observations, attempts to summarize the key issues related to the interpretation of polarized $\gamma\gamma \rightarrow l^+l^-$ measurements in high-energy experiments. To that end, we start by reviewing the historical context and essential theoretical developments, before focusing on the decades of progress made in high-energy collider experiments. Special attention is given to the evolution in experimental approaches in response to various challenges and the demanding detector capabilities required to unambiguously identify the linear Breit-Wheeler process and to detect the signatures of vacuum birefringence. We close the report with a discussion, followed by a look at near-future opportunities for utilizing these discoveries and for testing QED in previously unexplored regimes.
\end{abstract}

\maketitle
\newpage
\baselinestretch{0.5}
\setcounter{tocdepth}{0}
\tableofcontents


\section{Introduction}
    \label{sec:introduction}
    
\subsection{History}

The birth of quantum mechanics in the early part of the 20th century led to an explosion of progress in both experimental achievements and theoretical understanding. It was in the late 1920s that Dirac and Pauli undertook the work of unifying special relativity and quantum mechanics to obtain a relativistic equation of motion for the wave function of the electron. The solution, known for decades as the Dirac equation, describes all massive spin-1/2 particles.  The Dirac equation produces the two expected positive energy solutions, but also predicts two \textit{negative} energy states. The prediction of negative energy states was somewhat puzzling at the time, since their physical interpretation was not clear. Furthermore, available negative energy states should imply that electrons (with positive energy) could decay via a photon. In order to explain the apparent stability of the electron, Dirac proposed the Hole theory, which postulates that the quantum vacuum is a quantum many-body state in which all negative-energy ``holes'' are filled. The Hole theory, along with the Pauli exclusion principle, which forbids electrons with the same quantum numbers from occupying identical energy states, resolved the problem of electron decay. In the year 1930, Dirac built on his earlier work by proposing the annihilation process between positive energy states (electrons, $e^-$) and negative energy states (later known as positrons, $e^+$):
\begin{equation}
    e^- + e^+ \rightarrow \gamma + \gamma.
\end{equation}
This process, known as Dirac Annihilation, or simply as annihilation, describes the conversion of an electron and a positron into electromagnetic radiation in the form of two photons.  The physical interpretation of these negative energy states was resolved with the experimental discovery, by Anderson in 1932, of the positron in cosmic ray interactions~\cite{doi10.1126science.76.1967.238,PhysRev.43.491}. 
The discovery of the positron led to rapid progress in understanding and describing the fundamental interactions between charged particles and photons.  Around the same time, Sauter~\cite{sauterUeberVerhaltenElektrons1931} and Klein~\cite{kleinReflexionElektronenPotentialsprung1929} pioneered the concept of creating electron-positron pairs from the vacuum in the presence of a constant electromagnetic field above a critical value. 
This critical field strength corresponds to the value needed such that the energy gain of an electron accelerating over a Compton wavelength $(\lambda_C)$ is the rest mass of an electron.
They computed this critical value to be:
\begin{equation}
    \label{eq:crit}
    E_c \equiv \frac{m_e c^2}{e\lambda_C} = \frac{m_e^2c^3}{e \hbar} \simeq 1.3\times10^{16} {\rm\ V/cm},
\end{equation}
where $m_e$ is the electron mass, $c$ is the speed of light, $e$ is the charge of an electron, $\hbar$ is the Planck constant, and $\lambda_C \equiv \hbar/m_e c$ is the Compton wavelength of the electron. In static electric fields of this strength the quantum vacuum becomes unstable, spontaneously breaking down into a real electron-positron pair~\cite{sauterUeberVerhaltenElektrons1931}.
At that time, several others were also studying the creation of pairs from the vacuum via the collision of light quanta.
In the year 1934, Breit and Wheeler published their theory of pair creation via the inverse of the Dirac annihilation process, namely the creation of an electron-positron pair via the collision of two light quanta (i.e., photons):
\begin{equation}
    \gamma + \gamma \rightarrow e^+ + e^-
\end{equation}
\begin{figure}
    \centering
    \includegraphics[width=0.60\linewidth]{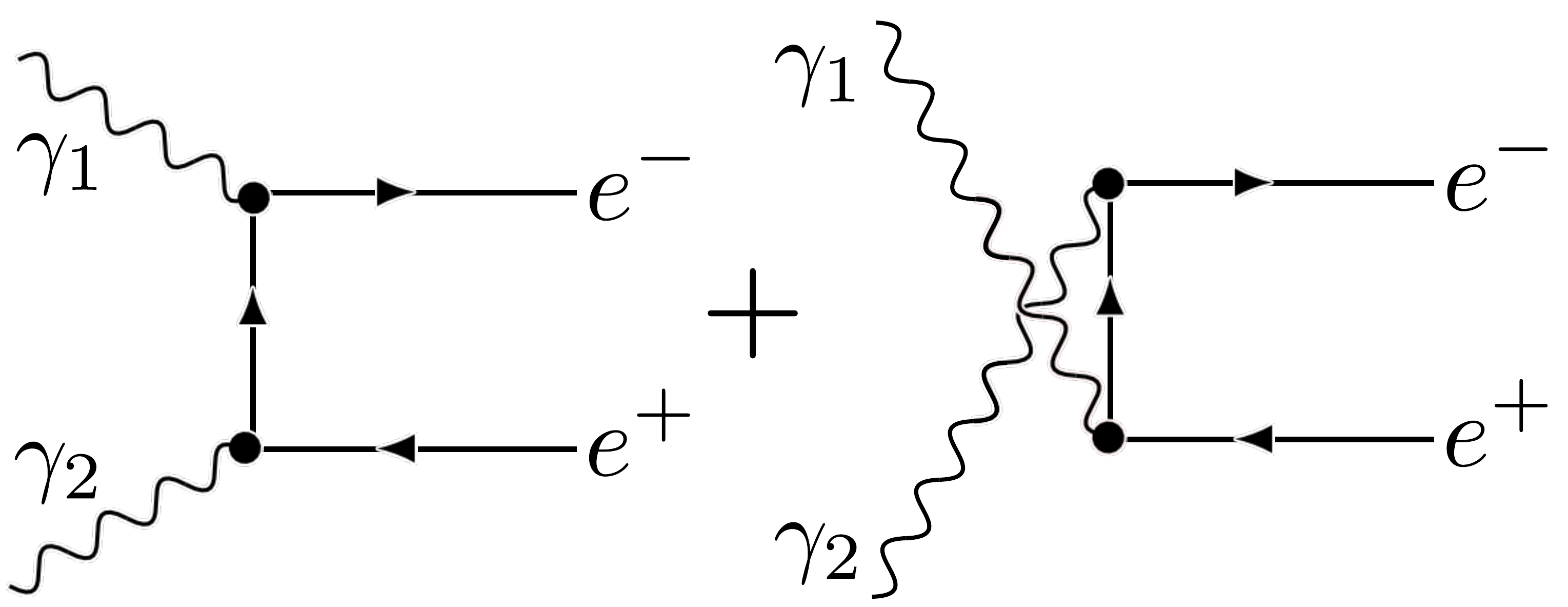}
    \caption{ The two Feynman diagrams contributing to pair creation in second order (the Breit-Wheeler process). The $\gamma_1$ and $\gamma_2$ denote the interaction with the photons 1 and 2, respectively. When realized in heavy-ion collisions, the photons are manifest from the external field of the heavy nuclei~\cite{PhysRevA.51.1874}. }
    \label{fig:bw_diagrams}
\end{figure}
This process, the time inverse of the Dirac annihilation process, has become known as the Breit-Wheeler process (BW, hereafter).
While the Dirac annihilation process was quickly realized\cite{Chao1930PhysRev.36.1519,klempererAnnihilationRadiationPositron1934,goworek201480} and has since become one of the most rigorously tested processes in physics~\cite{feynmanQuantumElectrodynamics1998}, experimental observation of the Breit-Wheeler process proved to be much more elusive. Indeed, Breit and Wheeler clearly recognized the difficulty of observing the proposed process of electron-positron pair creation, commenting in their 1934 paper~\cite{PhysRev.46.1087} that it is ``hopeless to try to observe the pair formation in laboratory experiments with two beams of x-rays or gamma-rays meeting each other''.
Their comment was an astute observation, considering the difficulty observing the process even after decades of technological advancements, including the invention of lasers~\cite{stricklandCompressionAmplifiedChirped1985}. 
However, Breit and Wheeler noted Fermi's~\cite{1924ZPhy29315F} method of equivalent photons and the recent work of Williams~\cite{PhysRev.45.729} and Weiszäcker\cite{weizsaeckerAusstrahlungBeiStoessen1934} which indicated that ultra-relativistic beams of highly charged nuclei may provide a viable photon source: 
\begin{quote}
    In the considerations of Williams, however, the large nuclear electric fields lead to large densities of quanta in moving frames of reference. This, together with the large number of nuclei available in unit volume of ordinary materials, increases the effect to observable amounts.
\end{quote}
Theoretical perspectives on the validity and viability of using Weiszacker-Williams (WW) photons in heavy-ion collisions to achieve the Breit-Wheeler process will be discussed further in ~\ref{sec:theory}.  

In addition to Breit and Wheeler, several others were also investigating the concept of pair production in the same year.  Unlike Breit and Wheeler, who were specifically interested in the collision of real photons, Bethe and Heitler~\cite{betheh.StoppingFastParticles1934} introduced the concept of pair production via a collision between one real photon and a virtual photon from the Coulomb field of a nucleus. Similarly, Landau and Lifshitz~\cite{196584,Landau1934} studied the total production cross section for electron positron pairs from the collision of two virtual photons from high-energy particle beams.

Over the next few decades, the work of Dirac~\cite{doi:10.1098/rspa.1928.0023}, Feynman~\cite{PhysRev.76.749}, Schwinger~\cite{PhysRev.76.790}, Bethe~\cite{relativisticLambShiftBaranger} and several others to describe the gauge invariant interactions of charged particles with photons grew into what we now call Quantum Electrodynamics (QED) – the fully covariant relativistic quantum field theory describing how light and matter interact.
The ability of QED to precisely predict the Dirac annihilation process and many other processes has led to it being regarded as the most rigorously tested theory in physics \cite{Sailer2022,Jegerlehner2018}.

\subsection{The Breit-Wheeler Process}
\label{sec:intro_bw}
Figure~\ref{fig:bw_diagrams} shows the two Feynman diagrams representing the Breit-Wheeler process at lowest-order (second-order) in QED. 
The potential interaction of two photons to produce an electron-positron pair is a clear violation of the superposition principle, an essential feature of the linear theory of classical electromagnetism. Several of the non-linear features of QED will be discussed in more detail in the section on vacuum polarization and vacuum birefringence (Sec. \ref{sec:intro_vb}).  The central result of Breit and Wheeler's investigation was the cross section for photon-photon collisions leading to electron-positron pair production:
\begin{align}
    \begin{split}
    \label{eq:bw}
        \sigma & (\gamma \gamma \rightarrow e^{+}e^{-}) =  \frac{4\pi \alpha_{em}^{2}}{W^{2}} \left[ \left(2+\frac{8m^{2}}{W^{2}} - \frac{16m^{4}}{W^{4}}\right) \right. \\
        & \times\text{ln}\left(\frac{W+\sqrt{W^{2}-4m^{2}}}{2m}\right) \left. -\sqrt{1-\frac{4m^{2}}{W^{2}}}\left(1+\frac{4m^{2}}{W^{2}}\right)\right].
  \end{split}
\end{align}
In this formula $\alpha_{em}$ is the electromagnetic coupling constant, $m$ is the lepton mass ($m=m_e$ for electron-positron production), and $W$ is the invariant mass of the produced pair. It is worthwhile to take a moment to note the key features of this cross section. First and foremost, the cross section is zero when the center-of-mass energy $s = W^2$ is less than twice the mass of the electron squared. This feature is crucial to understanding the near impossibility of achieving the Breit-Wheeler process in terrestrial laboratories, since the peak cross section, though small by everyday standards, is not much smaller than other more easily achievable processes (e.g., Compton scattering). This energy threshold further explains why a single photon can never decay into an electron-positron pair. A minimum of two photons are needed for the center of mass energy of the system to exceed the threshold and allow conversion of the initial massless state (two photons) into the pair of massive leptons. 
\begin{figure*}
    \centering
    \includegraphics[width=0.89\linewidth]{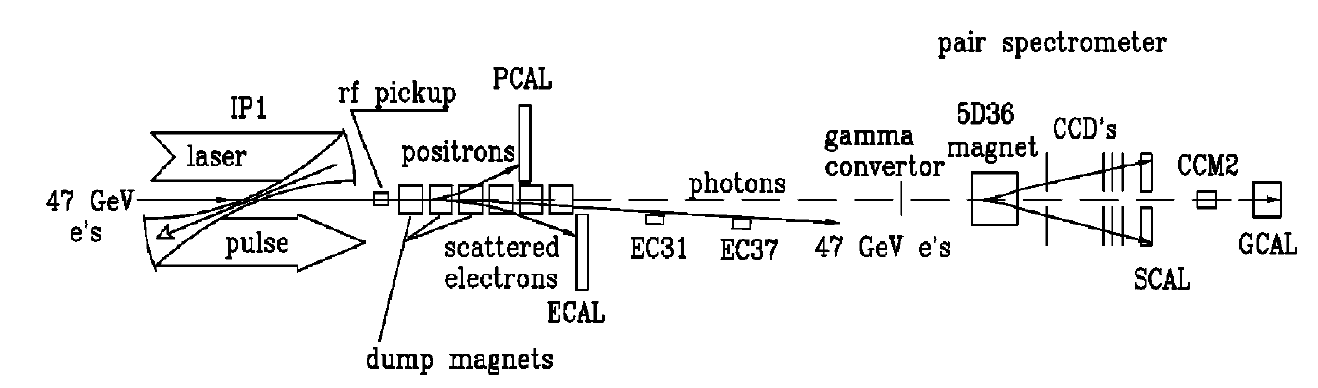}
    \caption{Schematic of the E-144 experimental setup. The laser pulses crossed through the electron beam at the interaction point, IP1. The scattered electrons were deflected by the dump magnets into the electron calorimeter (ECAL). Positrons were deflected into the positron calorimeter (PCAL). The scattered photons were detected in a Cerenkov counter (not shown), or converted to pairs which could be detected by the pair spectrometer. Reproduced from Ref.~\cite{PhysRevD.60.092004}.}
    \label{fig:e144_schematic}
\end{figure*}

\begin{figure}
    \centering
    \includegraphics[width=0.60\linewidth]{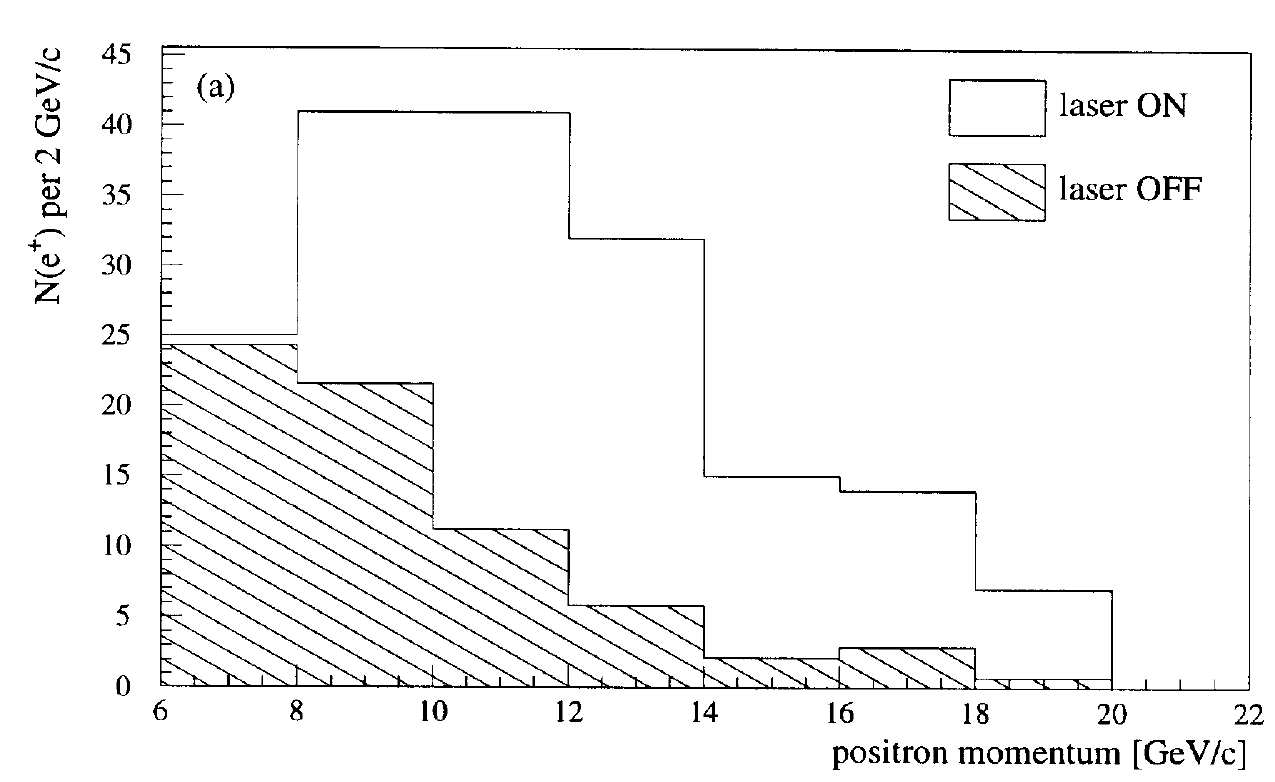}
    \caption{Results from the E-144 experiment conducted at SLAC in 1997. The number of positron candidates vs. momentum for laser-on pulses and for laser-off pulses (shown in the hatched distribution, scaled to the number of laser-on pulses). 
    Reproduced from Ref.~\cite{PhysRevLett.79.1626}.}
    \label{fig:e144_results}
\end{figure}

\begin{figure}
    \centering
    \includegraphics[width=0.60\linewidth]{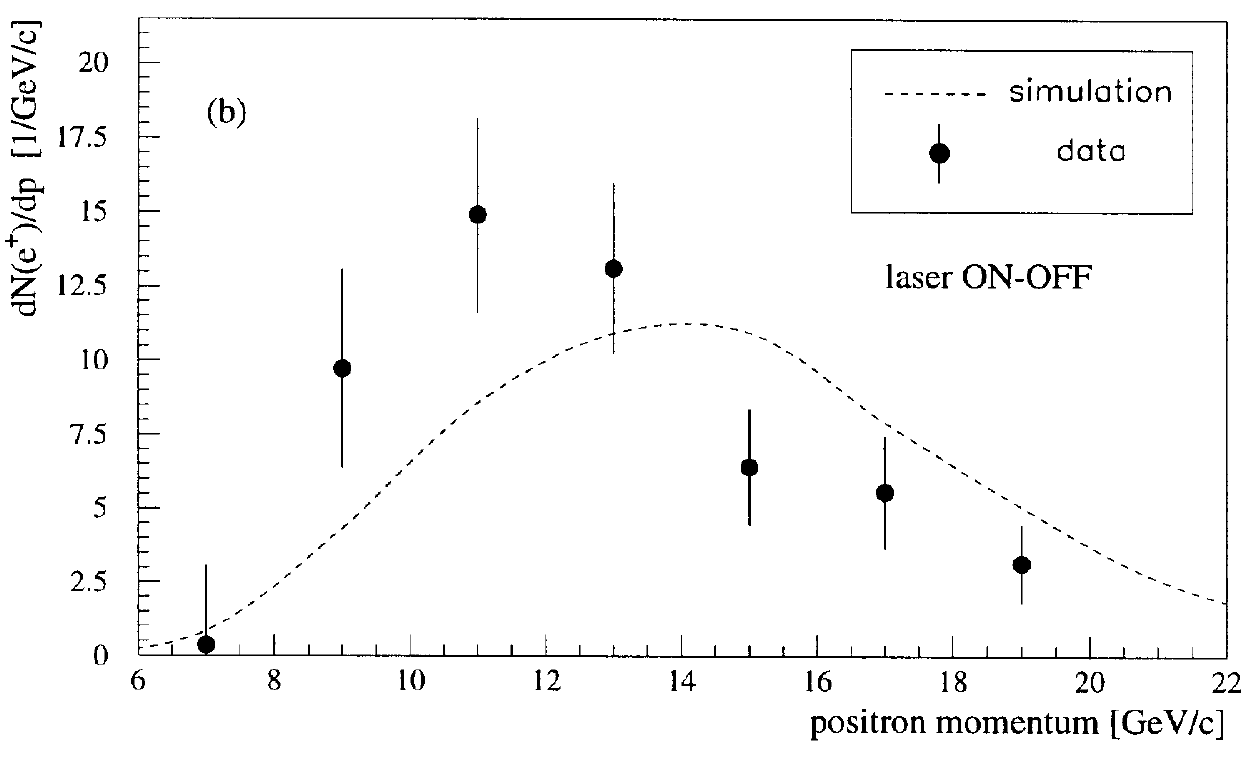}
    \caption{ Spectrum of signal positrons obtained by subtracting the laser-off from the laser-on distribution from the E-144 experiment. The dashed line shows the expected momentum spectrum from model calculations. PCAL cluster positrons have been converted to positron momentum via knowledge of the field in the magnetic spectrometer. 
    Reproduced from Ref.~\cite{PhysRevLett.79.1626}.}
    \label{fig:e144_results_signal}
\end{figure}

In addition to calculating the total cross section for $\gamma\gamma \rightarrow e^+e^-$, Breit and Wheeler also considered the photon polarization, a fundamental attribute of the photon's quantum nature. Since the photon is a massless spin-1 particle, it can have spin projections of only $J_z=\pm1$ ($J_z=0$ is forbidden by gauge invariance). Therefore, the real photon is necessarily transversely polarized~\cite{baymLecturesQuantumMechanics1969}. The general cross section for two polarized photons colliding can be expressed in terms of the cross section for two cases, when their polarization vectors are parallel $(\sigma_\parallel)$ or perpendicular $(\sigma_\perp)$:
\begin{equation}
    \label{eq:bw_pol}
    \begin{aligned}
    \sigma_{\parallel} & = \frac{4\pi \alpha_{em}^{2}}{s} \bigg[\left(2+\frac{8m^{2}}{s} - \frac{24m^{4}}{s^{2}}\right) \\ 
    & \times \text{ln}\left(\frac{\sqrt{s}+\sqrt{s-4m^{2}}}{2m}\right) -\sqrt{1-\frac{4m^{2}}{s}}\left(1+\frac{6m^{2}}{s}\right)\bigg]\\
    \sigma_{\perp} & = \frac{4\pi \alpha_{em}^{2}}{s} \bigg[\left(2+\frac{8m^{2}}{s} - \frac{8m^{4}}{s^{2}}\right) \\
    & \times\text{ln}\left(\frac{\sqrt{s}+\sqrt{s-4m^{2}}}{2m}\right) -\sqrt{1-\frac{4m^{2}}{s}}\left(1+\frac{2m^{2}}{s}\right)\bigg],
    \end{aligned}
\end{equation}
where again $s=W^2$ is the center of mass energy of the system and $m$ is the mass of the produced lepton. For randomly polarized photons, the total cross section can be recovered via a simple average over the two polarization dependent cross sections:
\begin{equation}
    \sigma(\omega_1, \omega_2; m) = \frac{\sigma_{\parallel}(\omega_1, \omega_2; m) + \sigma_{\perp}(\omega_1, \omega_2; m)}{2}.
\end{equation}
While $\sigma_\parallel$ and $\sigma_\perp$ share a similar form, the two cross sections differ significantly near the production threshold, but become nearly identical for $s\gg4m_e^2$. When the sum of the photon energies is comparable to the lepton masses, the cross section for perpendicular photon collisions is significantly larger than for parallel photon polarization.
Furthermore, these two independent cross sections are related to the  parity of the produced state such that the cross section for parallel (perpendicular) polarization can be identified with the cross section for producing scalar (pseudoscalar) final states. 
While a few others investigated the theory of photon-photon collisions in the 1930s, the polarization-dependent aspects of the process are unique to Breit and Wheeler's analysis of real photons. 

One can extend the cross-section level calculations from Breit and Wheeler to the more fundamental level of helicity amplitudes to more deeply investigate the effects of photon polarization. Such calculations were a major focus for John Toll, a student of John Wheeler's. The helicity structure~\cite{1952PhDT21T} of the $\gamma\gamma\rightarrow l^+l^-$ process can be expressed, using notation from Ref.~\cite{SuperChic3}, as:
\begin{align}
    \begin{split}
  & k_{1_\perp}^i k_{2_\perp}^j V_{ij} = \\
  & \,
  \begin{cases}     
    -\frac{1}{2} ({\bf k}_{1_\perp}\cdot {\bf k}_{2_\perp})(\mathcal{M}_{++}+\mathcal{M}_{--}) \quad\qquad (J^P_z=0^+)  \\ 
    -\frac{i}{2} |({\bf k}_{1_\perp}\times {\bf k}_{2_\perp})|(\mathcal{M}_{++}-\mathcal{M}_{--}) \qquad  (J^P_z=0^-) \\ 
    +\frac{1}{2}\bigg[(k_{1_\perp}^x k_{2_\perp}^x-k_{1_\perp}^y k_{2_\perp}^y) +i(k_{1_\perp}^x k_{2_\perp}^y+k_{1_\perp}^y k_{2_\perp}^x)\bigg]\mathcal{M}_{-+}  \\ 
    \qquad\qquad\qquad\qquad\qquad\qquad\qquad\qquad\  (J^P_z=+2^+)\\ 
    +\frac{1}{2}\bigg[(k_{1_\perp}^x k_{2_\perp}^x-k_{1_\perp}^y k_{2_\perp}^y) -i(k_{1_\perp}^x k_{2_\perp}^y+k_{1_\perp}^y k_{2_\perp}^x)\bigg]\mathcal{M}_{+-} \\ 
    \qquad\qquad\qquad\qquad\qquad\qquad\qquad\qquad\  (J^P_z=-2^+),     
  \end{cases}\label{eq:helicity_level}
  \end{split}
\end{align}
where $k_{1\perp}$ and $k_{2\perp}$ are the transverse momenta of photon 1 and 2, respectively, $\mathcal{M}_{\pm \pm}$ corresponds to the $\gamma(\pm) \gamma(\pm) \to X$ helicity amplitude, and $V_{\mu\nu}$ is the $\gamma\gamma \to X$ vertex. 
Since the photon is a spin-1 particle, two real photons can combine to form states with $J^P_z~=~0^+,0^-,+2^+,-2^+$ which are formed from admixtures of the individual helicity amplitudes.

\subsubsection{Observation of the Non-linear Breit-Wheeler Process at the SLAC E-144 Experiment}

The linear Breit-Wheeler process requires a minimum of two photons, since a single photon cannot overcome the invariant-mass threshold needed to produce a real electron-positron pair. 
However, when two sufficiently high-energy photons are not available for the linear Breit-Wheeler process, the non-linear Breit-Wheeler process can still be achieved through the fusion of multiple lower-energy photons:
\begin{equation}
    \gamma + n\gamma \rightarrow l^+l^-,
\end{equation}
where, in this case, one photon interacts with $n$ additional photons to produce a lepton anti-lepton pair. 
This process allows photons that do not satisfy the energy threshold of the linear Breit-Wheeler process to combine, and collectively overcome the threshold for lepton pair creation. 

The E-144 experiment at the Stanford Linear Accelerator Complex (SLAC) was built with the express purpose of exploring QED in the strong field regime~\cite{PhysRevLett.79.1626}. 
The E-144 experiment, illustrated in Fig.~\ref{fig:e144_schematic}, orchestrated a collision between a low-emittance 46.6 GeV electron beam and a (green) 527 nm wavelength pulse from a terawatt Nd:glass laser. 
With this setup they achieved a peak laser intensity of $\sim1.3\times10^{18}$ W/cm$^2$ corresponding to a value of $0.3$ in the parameter $\Upsilon$, where $\Upsilon = E^\star_{\rm rms} / E_c$, $E^\star_{\rm rms}$ is the rms electric field strength of the laser in the electron rest frame, and $E_c$ is the QED critical field strength for electron-positron production given in Eq.~\ref{eq:crit}.

After demonstrating their sensitivity to nonlinear QED effects via measurements of nonlinear Compton back-scattering~\cite{bulaPreliminaryObservationNonlinear1996,PhysRevD.60.092004}, the E-144 experiment made a discovery observation of $106 \pm 14$ positrons produced in the multiphoton Breit-Wheeler process~\cite{PhysRevLett.79.1626}. The observed positrons resulted from a two-step process used to achieve the multiphoton Breit-Wheeler process. In the first step of this process, an electron from the 46.6 GeV electron beam undergoes nonlinear Compton scattering by which laser photons (denoted $\omega$) are back-scattered, resulting in a single GeV energy photon ($\gamma$):
\begin{equation}
    \label{eq:nonlinear_compton}
    e + n\omega \rightarrow e^\prime + \gamma.
\end{equation}
Next, as the high-energy photon (maximum energy of 29.2 GeV for Compton-backscattering from a 46.6 GeV electron beam) travels through the laser field it may interact with additional photons, overcoming the energy threshold needed to produce an electron-positron pair:
\begin{equation}
    \label{eq:e144_bw}
    \gamma + n\omega_0 \rightarrow e^+ + e^-,
\end{equation}
where, $\omega_0$ denotes the 527 nm laser photons with an energy of $\hbar\omega_0 = 2.35$ eV, and where $\gamma$ denotes the GeV-energy photon produced in step 1 (Eq.~\ref{eq:nonlinear_compton}). Since the Compton scattering process leads to numerous electrons detected in the electron calorimeter (ECAL), the process was identified through the detection of positrons in the  positron calorimeter (PCAL). The measurement was conducted for laser-on and laser-off trials to establish the background level of positron production. Comparison of the two cases demonstrates a clear excess in positron production for the laser-on case, compared to the laser-off case, as shown in Fig.~\ref{fig:e144_results}. Figure~\ref{fig:e144_results_signal} shows the measured positron momentum spectrum, determined from the laser-on spectrum minus the normalized laser-off spectrum. The result is in good agreement with model calculations of the multi-photon Breit-Wheeler process, in terms of both the overall cross section and the differential distribution~\cite{SLACTridentQEDPRL2010}. Since the E-144 experiment identified only the positron from the process, and was not able to identify the electron counterpart, they did not obtain full information from the process. In contrast, the recent measurement of the linear Breit-Wheeler process by STAR~\cite{PhysRevLett.127.052302} included precise measurement of both the produced electron and positron, allowing polarization dependent effects to be observed. Further details of the STAR measurement will be discussed in Sec.~\ref{sec:hic_exp}. In the next subsection, the effects due to the polarization of photons will be discussed in more detail. 

\subsection{Vacuum Birefringence}
\label{sec:intro_vb}
In the 1930s, Dirac~\cite{doi:10.1098/rspa.1928.0023}, Euler, Heisenberg~\cite{heisenbergFolgerungenAusDiracschen1936}, and Weisskopf~\cite{weisskopfElectrodynamicsVacuumBased1936} all explored the lowest-order corrections to classical electromagnetism due to virtual electron-positron pairs. The Euler-Heisenberg Effective Lagrangian describes the \textit{non-linear} dynamics of electromagnetic fields – a clear deviation from the classical Maxwell theory of electromagnetism, which is manifestly linear. Later, Schwinger~\cite{PhysRev.82.664} derived the effective Lagrangian in the proper-time integral within QED~\cite{PhysRev.76.790}. 
The work of Euler, Heisenberg, and Schwinger showed that the presence of a background field polarizes the vacuum by interacting with the short-lived virtual particles that are continuously and nearly instantaneously created and annihilated. 

\begin{figure}
    \centering
    \includegraphics[width=0.60\linewidth]{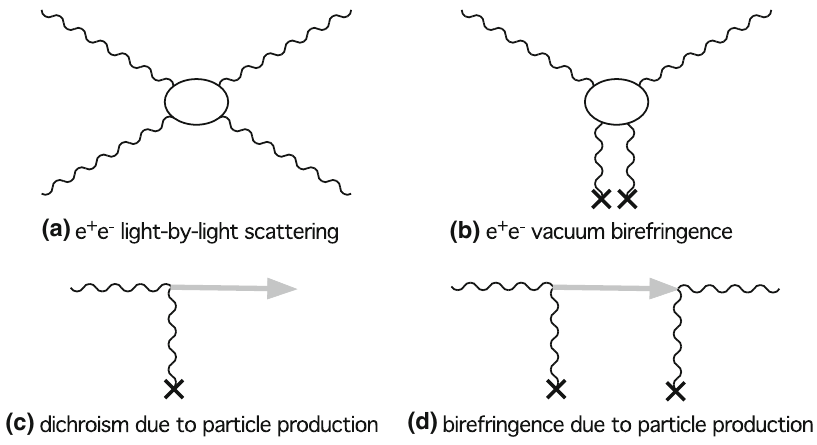}
    \caption{ The lowest order elementary processes contributing to vacuum birefringence and dichroism. Reproduced from Ref.~\cite{2016EPJC7624D}. }
    \label{fig:vb_pvlas_diagrams}
\end{figure}

Toll carried out pioneering work to understand vacuum polarization in the strong-field regime of QED~\cite{1952PhDT21T}. The essential insight in his work is that the vacuum itself behaves as a polarizable and magnetizable medium, which can lead to birefringence (i.e., double refraction) of light as it passes through a region of space in the presence of an external magnetic field.
This is known as vacuum magnetic birefringence (VMB), a phenomenon in which the QED vacuum gains distinct refractive indices, different from unity, for each polarization mode of the photon~\cite{baierVacuumRefractionIndex1967,heisenbergFolgerungenAusDiracschen1936}.
The effects of VMB can be concisely summarized as~\cite{EJLLI20201}:
\begin{itemize}
    \item Maxwell's equations are no longer linear and the superposition principle is violated;
    \item in vacuum Light-by-Light scattering can occur and the velocity of light is $v_{\rm light} < c$ in the presence of other electromagnetic fields;
    \item electromagnetism in vacuum is described by Maxwell’s equations in a medium.
\end{itemize}
In birefringent media, the strength of the birefringence effect is often characterized by the difference in refractive indices $(\Delta n)$ for light polarized parallel $(n_\parallel)$ vs. perpendicular $(n_\perp)$ to some axis:
\begin{align}
    \Delta n &= n_\parallel - n_\perp.
\end{align}

Vacuum birefringence is a purely quantum effect that becomes appreciable as the background field strength approaches the critical field strength, growing with the square of the magnetic field strength according to the Cotton-Mouton\footnote{The Cotton-Mouton effect is the magnetic analog to the Kerr effect} relation:
\begin{equation}
    \Delta n \equiv k_{CM,vac} B^2,
\end{equation}
where $k_{CM,vac}$ is the Cotton-Mouton coefficient for the vacuum. 

An optical medium can, in general, permit both transmission and absorption of electromagnetic waves. Both the transmission and absorption processes can be described by generalizing the (real valued) index of refraction ($n$) to a complex index of refraction ($\Tilde{n}$):
\begin{equation}
    \Tilde{n} = n + i \kappa,
\end{equation}
where $n$ and $\kappa$ are real-valued. In the complex index of refraction, the transmission process is characterized by the real part ($n$) while the absorption process is characterized by the imaginary part ($\kappa$). In general, a birefringent medium may have both $\Delta n \neq 0$ and $\Delta \kappa \neq 0$. When $\Delta \kappa \neq 0$, the medium's opacity varies with respect to the incident polarization. While the term, ``birefringent'' entails both the transmission and absorption phenomena, historically, vacuum magnetic birefringence has been used to describe primarily the transmission phenomena. The polarization-dependent absorption phenomena characterized by $\Delta \kappa \neq 0$ is known as vacuum magnetic dichroism~\cite{PhysRev.136.B1540,Heyl_1997}. Photon splitting, though forbidden for fewer than three couplings to the internal fermion loop, can also lead to a nonzero value of $\Delta\kappa$ and dichroism~\cite{PhysRevLett.25.1061,PhysRevD.2.2341,ADLER1971599}. The connection between vacuum birefringence, vacuum dichroism, and the Breit-Wheeler process will be discussed further in Sec.~\ref{sec:theory_hic} and Sec.~\ref{sec:discussion}.

Vacuum polarization was confirmed with the groundbreaking discovery of the Lamb shift, a purely quantum mechanical effect that leads to the splitting of the Hydrogen $^2S_{1/2}$ and $^2P_{1/2}$ energy levels~\cite{PhysRev.72.241}. 
However, vacuum birefringence has been much more difficult to confirm experimentally.  Terrestrial experiments are generally limited by the difficulty in obtaining sufficiently strong electromagnetic fields over macroscopic lengths, while cosmological experiments suffer from experimental uncertainties related to the photon sources and polarizing electromagnetic fields~\cite{EJLLI20201}.

\subsubsection{Earth and Space based searches for Vacuum Birefringence}

\begin{figure}
    \centering
    \includegraphics[width=0.60\textwidth]{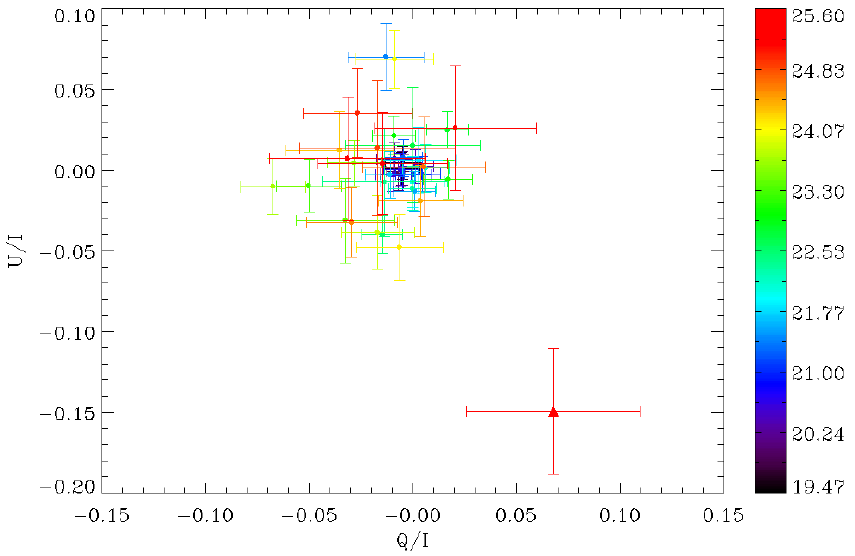}
    \caption{Plot of the normalized Stokes parameters U/I and Q/I for a selected sample of 42 field stars (filled point; top left) and for RX J1856.5-3754 (filled triangle, bottom right). 
    The values of U/I and Q/I plotted here are in unity
    fractions and, therefore, so would be the value of the degree of
    polarization computed from them. The color scale on the right corresponds to the stars' V-band magnitude. Reproduced from Ref.~\cite{mignaniEvidenceVacuumBirefringence2017}.}
    \label{fig:stokes_rx}
\end{figure}

Besides heavy-ion collisions (HICs)~\cite{kharzeevChiralMagneticVortical2015,kharzeevEffectsTopologicalCharge2008a,brandenburgMappingElectromagneticFields2021a}, no other terrestrial sources are known to create electromagnetic fields strong enough to experimentally realize VMB. 
However, some astronomical objects, such as magnetars, a class of neutron stars, are believed to produce magnetic fields on the order of $10^{12-15}$ Gauss~\cite{doi:10.1146/annurev-astro-081915-023329}. 
The presence of such strong fields over macroscopic lengths would be sufficient to make such quantum effects visible.
One of the expected signatures of vacuum birefringence acting on surface emission from magnetars is the apparent increase in polarization of light observed by a distant observer~\cite{heylHighenergyPolarizationlimitingRadius2003, tavernaPolarizationNeutronStar2015}. 
As shown in Fig.~\ref{fig:stokes_rx}, recent optical polarimetric measurements of the isolated neutron star RX J1856.5-3754 were consistent with such an increase and may be evidence for vacuum birefringence~\cite{mignaniEvidenceVacuumBirefringence2017}. However, due to the low significance of the result and uncertainties in the neutron star models and the direction of the neutron magnetization axis, the measurement from RX J1856.5-3754 is not able to confirm VMB unambiguously~\cite{EJLLI20201}.

\begin{figure*}
    \centering
    \includegraphics[width=0.89\linewidth]{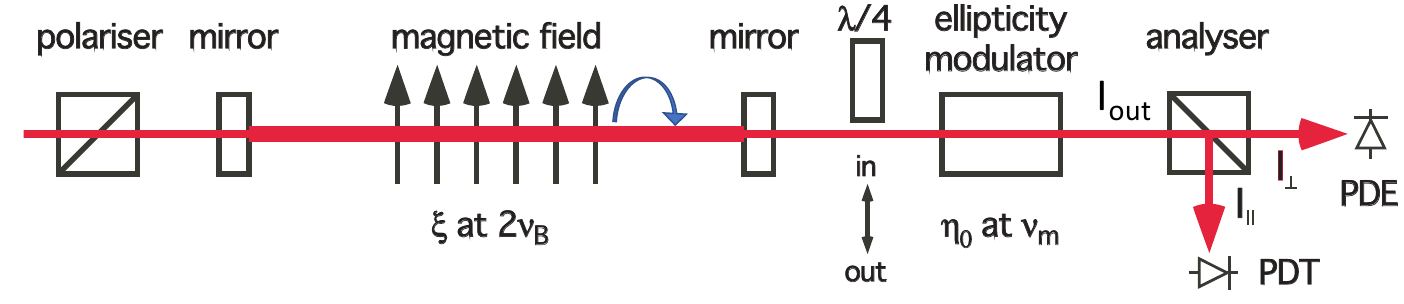}
    \caption{ Schematic of the PVLAS polarimeter experiment. A rotating magnetic field between the cavity mirrors generates a time dependent ellipticity. Reproduced from ~\cite{EJLLI20201}. }
    \label{fig:pvlas_schematic}
\end{figure*}

\begin{figure}
    \centering
    \includegraphics[width=0.60\textwidth]{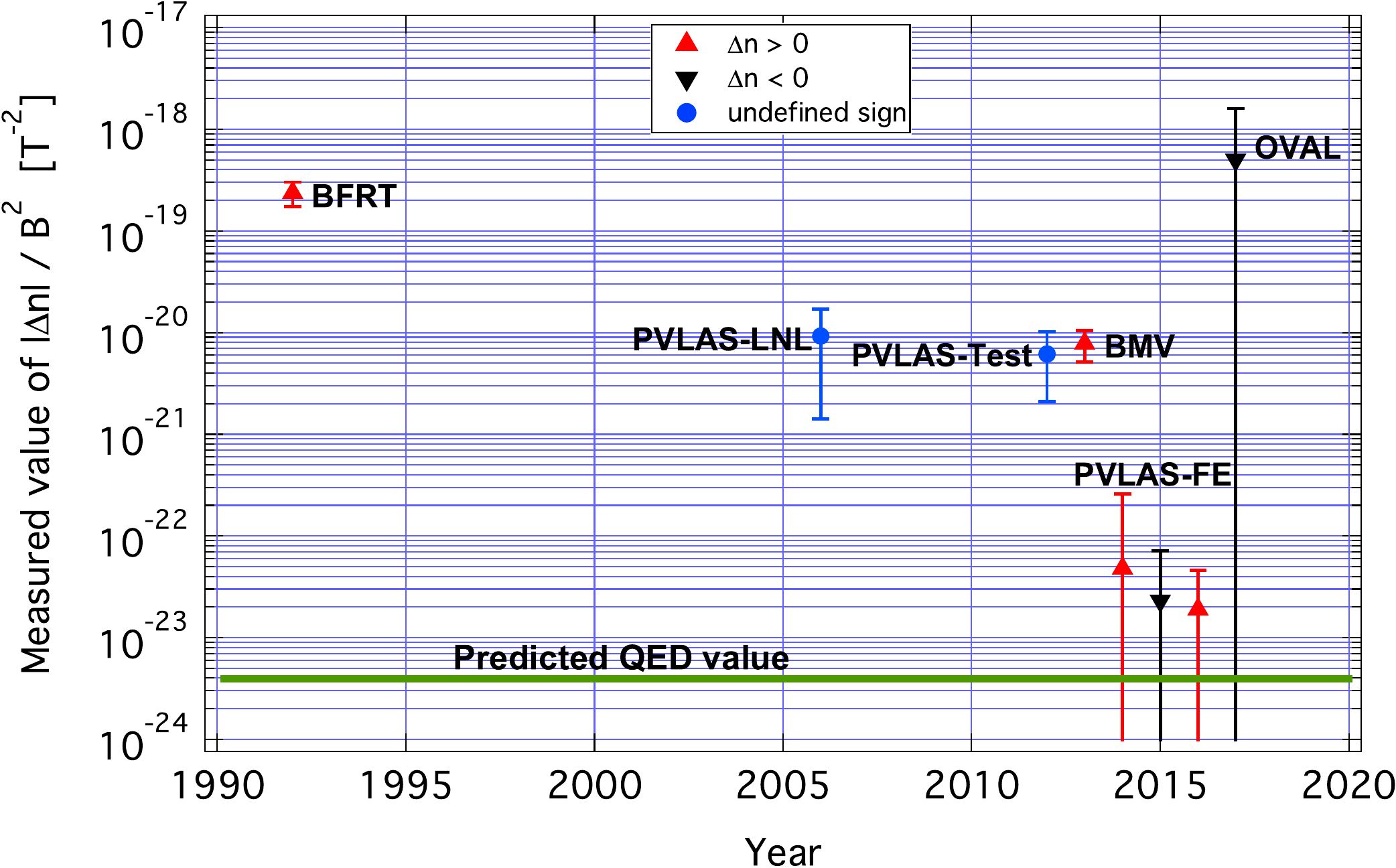}
    \caption{ Historical time evolution of the measurement of vacuum magnetic birefringence normalized to $B_{\rm ext}^2$. Error bars correspond to $1\sigma$ uncertainties. Values are taken from the following references: BFRT~\cite{PhysRevD.47.3707}, PVLAS-LNL~\cite{PhysRevD.77.032006,PhysRevD.78.032006}, PVLAS-Test~\cite{doi.10.1088/1367-2630/15/5/053026}, BMV~\cite{doi.10.1140/epjd/e2013-40725-9}, PVLAS-FE~\cite{doi.10.1140/epjc/s10052-015-3869-8,PhysRevD.90.092003}, and OVAL~\cite{fanOVALExperimentNew2017}. Reproduced from Ref.~\cite{EJLLI20201}.   }
    \label{fig:vb_pvlas_evo}
\end{figure}

Discovery of vacuum birefringence has also been sought through terrestrial experiments.
Such experiments are generally set up to search for a change in the polarization of light, since linearly polarized light may acquire an elliptical component after traveling though a birefringent medium. 
The PVLAS (Polarisation of Vacuum with LASer) experiment, shown schematically in Fig.~\ref{fig:pvlas_schematic}, is one such experiment that has employed high-powered lasers, strong magnetic fields, and an optical Fabry-Pérot cavity to search for VMB~\cite{2016EPJC7624D}. 
Their progress over the last 25 years and final results, which set the strongest limits on vacuum birefringence, are summarized in a comprehensive review~\cite{EJLLI20201}. 
Despite improvements in the measurement sensitivity over the years, the PVLAS experiment has not been able to demonstrate feasibility for performing a definitive measurement of 
VMB, as illustrated in Fig.~\ref{fig:vb_pvlas_evo}. 
On the other hand, their results have allowed exclusion of model-independent parameter space of axion-like and milli-charged particles, since the existence of such particles are expected to increase the VMB effect above that predicted for standard model particles alone. 

The Observing VAcuum with Laser (OVAL) experiment uses a similar apparatus as that of PVLAS, but with stronger, pulsed magnetic fields. The OVAL experiment has recently completed a calibration measurement to demonstrate the feasibility of observing VMB~\cite{fanOVALExperimentNew2017}. In addition to PVLAS and OVAL, there are other similar experiments that employ strong lasers and high magnetic fields~\cite{cadeneVacuumMagneticLinear2014}. As of 2022, none of these experiments have achieved the sensitivity required to observe VMB, though advances in laser power are expected to allow more stringent tests of QED in similar terrestrial experiments in the coming decade~\cite{PhysRevLett.119.250403,HEINZL2006318,Shen_2018}. To this end, significant theoretical progress has been made to identify the signatures of VMB in various experimental laser setups~\cite{denisovNonperturbativeQEDVacuum2017,PhysRevA.82.011803}. With all of the essential pieces in place and rapid progress being made, the next decade is sure to be an exciting time for laser-based QED experiments. 

\subsection{Context and Structure of this Report}
The STAR Collaboration recently reported observation of the linear Breit-Wheeler process and a novel vacuum polarization effect related to vacuum birefringence.  
This article presents the state of affairs, as of 2022, with special emphasis given to the discovery claims made by the STAR Collaboration (see Section~\ref{sec:discussion}) and the various challenges to such claims that have been raised (Section~\ref{sec:key_issues}) in past literature. 
While many reviews on the subject of electromagnetically produced lepton pairs in heavy-ion collisions exist\cite{BERTULANI1988299,BAUR20071,BALTZ20081,doi.10.1146/annurev-nucl-030320-033923}, we find that certain key issues and their physical interpretations are discussed in conflicting ways. 
We list these key issues and questions in Sec.~\ref{sec:key_issues}.
One of the primary goals of this Report is to rectify elements in the literature that have led to confusion and contradictions relevant to the Breit-Wheeler process and vacuum birefringence, specifically with respect to observations made in high-energy heavy-ion collisions. 
To this end, the major portion of the article is devoted to a review of the theoretical formulation (Sec.~\ref{sec:theory}) of these processes in heavy-ion collisions and the experimental progress that has been made over several decades (Sec.~\ref{sec:exp_elec_pos}-\ref{sec:hic_exp}), culminating in the STAR Collaboration's recent observations. 
We then continue with a discussion directly addressing the challenges and issues raised in Sec.~\ref{sec:key_issues}.
Finally, we close the article with a consideration of open problems of interest, perspectives for the future of the field, and with a few final conclusions.

\section{Key Issues}
    \label{sec:key_issues}
In this section, we summarize key issues and difficulties regarding the interpretation of photon-photon measurements from high-energy particle and heavy-ion collisions. We list a few important questions with respect to: a) the Breit-Wheeler process, and b) vacuum magnetic birefringence that will be addressed throughout the remainder of this Report.
\subsection{The Breit-Wheeler Process}
\label{sec:key_issues_bw}
Over the last many decades, high-energy particle colliders have made the creation of matter in the fusion of virtual photons a commonplace event~\cite{PhysRevSTAB.17.051003,Nisius:1999cv,Berger:2014rva}. In contrast, the Breit-Wheeler process has evaded observation for nearly a century. The essential difference is that the Breit-Wheeler process is the collision of real photons, not short-lived virtual quanta. Therefore, evaluating the recent claim made by the STAR collaboration that the Breit-Wheeler process has been observed in HICs requires evaluating the following key questions:
\begin{itemize}
    \item Do the highly-Lorentz contracted fields produced in heavy-ion collisions provide a valid source of photons for the Breit-Wheeler process?
    \item In heavy-ion collisions, can the Breit-Wheeler process be isolated from background?
    \item Are higher order effects present, and if so, are they inseparable from the lowest-order process (the Breit-Wheeler process)?
    \item Has the Breit-Wheeler process already been observed in HICs?
\end{itemize}

\subsection{Vacuum Birefringence}
\label{sec:key_issues_vb}
STAR's recent observation of polarization dependence in $\gamma\gamma \rightarrow e^+e^-$ production is a newly observed vacuum polarization effect. However, the relationship between this newly observed effect and vacuum birefringence is not straightforward since the experimental measurement is not directly comparable to traditional birefringence or dichroism experimental designs. In order to clarify the connection, we seek to answer the following questions:
\begin{itemize}
    \item What is novel about the STAR measurement, i.e., how does it differ from the SLAC E-144 measurement, with respect to polarization effects?
    \item How is the azimuthal angle $(\phi)$ modulation observed by STAR related to vacuum birefringence and/or dichroism?
\end{itemize}

The following sections of this Report will focus on addressing these questions through a review of the theoretical (Sec.~\ref{sec:theory}) and experimental (Sec.~\ref{sec:exp_elec_pos}-\ref{sec:hic_exp}) progress made over the  past several decades. 
Then, based on the review of theoretical and experimental progress, Sec.~\ref{sec:discussion} will revisit and address each of the key issues and questions listed here.

\section{ Theoretical Formulation of Photon Sources and Lowest-Order Polarization Effects }
\label{sec:theory}

It was Fermi in 1924 who first described electromagnetic fields in terms of their equivalent photon spectrum~\cite{1924ZPhy29315F}. 
The central idea in this equivalence is that the Lorentz-contracted electromagnetic fields of a fast-moving charged particle appear as a radial electric and circular magnetic field. 
At a distant point, such a field configuration resembles the electromagnetic wave carried by a real photon. 
Based on this insight, Williams and Weizsäcker later developed the method of equivalent photons, whereby the number of photons $(n)$ with a given energy $\omega$, may be related to the Fourier transform of the time-dependent electromagnetic field~\cite{weizsaeckerAusstrahlungBeiStoessen1934,PhysRev.45.729}. 
This concept has been employed to describe a wide range of physical phenomena in high-energy particle collisions. 
In this section we briefly review the application of the equivalent photon approximation (EPA) giving special attention to the question: under what conditions, if any, do WW photons from highly-Lorentz contracted Coulomb fields provide a valid and viable source for achieving the Breit-Wheeler process?
We then close this section with a brief look at the connection between the polarization dependent aspects of the Breit-Wheeler process and vacuum birefringence and dichroism.

\subsection{ Theoretical Description of Equivalent Photons from High-Energy Electron Beams}
\label{sec:theory_particle}
The scattering of a high-energy electron beam can be described in terms of a virtual photon spectrum characterized by photon energy $\omega$ and invariant-mass squared of the four-momentum transfer $q^2 = -Q^2$. This paradigm is described in great detail in Ref.~\cite{BUDNEV1975181} and summarized by the Particle Data Group report on cross section formula~\cite{PDGZyla:2020zbs}, in which the number density of virtual photons for an electron (positron) beam with energy $E$ is given by:
\begin{equation}
    \label{eq:photon_pdg}
    dn = \frac{\alpha_{em}}{\pi}\left[  1 - \frac{\omega}{E} + \frac{\omega^2}{E^2} - \frac{m^2_e \omega^2}{Q^2 E^2} \right] \frac{d\omega}{\omega} \frac{ dQ^2 }{ Q^2 }.
\end{equation}
This description of the equivalent photon spectrum resulting from a high energy electron beam has proven a convenient tool for many calculations over the decades even though its applicability has some known limitations\cite{BUDNEV1975181}.  

In this formulation, the cross section for a process of the form $e^+e^- \rightarrow e^+e^- + X$ can be computed by the convolution of the photon number density with the appropriate photoproduction cross section $( \sigma_{\gamma\gamma\rightarrow X} )$, i.e. $d\sigma_{X} (s) = dn_1dn_2 d\sigma_{\gamma\gamma\rightarrow X}(W^2)$,
where $W$ is the invariant mass of the produced system $X$. The total production cross section can be obtained by integrating over the $Q^2$ range from $Q^2_{\rm min} = m^2_e \omega_i^2/[E_i ( E_i - \omega_i )]$ to $Q^2_{\rm max}$, which is determined by the details of the  produced system.

The resulting total cross sections for the production of various final states can be found in Ref.~\cite{PDGZyla:2020zbs}. 
The minimum cutoff is especially important, as the photon flux shown in Eq.~\ref{eq:photon_pdg} diverges for $Q^2\rightarrow 0$. 
For this reason, electron beams cannot provide a source of real photons which interact without significant effects due to finite virtuality. 
Therefore, electron beams have never been considered a viable source of photons for the Breit-Wheeler process, even though they have provided valuable experimental information about virtual photon interactions, even in the limit of $Q^2 \rightarrow 0$~\cite{abbiendi_total_2000}.

\subsection{ Theoretical Description of Equivalent Photons in Heavy-Ion Collisions}
\label{sec:theory_hic}
        
The photon flux provided by the Lorentz-boosted Coulomb field of an ultra-relativistic heavy nucleus shares some similarities with the virtual photon flux of a high-energy electron beam. However, there are a number of qualitative differences between the two cases. One important difference is due to the point-like structure of the electron compared to the diffuse charge distribution within a heavy nucleus. Another practical difference arises from the significantly heavier mass of highly-charged nuclei, making them less easily deflected. The implications of these differences will be further expounded below.
In the EPA, two photon interactions are computed by factoring the calculation into a semi-classical and a quantum component. The equivalent flux of photons is dealt with in terms of an external classical field, while the quantum part of the calculation deals with the elementary cross section for the $\gamma\gamma \rightarrow X$ cross section (where, e.g., $X = e^+e^-$ for the Breit-Wheeler process).

According to the EPA, the number density spectrum of photons with energy $\omega$~\cite{Krauss:1997vr} manifest by the field of a single nucleus is:
    \begin{equation}
    \label{eq:photon_density}
    n(\omega) = \frac{(Ze)^{2}}{\pi\omega}\int_{0}^{\infty}\frac{d^{2}k_{\perp}}{(2\pi)^{2}}\left[\frac{F\left(\left(\frac{\omega}{\gamma}\right)^{2}+\overrightarrow{k}_{\perp}^{2}\right)}{\left(\frac{\omega}{\gamma}\right)^{2}+\overrightarrow{k}_{\perp}^{2}}\right]^{2}\overrightarrow{k}_{\perp}^{2},
    \end{equation}
where $Z$ is the nuclear charge number, $\gamma$ is the Lorentz factor, $\overrightarrow{k}_{\perp}$ is the photon transverse momentum, 
and $F(...)$ 
is the nuclear electromagnetic form factor. The nuclear electromagnetic form factor for a spherically symmetric nucleus can be obtained via the Fourier transform of the charge distribution as:
    \begin{equation}
    \label{eq:formFactor}
    F(k^{2}) = \int d^{3}re^{ik.r}\rho_{A}(r).
    \end{equation}
Generally, a Woods-Saxon distribution is assumed to describe the charge distribution of heavy nuclei ~\cite{doi.10.1103/PhysRev.95.577} without any fluctuations or point-like structures:
    \begin{equation}
    \label{eq:charge_density}
    \rho_{A}(r)=\frac{\rho_{0}}{1+\exp[(r-R_{\rm{WS}})/d]},
    \end{equation}
where the radius $R$ and skin depth $d$ are based on fits to low energy electron scattering data such that all deformations are assumed to be higher order and are ignored~\cite{doi.10.1016/0092-640X(87)90013-1}, and $\rho_{0}$ is the normalization factor for the distribution. Since the Fourier transform of the Woods-Saxon distribution does not have an analytic form, it is commonly approximated with a hard sphere, with radius $R_A$, convolved with a Yukawa potential with range $a$~\cite{daviesCalculationMomentsPotentials1976}. The resulting charge distribution form factor provides a good approximation to the Woods-Saxon distribution and has an analytical form. See Ref.~\cite{PhysRevC.60.014903} for a direct comparison between this form factor and numerical calculations from the exact Wood-Saxon distribution.

With these, the cross section for the polarization averaged two-photon process in heavy-ion collisions can be computed as
    \begin{align}
    \label{eq:two_lepton_xs}
    \begin{split}
    & \sigma_{\mathrm{A + A}\rightarrow\mathrm{A + A} + l^+l^-}  = \\ & \int  d\omega_1 d\omega_2  n_1(\omega_1) n_2(\omega_2) \times \sigma_{\gamma\gamma\rightarrow l^+l^-}(W; m), 
    \end{split}
    \end{align}
where $W$ is the invariant mass of the produced lepton pair and $n_1(\omega_1)$ and $n_2(\omega_2)$ are the equivalent number of photons with energies $\omega_1$ and $\omega_2$ from the field of nucleus 1 and 2, respectively. 
In the next sections, we review various theoretical approaches for computing the photon-photon fusion process in HICs. The following subsections review various theoretical approaches: 1) the traditional EPA approach, 2) lowest-order QED, 3) Wigner quasi-probability distributions, and 4) the exact solution to the time-dependent Dirac equation. Then we discuss the issue of photon virtuality and present a well-defined criterion for the domain of applicability of the Breit-Wheeler process in heavy-ion collisions. Finally, the potential for higher order effects are discussed.

\subsubsection{Traditional EPA Calculations}
The above description provides a fairly direct and straightforward method for calculating the total cross section of the Breit-Wheeler process in heavy-ion collisions. However, the differential cross sections reflecting the detailed kinematic distributions of the produced electron and positron are less straightforward to compute. 
When the pair transverse momentum, $P_\perp$, is small compared to the lepton pair invariant mass $W$, the photon energies are related to the lepton pair invariant mass and rapidity $y$ as
\begin{equation}
    \omega_{1,2} = \frac{W}{2}e^{\pm y},
\end{equation}
and
\begin{equation}
    y=\frac{1}{2}\text{ln}\frac{\omega_{1}}{\omega_{2}}.
\end{equation}
Furthermore, the traditional EPA calculations determine the photon transverse momentum $(k_\perp)$ distribution using the so-called $k_\perp$-factorization method (throughout this review we use the term ``$k_\perp$-factorization'' as defined in Refs. ~\cite{Klusek-Gawenda:2018zfz,Mazurek:2021ahz}). In this approach the one-photon distribution is integrated over all transverse distances, i.e. $0 < b_\perp < \infty$ to obtain the $k_\perp$ distribution (Eq.~16 of Ref.~\cite{doi.10.1016/j.cpc.2016.10.016}) of the interacting photons. From this, the transverse momentum of the produced pair is determined via a direct sum of the two colliding photon momenta. Some experiments cannot measure the transverse momentum accurately and instead characterize the kinematics of the process via the accoplanarity, which is straightforwardly related as $\sqrt{2}P_{\perp}\simeq\pi\omega\alpha/2$~\cite{ATLAS:2018pfw,ZHA2020135089}.

While the above treatment provides an adequate description of the processes at the level of the total cross section, it suffers several shortcomings when describing the differential distributions. However, this traditional EPA approach was able to successfully described experimental results for nearly two decades before its shortcomings were fully uncovered. Section~\ref{sec:hic_exp} will review in detail the experimental progress that has challenged the traditional EPA description of the two photon process and encouraged a more rigorous theoretical treatment. 

\subsubsection{ Lowest-order QED }
Pair creation via the two photon interaction at lowest order can be depicted as a process with two Feynman diagrams contributing, as shown in Fig.~\ref{fig:bw_diagrams}. There is an approximation commonly used for describing events: that of external fields generated by nuclei that are undeflected by the collision and travel along straight-line trajectories. 
Following the derivation of Ref.~\cite{PhysRevA.51.1874,PhysRevA.55.396}, the cross section for pair production of leptons is given by
\begin{equation}
\label{eq:cross_section_pair}
    \sigma = (Z\alpha)^{4} \frac{4}{\beta^{2}} \frac{1}{(2\pi)^{6}2\epsilon_{+}2\epsilon_{-}} \int d^{2}q \frac{d^{6}P(\vec{q})}{d^{3}p_{+}d^{3}p_{-}} \int d^{2}b e^{i {\vec{q}} \cdot  {\vec{b}}},
\end{equation}
where $b$ is the nucleus-nucleus impact parameter and $p_{+}$($\epsilon_{+}$) and $p_{-}$($\epsilon_{-}$) are the momentum (energy) of the created leptons. The full form of the differential probability $P(\vec{q})$ for QED at the lowest order is given in Ref.~\cite{ZHA2020135089}. The differential probability depends on $F(...)$, the nuclear electromagnetic form factor, $p_{+}$ and $p_{-}$, the mass of the produced leptons, and the photon momenta $q_1$ and $q_2$ where the longitudinal components of $q_{1}$ are given by $q_{10} = \frac{1}{2}[(\epsilon_{+} + \epsilon_{-}) + \beta(p_{+z}+p_{-z})]$, $q_{1z} = q_{10}/ \beta$. 
Unlike the $k_\perp$-factorization method, the lepton pair kinematics computed in lowest order QED clearly depends on the nucleus-nucleus impact parameter, evident in the $\int d^{2}b e^{i {\vec{q}} \cdot  {\vec{b}}}$ term.
In order to compute results at all impact parameters, where in general no simple analytical form is available, the multidimensional integration should be performed with numerical algorithms such as the VEGAS Monte Carlo integration routine~\cite{peterlepageNewAlgorithmAdaptive1978}. 

\subsubsection{Photon Wigner Function Formalism}
The Wigner quasi-probability distribution, often shortened as the Wigner function, is a method of relating a system's quantum wave function to a quasi-probability function in phase space~\cite{PhysRev.40.749,Hillery1984,caseWignerFunctionsWeyl2008}. The Wigner function is similar to a classical probability distribution, in that it describes the phase space distribution of a process. The Wigner function is a quasi-probability distribution because, unlike its classical analog, it does not satisfy all properties of a probability function. The most obvious difference is that Wigner functions can take on negative values for states of the quantum system that have no classical analogue, e.g., states characterized by quantum interference of wave functions. 
To our knowledge, the authors of Ref.~\cite{PhysRevD.101.034015} first suggested that the full spatial dependence of the $\gamma\gamma$ processes, including the impact parameter dependence, may be expressed in terms of photon Wigner distributions. 

Since then, several groups have investigated the photon Wigner function (PWF) formalism for describing the $\gamma\gamma$ processes in heavy-ion collisions~\cite{PhysRevD.102.094013,PhysRevD.104.056011,KLUSEKGAWENDA2021136114}.
Following the notation of Ref.~\cite{KLUSEKGAWENDA2021136114}, the photon Wigner function can be expressed as
\begin{align}
  \begin{split}
    \label{eq:Wigner}
    N_{ij} (\omega,b,q) & = \int {d^2 Q \over (2 \pi)^2} \exp[-i b Q] \\ 
    & \times E_i \Big(\omega, q + {Q \over 2} \Big) E^*_j \Big(\omega, q - {Q \over 2} \Big) \, 
    \\
    & = \int d^2s \,  \exp[i q s] \\ 
    & \times E_i \Big(\omega, b + {s \over 2} \Big) E^*_j \Big(\omega, b - {s \over 2} \Big),
  \end{split}
\end{align}
where $E_{i,j}$ are the electric field vectors expressed in terms of the nuclear charge form factors ($F$, see Eq.~\ref{eq:formFactor}) as
\begin{eqnarray}
    E (\omega,q) = Z \sqrt{\frac{\alpha_{em}}{\pi}} \,  {q F(q^2+q^2_\parallel) \over q^2 + q^2_\parallel} \,
\end{eqnarray}
where, $q_\parallel = {\omega / \gamma}$.

By virtue of the Wigner function definition, Eq.~\ref{eq:Wigner} is a function of both the spatial location ($b$) and transverse momentum ($q$) coordinates.
The standard formula for the photon flux in either momentum space or position space may be obtained by integration over $b$ or $q$, respectively. The differential cross section for lepton pair production can then be expressed in terms of the photon Wigner functions as a convolution over the transverse momenta and transverse positions~\cite{KLUSEKGAWENDA2021136114}
\begin{equation}
    \begin{split}
        \label{eq:mat_space}
        {d\sigma \over d^2b d^2P} &= \int d^2b_1 d^2b_2 \, \delta^{(2)}(b-b_1 + b_2)  \\
        &\times \int {d^2q_1 \over \pi} {d^2q_2 \over \pi} \, \delta^{(2)}(P-q_1 - q_2) \\
        &\times
        \int {d \omega_1 \over \omega_1} {d \omega_2 \over \omega_2} N_{ij} (\omega_1,b_1,q_1) N_{kl} (\omega_2,b_2,q_2)   \\ 
        &\times \, \, {1 \over 2 \hat s} \sum_{\lambda \bar \lambda} M^{\lambda \bar \lambda}_{ik} M^{\lambda \bar \lambda \dagger}_{jl} \, d\Phi(l^+ l^-),
    \end{split}
\end{equation}
where the final term (Eq.~\ref{eq:mat_space}) is a sum over the helicity amplitudes for the $\gamma\gamma \rightarrow l^+l^-$ process and $d\Phi(l^+ l^-)$ is the invariant phase space for the leptons~\cite{KLUSEKGAWENDA2021136114}. 

\begin{figure}
    \centering
    \includegraphics[width=0.60\linewidth]{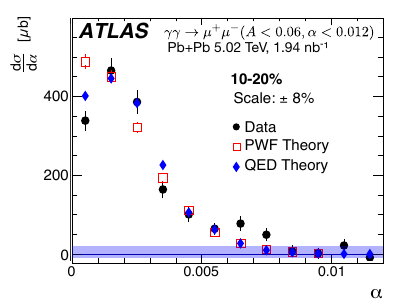}
    \caption{ Differential cross-sections as a function of $\alpha$ for $\gamma\gamma \rightarrow \mu^+\mu^-$ pairs passing the ATLAS fiducial acceptance selection. The data are shown together with theoretical predictions from lowest-order QED (QED) and the photon Wigner function (PWF) formalism. Reproduced from Ref.~\cite{atlascollaborationMeasurementMuonPairs2022}. }
    \label{fig:two_photon_alpha}
\end{figure}

\subsubsection{Exact solution of the time-dependent Dirac equation}
The Breit-Wheeler process in relativistic heavy-ion collisions could also be calculated and related to the solution of the (single-particle) Dirac equation. In the limit of ultra-relativistic nuclei with an appropriate gauge, the expression for the electromagnetic interaction simplifies and the Dirac equation can be solved analytically. In principle, Breit-Wheeler pair production is a problem in relativistic quantum field theory, which involves particle creation and annihilation. In contrast to the similar problem of ionization of an atom by a non-relativistic charged particle, it can be dealt with by the time-dependent Schrödinger equation. The atomic electron can be described by a Hamiltonian $H_{0} = T +e^2/r$ and the problem of the ionization process is to solve the time-dependent Schrödinger equation for the wave function $\Psi$ of the electron
\begin{equation}
\label{Ionization_schrodinger}
i\frac{\partial \Psi}{\partial t}= (H_0 + V(t))\Psi (t),
\end{equation}
where $V(t)$ is a time-dependent perturbation caused by the classical motion of the charged particle. This equation can be solved by a number of well-established methods like perturbation theory~\cite{sakurai_2004}.

In the case of vacuum pair production in relativistic heavy-ion collisions, Eq.~\ref{Ionization_schrodinger} should be modified by the Dirac equation for an electron in the time-dependent external field of the colliding nuclei~\cite{BAUR20071}:
\begin{equation}
\label{pair_production_Dirac}
i\frac{\partial}{\partial t}|\Psi(\vec{r},t\rangle= (H_0 + H_A (t)+H_B (t))|\Psi(\vec{r},t\rangle,
\end{equation}
where $|\Psi(\vec{r},t\rangle$ is the Dirac spinor wave function of the electron, $H_0 = -i \vec{\alpha} \cdot \vec{\nabla} + \gamma^0$ is the free Dirac Hamiltonian and $H_A (t)$ and $H_B (t)$ are the interactions with nucleus A and B respectively, 
\begin{equation}
    \label{pair_production_Dirac_2}
    H_{A,B}(t) = (I_4 \mp \beta\alpha_z) \frac{-Z_{A,B} \alpha}{\sqrt{(\vec{\rho} \mp \vec{b})^{2}/\gamma^{2}+(z \mp \beta t)^{2}}}.
\end{equation}
As derived in Refs.~\cite{Baltz:1998zb,Segev:1997yz}, a gauge transformation can be used to analytically solve the two-center Dirac equation in the ultra-relativistic limit. The expression for the cross section of the Breit-Wheeler process reads
 \begin{equation}
\label{cross_section_Dirac}
\begin{split}
 d\sigma = &\frac{m^{2}d^{3}pd^{3}q}{(2\pi)^{6}\epsilon_{p}\epsilon_{q}}\int \frac{d^{2}k}{(2\pi)^{2}}|F_{B}(\vec{k})|^{2} \times \\
 &|F_{A}(\vec{q}_{\perp}+ \vec{p}_{\perp}-\vec{k})|^{2}|M(\vec{k})|^{2},
 \end{split}
 \end{equation}
 with 
 \begin{equation}
 \begin{split}
  M(\vec{k}) = &\bar{\mu}(p)  \frac{\vec{\alpha} \cdot (\vec{k} - \vec{p}_{\perp})+\gamma_{0}m}{-p_{+}q_{-}-(\vec{k}-\vec{p}_{\perp})^{2}-m^{2}+i\epsilon}\gamma_{-}\mu(-q) +\\
 &\bar{\mu}(p)  \frac{\vec{\alpha} \cdot (\vec{k} - \vec{p}_{\perp})+\gamma_{0}m}{-p_{-}q_{+}-(\vec{k}-\vec{p}_{\perp})^{2}-m^{2}+i\epsilon}\gamma_{+}\mu(-q),
\end{split}
\end{equation}
where $p$ and $\epsilon_p$ ($q$ and $\epsilon_q$) are the momentum and energy of the electron (positron); $\mu(p)$ and $\mu(-q)$ are positive- and negative-energy Dirac spinors; $\vec{\alpha} = \gamma_0 \vec{\gamma}$; $\gamma_{\pm} = \gamma_0 \pm \gamma_z$; $\gamma_\mu$ are the Dirac matrices; $p_{\pm} = \epsilon_p \pm p_z$; $q_\pm = \epsilon_q \pm q_z$; $m$ is the electron mass; $\vec{k}$ is the momentum vector in the transverse plane; and the function $F(\Delta)$ is proportional to the electron eikonal scattering amplitude in the potential $V(\vec{r})$ of the colliding nuclei:
 \begin{equation}
\label{photon_propagator}
\begin{split}
&F(\Delta) = \int d^2 \rho \rm{exp}[-i\vec{\rho} \cdot \Delta]\{\rm{exp}[-i\chi(\rho)] -1\},\\
&\chi(\rho) = \int^{\infty}_{-\infty}dz V(z,\vec{\rho}).
\end{split}
\end{equation}
For the Coulomb potential $V(\vec{r}) = -Z\alpha/r$, the integral in $\chi(\vec{\rho})$ becomes divergent, which needs to be regularized. To match with the known perturbative result, $F(\Delta)$ should be regularized by putting a $\Delta$ cutoff at $\omega/\gamma$, which leads to 
\begin{equation}
\label{photon_propagator_perturbative}
F(\Delta) = \frac{4\pi \alpha Z}{(\Delta^{2}+\omega^{2} / \gamma^{2})}.
\end{equation}

\subsubsection{Photon Virtuality and a Criterion for the Breit-Wheeler Process}
\label{sec:Virtuality}
It is often considered that the transverse momentum of the photons in HICs is related to the transverse dimensions of the nuclei and the photon virtuality, according to the uncertainty principle~\cite{aaboud_evidence_2017}. 
This has been used as an argument~\cite{PhysRevC.70.031902,doi.10.1016/j.cpc.2016.10.016,Esnault_2021,Zhao:2021ynd} that the $e^+e^-$ pair production from HICs is not the Breit-Wheeler process, despite the original proposal from Breit and Wheeler in their seminal paper~\cite{PhysRev.46.1087}. In this section, we follow Ref.~\cite{PhysRevC.47.2308} via the S-Matrix derivation to illustrate the approximation which results in the EPA and propose a well-defined criterion for the Breit-Wheeler process in relativistic heavy-ion collisions to clearly separate it from the Landau-Lifshitz~\cite{Landau1934} and Bethe-Heitler~\cite{betheh.StoppingFastParticles1934} processes involving virtual photons. 

Since the Coulomb field of a nucleus is a pure electric field, the Lorentz boost does not change the fact that real photons cannot be generated by the field itself. 
The resulting quantization as photons from the Poynting Vector would have the form shown in Eq.~\ref{eq:photon_density}, with a spacelike Lorentz vector and a ``negative squared mass'' (imaginary mass) of $-((\omega/\gamma)^2+k_{\perp}^2)$. 
It was argued that if one were to define the process in UPCs as the Breit-Wheeler process, the virtuality would simply have to be ignored. In fact, this is not the case. Simply ignoring the virtuality by setting this term to zero would result in infrared divergence of the photon flux. 
Instead, Eq.~25 from Ref.~\cite{PhysRevC.47.2308} shows the approximation required for the conserved transition current to behave as real photon interactions in the S-Matrix. The requirement in the center-mass-frame of the heavy-ion collision is that both photons satisfy the following condition: 
    \begin{equation}
    \label{equation_BW_criterion1}
    \omega/\gamma\, \lesssim \, k_{\perp}\, <<\, \omega  .   \end{equation}

With this criterion and subsequent omission of the higher order second and third terms (in Eq.~25 from Ref.~\cite{PhysRevC.47.2308}) of the order of $1/\gamma^2$, the vertex function of the two-photon process in relativistic heavy-ion collisions in Eq.~28 of Ref.~\cite{PhysRevC.47.2308} is identical to that of the real-photon interaction, e.g., Eq.~19 of Ref.\cite{PhysRevC.47.2308}.

The interpretation is therefore that the single photon flux of the virtual states from the Lorentz boosted field is given by Eq.~\ref{eq:photon_density} and that the interaction is only relevant for (or behaves as) photons with real-photon states characterized by energy of $\omega$ and transverse momentum of $k_{\perp}$, validating the implementation of the process via the photon Wigner function (PWF) formalism~\cite{PhysRevD.102.094013,PhysRevD.101.034015,KLUSEKGAWENDA2021136114,PhysRevD.104.056011}. 
In this sense, STAR's observation of the linear Breit-Wheeler process~\cite{PhysRevLett.127.052302} is concisely and accurately described as: the linear Breit-Wheeler process has only been recently observed with quasi-real photon sources~\cite{doi.10.1063/5.0086577}.
The form factor (field strength) in the photon flux limits the photon transverse momentum to be $k_{\perp} \lesssim 1/R$. In the regime of much higher $k_{\perp}$ ($k_{\perp}\gtrsim1/R$ and/or $\omega\gtrsim\gamma/R$) there are significant contributions from the ``semi-coherent'' process~\cite{Staig:2010by} due to photons scattering off constituent nucleons and quarks inside the target nucleus, which may invalidate EPA assumptions. This puts a further constraint on the available phase space for the photons that may participate in the Breit-Wheeler process~\cite{Wang:2022ihj}: 
    \begin{equation}
    \label{equation_BW_criterion2}
    \omega/\gamma\, \lesssim\, k_{\perp}\, \lesssim\, 1/R\, \ll\, \omega     \end{equation}
With decreasing beam energy ($\gamma$) in the same kinematic acceptance, the phase space for the Breit-Wheeler process decreases, and we would expect that the photons outside this valid range ($k_{\perp}\lesssim\omega/\gamma$) to contribute substantially to the interaction cross section at low beam energy.  At extremely high energy, there are constraints on the validity of the Breit-Wheeler process as well. 
We note that in addition to the lepton pair momentum, the acoplanarity ($\alpha$) has been used in literature~\cite{ATLAS:2018pfw}. 
The criterion can be readily defined in terms of acoplanarity since it is straightforwardly related as  $\sqrt{2}P_{\perp}\simeq\pi\omega\alpha/2$~\cite{ATLAS:2018pfw,ZHA2020135089}. Therefore, the criterion of the Breit-Wheeler process in terms of acoplanarity reads: 
    \begin{equation}
    \label{equation_BW_criterion3}
    {\frac{\sqrt{2}}{\gamma}} \lesssim{\frac{\pi}{2}}\alpha\lesssim{\frac{\sqrt{2}}{\omega R}}<<1    \end{equation}
For the kinematics of the ATLAS experiment at the LHC~\cite{ATLAS:2018pfw,atlascollaborationMeasurementMuonPairs2022} with $\gamma=2500$ and $\omega \gtrsim 10$ GeV and shown in Fig.~\ref{fig:two_photon_alpha}, the real-photon criterion becomes $4\lesssim k_{\perp}\lesssim30$ MeV or, in terms of acoplanarity, $0.0004\lesssim\alpha\lesssim0.003$. 

\subsubsection{Higher order QED Corrections}
\label{sec:higher_order}
Due to the large charge carried by a heavy ion, the effective coupling $Z\alpha_{em}$ ($\sim$ 0.6 for e.g., gold and lead nuclei) is sufficiently close to unity that higher-order terms may not be negligible. This suggests the potential for higher order effects, which signify the crossover from the perturbative to the nonperturbative regime of QED. In 1954, the pioneering studies~\cite{PhysRev.93.768,PhysRev.93.788} of higher order QED effects were made by Bethe, Maximon and Davies in a similar process -- the Bethe-Heitler process~\cite{PhysRev.93.768} (the photoproduction of electron-positron pairs in a nuclear Coulomb field at rest). Higher order effects were treated using the Sommerfeld-Maue wave functions, which are appropriate solutions of the Dirac equation at high energy. This approach takes higher order effects into account to all orders and can be related to the usual Feynman graph technique~\cite{Ivanov:1998PRD}. 

A sizeable negative correction (compared to the lowest-order result) was found for the Bethe-Heitler formula~\cite{klein.10.1016/j.radphyschem.2005.09.005}. For the Breit-Wheeler process in heavy-ion collisions, the correction might be larger, since the quasi-real photon from the projectile is also attached to the field of the source nucleus, in contrast to the Bethe-Heitler process. It was pointed out by Ivanov, Schiller, and Serbo~\cite{Ivanov:1998ru} that the higher-order effect (or Coulomb correction) of lepton pair production in heavy-ion collisions with photon energy in the range of $m_e<<\omega<<\gamma m_e$ (for example, $0.5<<\omega<<50$ MeV for $e^+e^-$ at RHIC) was analogous to the well-known Bethe-Heitler process on a heavy target~\cite{betheh.StoppingFastParticles1934}. They established the equivalence between the calculations using Sommerfeld-Maue wave function and the standard higher-order QED approach. It is, however, important to realize that most of those Coulomb corrections are performed for photon energy of a few MeV in the rest frame of the target nucleus. Therefore, 
none of the conditions presented in Eq.~4 of Ref.~\cite{Lee:2009vu} for such a correction are satisfied in the situation under discussion in this review. It has been argued and demonstrated that Coulomb corrections in the kinematics relevant to the Breit-Wheeler process under discussion ($\gamma_{1,2}>>1$ and $\gamma/R>>\omega_{1,2}>>1/R$) are vanishingly small ($<1\%$)~\cite{2001NuPhA695395B,Hencken:2006ir,Sun:2020ygb,Ivanov:1998PRD,BAUR20071}. This result is intuitive to understand, since, in the center-of-mass frame, an $e^+e^-$ pair within these kinematics behave as a neutral object in the Coulomb field~\cite{Sun:2020ygb} (i.e., Coulomb effects cancel exactly). On the other hand, it may be possible that such Coulomb corrections are indeed present at the LHC through the dimuon channel~\cite{Zhao:2021ynd,PhysRevC.80.034901} with photon energy of $m_{\mu}<<\omega\simeq10$ GeV $<<\gamma m_{\mu}\simeq250$ GeV, in the expected range for significant Coulomb corrections~\cite{Lee:2009vu} but may be absent in the $e^+e^-$ channel~\cite{ATLAS:2022srr}. In this context of a negligible non-perturbative QED effect, some experts~\cite{BAUR20071} concluded in dismay 15 years ago: ``In April 1990 a workshop took place in Brookhaven with the title ‘Can RHIC be used to test QED?’~\cite{fatyga1990can}. We think that after about 17 years the answer to this question is ‘no’. However, many theorists were motivated to deal with this topic. The gradual progress, which was sometimes quite tortuous...''

Another type of higher order QED effect is internal photon radiation, which leads to the appearance of a tail in the transverse momentum distributions of the produced lepton pairs. 
The authors of Ref.~\cite{PhysRevLett.122.132301} study the QED showering effect utilizing a Sudakov formalism, which qualitatively described the notable tail in acoplanarity distributions observed by the ATLAS collaboration~\cite{PhysRevLett.121.212301,PhysRevD.102.094013}. The QED showering process does not affect the overall cross section, however it can modify the transverse momentum (acoplanarity) distribution of the produced vacuum pair, especially at higher transverse momentum (acoplanarity). There also exists a problem in that calculations via lowest order perturbation theory may violate unitarity~\cite{BERTULANI1988299}. 
Unitarity can be restored by introducing the production of multiple pairs~\cite{PhysRevA.42.5736,Rhoades-Brown:1991txg,Best:1991uw,PhysRevA.55.396}, which serves as an additional source of higher order correction. 
However, this source of higher order corrections is exceedingly negligible in the production phase space discussed in this Report.

\subsection{Photons from Lasers and Other Sources}
    
Many publications in the last three decades prior to 2021 explicitly stated that the Breit-Wheeler process ``is still today waiting a direct observational verification''~\cite{RUFFINI20101}. 
On the other hand, the nonlinear Breit-Wheeler process (multiple photon version) observed in an experiment at SLAC~\cite{PhysRevLett.79.1626} in 1997 has been widely recognized as a discovery of that related process. 
In the last ten years, there have been theory proposals on how to discover the Breit-Wheeler process in photon-photon collisions using high-intensity lasers~\cite{PhysRevE.93.013201} and a vacuum hohlruam~\cite{doi.10.1038/nphoton.2014.95}, which explicitly stated that the Breit-Wheeler process ``has never been observed in the laboratory''. Creative experimental designs and increasing laser power may render the exclusive Breit-Wheeler process achievable at laser facilities~\cite{RUFFINI20101,PhysRevE.93.013201,Esnault_2021,ELIERIC,ELINATURE} in the near future.

\subsection{ Vacuum Birefringence and Polarization effects }
Vacuum birefringence has received significant theoretical attention ever since it was recognized as an implication of the Euler-Heisenberg Lagrangian~\cite{heisenbergFolgerungenAusDiracschen1936}. 
Several reviews exist that discuss the essential features of QED in electric and magnetic fields~\cite{Fedotov:2022ely,Battesti2012.10.1088/0034-4885/76/1/016401,Dunne2014.10.1140/epjst/e2014-02156-4} and a considerable amount of work has gone into exploring methods for computing VMB effects in perturbative and nonperturbative regimes~\cite{Denisov2017.10.1007/JHEP05(2017)105,Hattori2013.10.1016/j.aop.2012.11.010,Hattori2013.10.1016/j.aop.2013.03.016,PhysRevD.92.071301}. 
Significant theoretical effort has also been devoted to understanding the signatures of VMB in various experimental setups~\cite{Heinzl2006.10.1016/j.optcom.2006.06.053,Battesti2018.10.1016/j.physrep.2018.07.005,Adler2007.10.1088/1751-8113/40/5/f01,Iacopini1979.10.1016/0370-2693(79)90797-4} based on approaches including laser-electron collisions~\cite{PhysRevA.94.062102}, x-ray free electron lasers~\cite{shenExploringVacuumBirefringence2018}, and utilizing ultra-strong laser fields~\cite{PhysRevLett.119.250403}.

The study of vacuum magnetic birefringence and dichroism have been intimately connected with the Breit-Wheeler process since they were first predicted. Wheeler's student, John Toll, first studied the dispersion relationships for light traveling through various combinations of background electric and magnetic fields~\cite{1952PhDT21T}. 
Toll studied the behavior of the absorption and forward transmission of polarized light in various special cases. In Section~4.3 of Ref.~\cite{1952PhDT21T}, Toll specifically considered the effects on polarized light undergoing Delbrück scattering~\cite{delbruck1933zusatz,Meitner1933,Milstein1994} (the forward transmission process) and the Breit-Wheeler process (the absorption process). In Toll's calculation, the Weiszäcker~\cite{weizsaeckerAusstrahlungBeiStoessen1934}-Williams~\cite{PhysRev.45.729} method was applied to highly-Lorentz contracted heavy-nuclei as a photon source for the Breit-Wheeler process. It was further noted that to investigate the polarization dependence, one must consider the photon position relative to the source nucleus, not just the total WW photon cross section, since ``The polarization of these photons is with the electric vector directed radially with respect to the axis of nuclear motion.''~\cite{1952PhDT21T}. 
This insight and its implications were only recently rediscovered~\cite{doi.10.1016/j.physletb.2019.07.005} and confirmed by further theoretical     investigations~\cite{PhysRevD.101.034015,KLUSEKGAWENDA2021136114}. 
Following this, Toll calculated the polarization-dependent photon-photon absorption cross section as a function of the photons' relative polarization angle $\phi$ (see Eq.~4.3-7 in Ref.~\cite{1952PhDT21T}):
\begin{align}
\begin{split}
    \label{eq:bw_phi}
    \sigma_{\rm BW} \left( \omega; \phi_{\rm lab} \right) =  &\frac{ \sigma_{\perp}(\omega) + \sigma_{\parallel}(\omega) }{2} \\ 
    & -  \left[\frac{ \sigma_{\perp}(\omega) - \sigma_{\parallel}(\omega) }{2} \right] \cos 2\phi_{\rm lab}.
\end{split}
\end{align}
The angular dependence due to the effect of photon polarization is entirely contained in the $\cos 2\phi_{\rm lab}$ modulation term, with a prefactor proportional to the difference in the absorption cross sections  $\tau_{\gamma\gamma} \equiv \sigma_{\perp}(\omega) - \sigma_{\parallel}(\omega)$.
At the cross section level, the difference in the absorption for parallel vs. perpendicular polarized photons is directly related to vacuum dichroism and can be interpreted as the preferential absorption of a photon with respect to a given external electromagnetic field orientation~\cite{PhysRev.136.B1540,Heyl_1997,doi.10.1140/epjc/s10052-015-3869-8,PhysRevLett.119.250403,hattoriDileptonProductionSingle2021}.

The angle $\phi_{\rm lab}$ in Eq.~\ref{eq:bw_phi} is determined with respect to the photon's polarization as measured in the laboratory coordinate system. 
However, in the case of heavy-ion experiments, every nucleus-nucleus collision occurs with an impact parameter vector randomly oriented in $\phi_{\rm lab}$, naturally making measurement with respect to $\phi_{\rm lab}$ isotropic, since it effectively averages out the photon polarization. For this reason, when calculating the total photon-photon fusion cross section, most theoretical calculations implicitly or explicitly integrated over the photon polarization and the azimuthal angle dependencies of the cross section~\cite{SuperChic3}.

In 2019, the authors of Ref.~\cite{doi.10.1016/j.physletb.2019.07.005} 
predicted a unique spin interference effect driven by the linear photon polarization of WW photons. This spin interference effect results when the initial quantized spin of the colliding polarized photons is transferred into 
$e^+e^-$ as orbital angular momentum. The predicted signature for linearly polarized photon-photon collisions is a $\cos 4\phi$ modulation in the azimuthal angle ($\phi)$, which is approximately the angle of the electron (or positron) transverse momentum with respect to the $e^+e^-$ pair momentum.  
More precisely, the modulation occurs in the angle $\phi$, defined as
\begin{align}
    \cos{\phi} = \frac{(\vec{p_{T1}} + \vec{p_{T2}})\cdot{(\vec{p_{T1}} - \vec{p_{T2}})}}
    {
    |\vec{p_{T1}} + \vec{p_{T2}}|\times|\vec{p_{T1}} - \vec{p_{T2}}|
    },
\label{eq:phiangle}
\end{align}
where $\vec{p_{T1}}$ and $\vec{p_{T2}}$ are the 2D momentum vectors of the daughter electron and positron in the plane transverse to the beam direction.
Since this modulation is not measured with respect to impact parameter direction, it is not washed out over many events.
For this reason, the STAR measurement~\cite{PhysRevLett.127.052302} is conducted with respect to the angle $\phi$, equivalently defined as the angle between $q_\perp \equiv p_{1\perp} + p_{2\perp} $ and $P_\perp \equiv (p_{1\perp} - p_{2\perp})/2$, where $p_{1\perp},\ p_{2\perp}$ are the transverse momentum of the $e^+$ and $e^-$, respectively~\cite{doi.10.1016/j.physletb.2019.07.005}.

An essential feature of the EPA is that the polarization vector $(\vec{\xi})$ and transverse momentum vector $(\vec{k_\perp})$ are parallel ($\xi \parallel k_\perp$) for photons manifest from highly-Lorentz contracted fields. 
The conversion of the initial photon polarization into the orbital angular momentum of the produced $e^+$ and $e^-$, along with this correlation between the $\vec{\xi}$ and, $\vec{k_\perp}$ results in a final-state correlation between the photon polarization angle $\phi_{\rm lab}$ and $\phi$. 
However, the angle $\phi$ computed with the $e^+$ and $e^-$ momentum vectors is different in two important ways: 1) the angle $\phi$ has a finite resolution ($<100\%$) with respect to the true angle $\phi_{\rm lab}$, and 2) using the orbital angular momentum direction introduces an ambiguity in the direction of the spin. The former effect reduces only the magnitude of the modulation (compared to measurement with $\phi_{\rm lab}$) while the latter results in a transformation of the $\cos2\phi_{\rm lab}$ into a $\cos2\phi$ and a $\cos4\phi$ modulation. All of these effects are easily computed for the lowest order QED process, as was done in Ref.~\cite{PhysRevD.101.034015}, allowing a direct comparison between the STAR measurement and the predictions from lowest order QED.

\subsubsection{Polarized Photonuclear Interactions and Quantum Interference}
\label{sec:photoprod}
In addition to photon-photon interactions, the quasi-real photons generated by the projectile nucleus can also interact directly with the target nucleus in so-called photonuclear interactions. While not directly related to the Breit-Wheeler process or vacuum birefringence, we briefly mention the theoretical formulation of photonuclear interactions since they provide another avenue for utilizing and investigating the wavefunction of the incident linearly polarized photon. We will revisit this topic in Sec.~\ref{sec:future}, with future applications in photonuclear interactions that allow precise tomography of the gluon distribution within heavy nuclei and the measurement of the neutron skin of nuclei at high energy. 

The photonuclear production process can be explained by the photon first fluctuating into a $q\bar{q}$ pair, which may then interact with the target via the strong nuclear force. 
This provides an effective tool to further validate the photon kinematics and polarization discussed in the previous sections in addition to the measurements of nuclear geometries. In this subsection, we discuss how the newly discovered photon polarization could be used to enable precise nuclear tomography. 
The wave function of a quasi-real photon can be written as a Fock decomposition~\cite{Schuler:1993td}:
\begin{equation}
\label{Fock_photon_wave}
|\gamma \rangle = C_{\rm{bare}}|\gamma_{\rm{bare}}\rangle + C_{V}|V\rangle + ... + C_{q}|q\bar{q}\rangle.
\end{equation}
Here $C_{bare} \approx 1$ and $C_{V} \sim \sqrt{\alpha}$  ($V = \rho, \omega, \phi, J/\psi, ...$). The coefficient $C_V$ is determined by the photon-vector meson coupling, $f_V$, through $C_V = \sqrt{4\pi \alpha}/f_V$, 
where the coupling $f_V$ can be extracted from the vector meson leptonic decay width $\Gamma(V \rightarrow e^{+}e^{-})$.
The photon has quantum numbers $J^{PC} = 1^{--}$, which makes it preferentially fluctuate to a vector meson. According to the Vector Meson Dominance Model (VMD), the scattering amplitude for the process $\gamma + A \rightarrow B$ is the sum over the corresponding vector meson scattering amplitudes~\cite{Sakurai1960}:
\begin{equation}
\label{VMD_defintion}
A_{\gamma + A \rightarrow B} (s,t) = \sum_{V}C_{V} A_{V + A \rightarrow B}(s,t).
\end{equation}
For coherent photoproduction, $\gamma + A \rightarrow V + A$, the off-diagonal terms ($V^\prime + A \rightarrow V + A$) can be neglected. The elastic vector meson scattering on nuclei ($A_{V + A \rightarrow V + A}(s,t)$) can be related to the vector meson-nucleon cross section, $\sigma_{VN}$, via Quantum Glauber~\cite{Miller:2007ri} and the optical theorem relation:
\begin{equation}
\begin{split}
& A_{\gamma + A \rightarrow B} (s,t) =  \\
\frac{2}{\hbar} & \int 
e^{ip_{\perp}x_{\perp}}  2\left[1-\rm{exp}(-\frac{\sigma_{VN}}{2}T^{\prime}(\vec{x}_{\perp}))\right]d^{2}x_{\perp}.
\end{split}
\end{equation}
 $T^{\prime}(\vec{x}_\perp)$ is the modified thickness function accounting for the coherent length effect:
\begin{equation}
\begin{split}
	T^{\prime}(\vec{x}_\perp) &=\int_{-\infty}^{+\infty}\rho(\sqrt{{\vec{x}_\perp}^2+z^2})e^{iq_Lz}dz,\\
	\quad q_L &=\frac{M_Ve^y}{2\gamma_c}, \label{6}
	\end{split}
\end{equation}
where $q_L$ is the longitudinal momentum transfer required to produce a real vector meson.The vector meson-nucleon cross section,$\sigma_{VN}$ can be determined from the measurements of the forward-scattering cross section $\frac{d\sigma_{{\gamma}N{\rightarrow}VN}}{dt}|_{t=0}$ through VMD relation, which is well parameterized in Ref.~\cite{PhysRevC.60.014903}. In the coherent photoproduction process, the produced vector meson inherits the quantum state of the photon, which is fully linearly polarized. This leads to the preferential orientation of the decay angle along the direction of polarization. Here, we take the process of $\gamma + A \rightarrow \rho^{0} + A \rightarrow \pi^{+} + \pi^{-} +A$ as an example.  Following the derivation in Ref.~\cite{Schilling.10.1016/0550-3213(70)90070-2}, the decay angular distribution of vector meson to two spinless daughters  ($\rho^{0} \rightarrow \pi^{+} + \pi^{-}$) is   
\begin{equation}
\label{decay_angle}
	\frac{d^2N}{d\mathrm{cos}{\theta}d\phi}=\frac{3}{8\pi}\mathrm{sin}^2\theta\left[1+\mathrm{cos}(2\phi)\right],
\end{equation}
where the decay angles $\theta$ and $\phi$ are the polar and azimuthal angles, respectively, which denote the direction of one of the decay daughters in the vector meson rest frame.

In relativistic heavy-ion collisions, the coherent vector meson photoproduction consists of two indistinguishable processes: either nucleus 1 emits a photon and nucleus 2 acts as a target, or vice versa. The two processes interfere with each other, forming a Young’s double-slit experiment at the Fermi scale. Destructive interference of the cross section was first proposed in Ref.~\cite{Klein:1999gv} and confirmed by the STAR collaboration~\cite{STAR:2008llz}. 
Due to the linear polarization of the interacting photon, the destructive interference can also reveal itself in polarization space, which leads to a periodic oscillation pattern for the asymmetries of the decay angular distribution~\cite{Zha:2020cst,STAR:2022wfe}. 
Alternatively, the diffractive process can be implemented in terms of polarized photon-gluon interactions~\cite{Xing:2020hwh}, employing the gluon saturation mechanism relevant at small-$x$. The $\rho^{0}$ wave function cannot be implemented in perturbative QCD, but has been measured experimentally with good precision and been adapted~\cite{Forshaw:2010py,Xing:2020hwh,STAR:2022wfe}. Regardless of the exact theoretical prescription used, this effect is remarkable since it demonstrates quantum interference between distinguishable particles.

\begin{figure}
    \centering
    \includegraphics[width=0.60\linewidth]{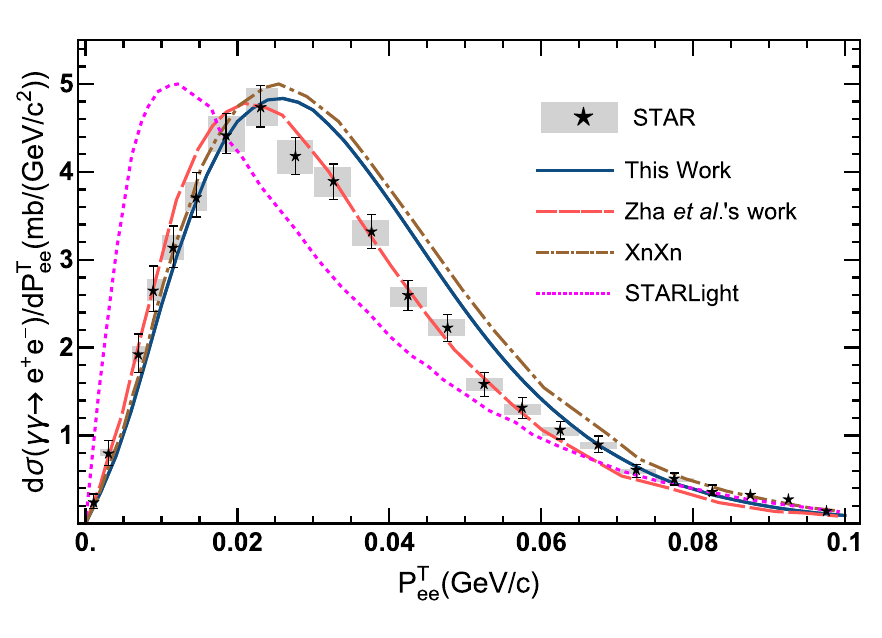}
    \caption{ Differential cross section as a function of the pair’s transverse momenta. The blue-solid line shows results from Eq.~20 of Ref.~\cite{PhysRevD.104.056011}. The red-dashed, brown-dot-dashed, and magenta-dotted lines are the results from Ref.~\cite{ZHA2018182} (times a factor of 9.1), Ref.~\cite{PhysRevLett.127.052302}, and STARLight~\cite{doi.10.1016/j.cpc.2016.10.016}, respectively. The data points are from STAR measurements~\cite{PhysRevLett.127.052302}, while the shaded areas stand for the experimental uncertainty. Reproduced from Ref.~\cite{PhysRevD.104.056011}. }
    \label{fig:two_photon_compare_pTee}
\end{figure}

\subsection{ Available Theoretical Calculations and Monte Carlo Generators }

Now that a general overview of the theory for describing photon-photon processes in heavy-ion collisions has been given, we list below the most commonly used theoretical calculations and Monte Carlo generators, along with notes on the methodology each implements.

\begin{itemize}
    \item \textbf{STARLight}~\cite{doi.10.1016/j.cpc.2016.10.016}: Monte Carlo generator which exemplifies the traditional EPA approach. Does not include any photon polarization effects. Another essential feature is the treatment of the photon $k_\perp$ independently of the nucleus-nucleus impact parameter, and the cross section and transverse momentum distribution are calculated separately. By default, STARLight does not include pair production within the geometrical radius of the nuclei. Calculations with STARLight have been carried out taking into account production within the nuclei, e.g. important for peripheral collisions ~\cite{PhysRevC.97.054903}.   
    \item \textbf{SuperChic3}~\cite{SuperChic3}: Monte Carlo generator which implements photon-photon processes based on Ref.~\cite{BUDNEV1971470,BUDNEV1975181}. SuperChic3 implements the EPA at the level of helicity amplitudes such that the photon polarization is taken into account. The code only provides calculations of the exclusive photon-photon process. Therefore, it cannot provide impact parameter dependent calculations for events with hadronic overlap or neutron tagging. 
    \item \textbf{Lowest-Order QED} (Ref.~\cite{doi.10.1016/j.physletb.2019.07.005,PhysRevD.101.034015,ZHA2020135089}) : Implementation of the lowest-order QED calculations following the prescription developed in Ref.~\cite{PhysRevA.51.1874,PhysRevA.55.396} and via low-x formalism in Ref.~\cite{doi.10.1016/j.physletb.2019.07.005} These calculations perform numerical computation of the process at all impact parameters. Reference~\cite{ZHA2020135089} primarily investigates the impact parameter dependent photon $k_\perp$ distribution, while Ref.~\cite{doi.10.1016/j.physletb.2019.07.005} primarily investigates polarization driven effects in photon-photon interactions.
    \item \textbf{All-order QED}~\cite{Zha2021.10.1007/JHEP08(2021)083}:  Investigation of the potential influence of higher-order QED effects for photon-photon interactions in heavy-ion collisions. Only calculations of the total cross section (within experimental acceptance where applicable) are currently supplied. Differential cross sections and polarization dependent effects have not yet been explored. 
    \item \textbf{Photon Wigner Formalism} (Refs.~\cite{PhysRevD.101.034015,KLUSEKGAWENDA2021136114,PhysRevLett.122.132301,PhysRevD.102.094013}): A complete formalism for the calculation of differential distributions of photon-photon fusion to dileptons via the Wigner function formalism~\cite{PhysRev.40.749}. Provides invariant mass, dilepton transverse momentum, and dilepton acoplanarity for arbitrary nucleus-nucleus impact parameter ranges.
    \item \textbf{Classical Field Approximation}~\cite{PhysRevD.104.056011}: Derivation of a general form of the cross section in terms of photon distributions, which depend on the transverse momentum and coordinates of the wave packet form of nuclear wave functions. Connections to the EPA and corrections in the Born approximation are clearly indicated. Connections to the PWF approach and those utilizing transverse momentum dependent photon distributions~\cite{doi.10.1016/j.physletb.2019.07.005} are also explored.
\end{itemize}

The connection among the various formalism utilized to compute the production of $e^+e^-$ via the two-photon interaction is explored most thoroughly in Ref.~\cite{PhysRevD.104.056011}. 
Figure~\ref{fig:two_photon_compare_pTee} shows STAR measurement of $e^+e^-$ transverse momentum with comparison between theoretical calculations from STARLight~\cite{doi.10.1016/j.cpc.2016.10.016}, lowest order QED~\cite{ZHA2020135089}, and classical field approach~\cite{PhysRevD.104.056011}.
The differential cross section with respect to $\alpha$ for the $\gamma\gamma \rightarrow \mu^+\mu^-$ process computed via the photon Wigner function formalism is shown in Fig.~\ref{fig:two_photon_alpha} and compared to data from ATLAS. The Wigner function calculation describes the data reasonably and reproduces the qualitative features of the data. Furthermore, the lowest order QED and Wigner function formalism render equivalent results in the region of applicability for the Breit-Wheeler process. Indeed, based on the comparisons shown in Ref.~\cite{KLUSEKGAWENDA2021136114,PhysRevD.104.056011}, it appears that the Wigner function calculations can recover the full impact parameter dependence of the lowest order QED result except at the very small values of acoplanarity~\cite{Wang:2022ihj}, where the processes may no longer be due to the pure Breit-Wheeler process, as discussed in Sec.~\ref{sec:Virtuality}.


\section{Electroproduction Measurements from electron-positron and hadron colliders}
    \label{sec:exp_elec_pos}

\begin{figure}
    \centering
    \includegraphics[width=0.60\linewidth]{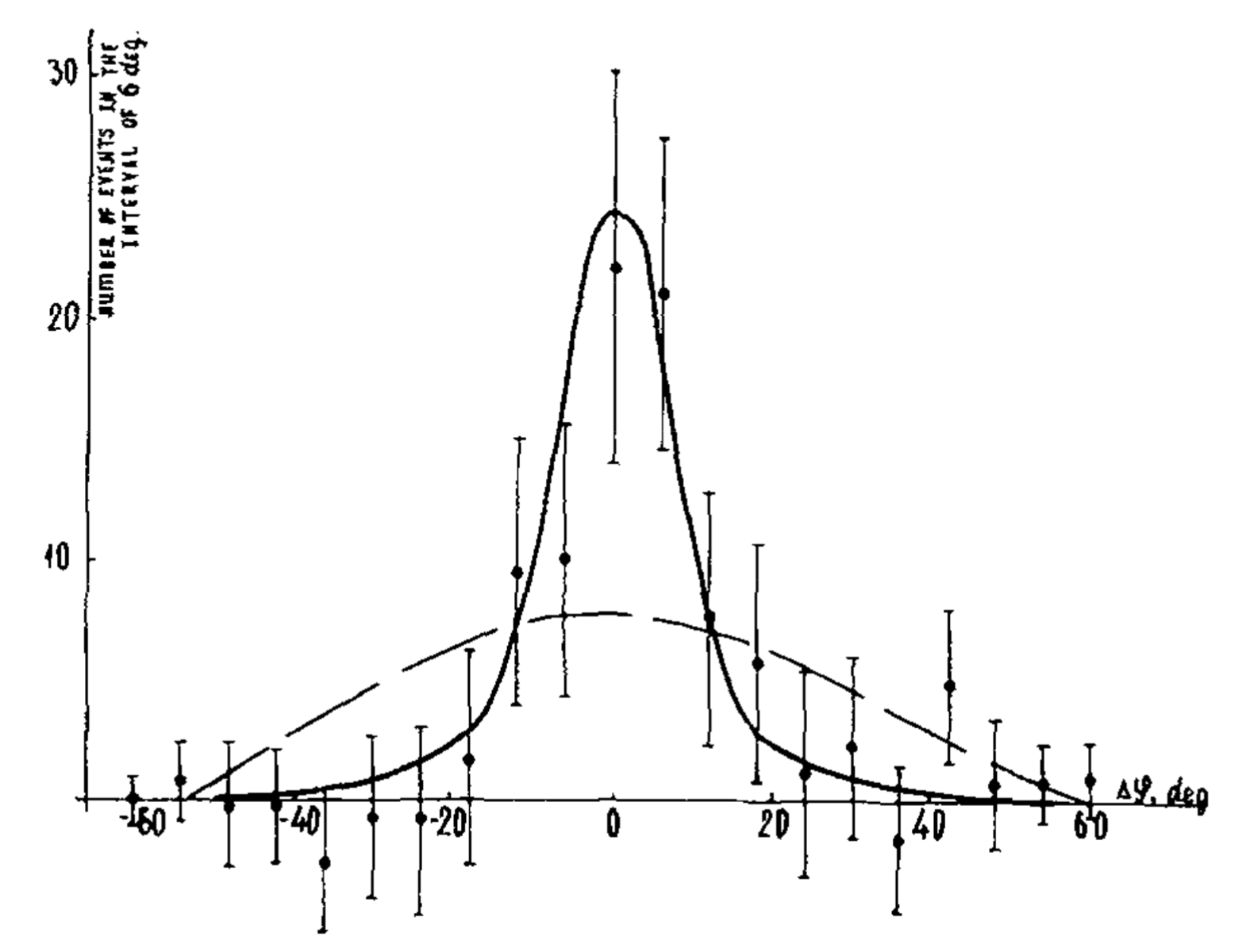}
    \caption{First electroproduction measurement at the VEPP-2 collider. The plot shows the distribution of electroproduction events in the angle $\Delta\phi$ - showing that the $e^+$ and $e^-$ are nearly back-to-back in the plane transverse the beam. The solid curve is a theoretical prediction for the $e^+e^-\rightarrow e^+e^- + e^+e^-$ by Baier and Fadin~\cite{BAIER1971156}. The dashed curve illustrates the expected distribution for the same process with independent and isotropic particle distributions. Reproduced from ~\cite{BALAKIN1971663}    }
    \label{fig:ee_to_ee_vepp2}
\end{figure}

\subsection{Early observations from electron-positron Colliders }
Significant attention was devoted to electroproduction ($e^+e^-\rightarrow e^+e^- + e^+e^-$) at the XVth International Conference on High Energy Physics held in Kiev in 1970~\cite{osti_4089925,PhysRevLett.25.972} in anticipation of powerful electron-positron colliders about to come online in the Soviet Union, the United States, and Europe. 
The electroproduction process was first observed in 1971 at the VEPP-2, a 2000 MeV electron-positron collider~\cite{BALAKIN1971663} in Novosibirsk, Soviet Union. That observation consisted of only about 100 events found to be nearly coplanar with the colliding beams and with the peculiar feature that the produced $e^+$ and $e^-$ were nearly back to back in the plane transverse to the beam, as shown in Fig.~\ref{fig:ee_to_ee_vepp2}. 
These characteristics distinguished the events from other known processes and supported the electroproduction hypothesis~\cite{BAIER1971156}: the logarithmic contribution to the back-to-back azimuthal correlation of $\gamma^*\gamma^*\rightarrow e^+e^-$ from the integration over the four-momentum transfer of the two virtual photons. Shortly after the ACO collider came online in France, the ADONE collider came online in Italy, and DORIS came online at DESY in Germany. Around the same time in the United States, SPEAR became operational at Stanford and CESR became operational at Cornell. With a host of new colliders online, the 1970s were well positioned to be a productive decade for high-energy particle physics in general and for further tests of electroproduction in particular~\cite{PhysRevD.4.1531,BUDNEV1971470}. 
The rapid experimental progress spurred significant theoretical progress, especially in understanding electroproduction, which laid the foundation for future decades. For instance, the review article by Budnev {\it et al.}~\cite{BUDNEV1975181} in 1975 outlined the fundamental calculations for photon-photon processes with finite four-momentum transfer at electron-positron colliders and its formulation in the Weizsäcker-Williams picture has been a standard in the Particle Data Book (Section 51.7)~\cite{PDGZyla:2020zbs}. 

\subsection{Measurements from electron-positron Colliders }

The success of the colliders in the 1970s motivated even more powerful machines with a thousand times higher reach in center of mass energy to be built in the 1980s and 1990s. At CERN in Europe, the Large Electron-Positron (LEP) collider was commissioned and brought online in 1989. Upgrades to the Stanford accelerator in the United States and to DESY in Germany provided several machines of comparable power around the world. Many more measurements of $e^+e^-$ production from two photons were performed at these machines. The increased energy and luminosity provided substantially more pairs and allowed more differential measurements.

The $e^+e^-$ pair production from the two-photon process was studied in high-energy $e^+e^-$ collisions at LEP~\cite{ACCIARRI1997341,abbiendi_total_2000}, PETRA~\cite{PhysRevD.38.2665,CELLO.10.1007/BF02430605,JADE.10.1007/BF01571802}, and SLAC~\cite{PhysRevD.42.2171,PhysRevD.42.2180} but with photons that were significantly virtual, except in a set of so-called un-tagged events. For instance, the OPAL~\cite{abbiendi_total_2000} experiment used un-tagged $e^+e^-$ events to study the total cross section for producing hadronic particles from real photon collisions. 
We note that high-energy electrons can emit real photons in the vacuum (under a magnetic field), and a classic example is synchrotron radiation. The study of two-photon physics processes has been an active field, using both real and virtual photons to study hadron production and photon structure functions (see~\cite{przybycien_two-photon_2008,SuperChic3,Nisius:1999cv,PDGZyla:2020zbs} and references therein).

\section{ Experimental Progress in High-Energy Collisions }
    \label{sec:hic_exp}
The first heavy-ion collision experiments were envisaged as a tool for studying nuclear matter in a previously unexplored regime. Unlike high-energy $e^++e^-$ and hadron colliders, heavy-ion collisions produce a relatively larger and hotter system capable of melting the quarks and gluons composing protons and neutrons within nuclei -- reaching temperatures of $\sim4$ trillion Kelvin\cite{Arsene2005.10.1016/j.nuclphysa.2005.02.130,Back2005.10.1016/j.nuclphysa.2005.03.084,Adams2005.10.1016/j.nuclphysa.2005.03.085,Adcox2005.10.1016/j.nuclphysa.2005.03.086,PHENIX:2009gyd}. The first manifestation of heavy-ion collisions were provided by accelerators with a fixed-target configuration, such as the BEVALAC at the Lawrence Berkeley National Laboratory~\cite{doi:10.1126/science.224.4651.857}. Despite the focus on nuclear physics, heavy-ion collisions have been understood as an opportunistic tool for studying QED in unique regimes for nearly a century already~\cite{1952PhDT21T,PhysRev.45.729}, with significant interest in the last 40 years~\cite{EICHLER1990165,PhysRevA.42.5736,2001NuPhA695395B}. Some of the earliest measurements of electromagnetic processes in heavy-ion collisions were studied by the Berkeley BEVALAC\cite{PhysRevA.56.2806}, the Alternating Gradient Synchrotron at BNL~\cite{PhysRevA.58.1253}, and by the CERN SPS~\cite{PhysRevLett.69.1911}.

In the year 2000, the Relativistic Heavy Ion Collider (RHIC) began operation and ushered in a new era of high-energy nuclear physics with ultra-relativistic Au beams providing collisions with a center-of-mass energy per nucleon pair $(\sqrt{s_{NN}} = 200 {\rm\  GeV})$. The payoff from this next-generation facility was almost immediate, with all four RHIC experiments reporting observation of a new state of nuclear matter in 2004: the Quark Gluon Plasma (QGP)~\cite{Arsene2005.10.1016/j.nuclphysa.2005.02.130,Back2005.10.1016/j.nuclphysa.2005.03.084,Adams2005.10.1016/j.nuclphysa.2005.03.085,Adcox2005.10.1016/j.nuclphysa.2005.03.086}. It should be emphasized that in an influential workshop in 1990 titled "Can RHIC be used to test QED?"~\cite{fatyga1990can}, three general subjects were addressed with the top priority as "to understand the validity of the best available descriptions of $e^+e^-$ pair production in peripheral heavy-ion collisions, especially for the domain where this process is known to be non-perturbative (multiple pair production)". It concluded in a positive perspective that "A study of electromagnetic phenomena in extremely peripheral
collisions of relativistic heavy ions can become a rich and exciting field that will complement studies of central collisions.".  However, none of the three (Landau-Lifschtz, Bethe-Heitler and Breit-Wheeler) processes was explicitly mentioned. In the next sections, the relevant detectors are described with special attention given to their unique capabilities for and limitations with respect to measuring novel QED phenomena, especially the Breit-Wheeler process and vacuum birefringence.    


\subsection{Detectors, Requirements, and Limitations}

\subsubsection{Detectors}
There were four experiments in operation shortly after RHIC began collisions in the year 2000~\cite{LUDLAM2003428}, two large and two smaller detectors. The Solenoidal Tracker at RHIC (STAR~\cite{ANDERSON2003659}) and the Pioneering High Energy Nuclear Interaction eXperiment (PHENIX~\cite{ADCOX2003469}) were the larger general purpose detectors. In contrast, the BRAHMS~\cite{ADAMCZYK2003437} and PHOBOS~\cite{BACK2003603} detectors were significantly smaller, more specialized detectors. Here we concentrate on the STAR and PHENIX detectors since these were used to test various aspects of QED through measurements of electromagnetically produced $e^+e^-$. 

\begin{figure}
    \centering
    \includegraphics[width=0.50\textwidth]{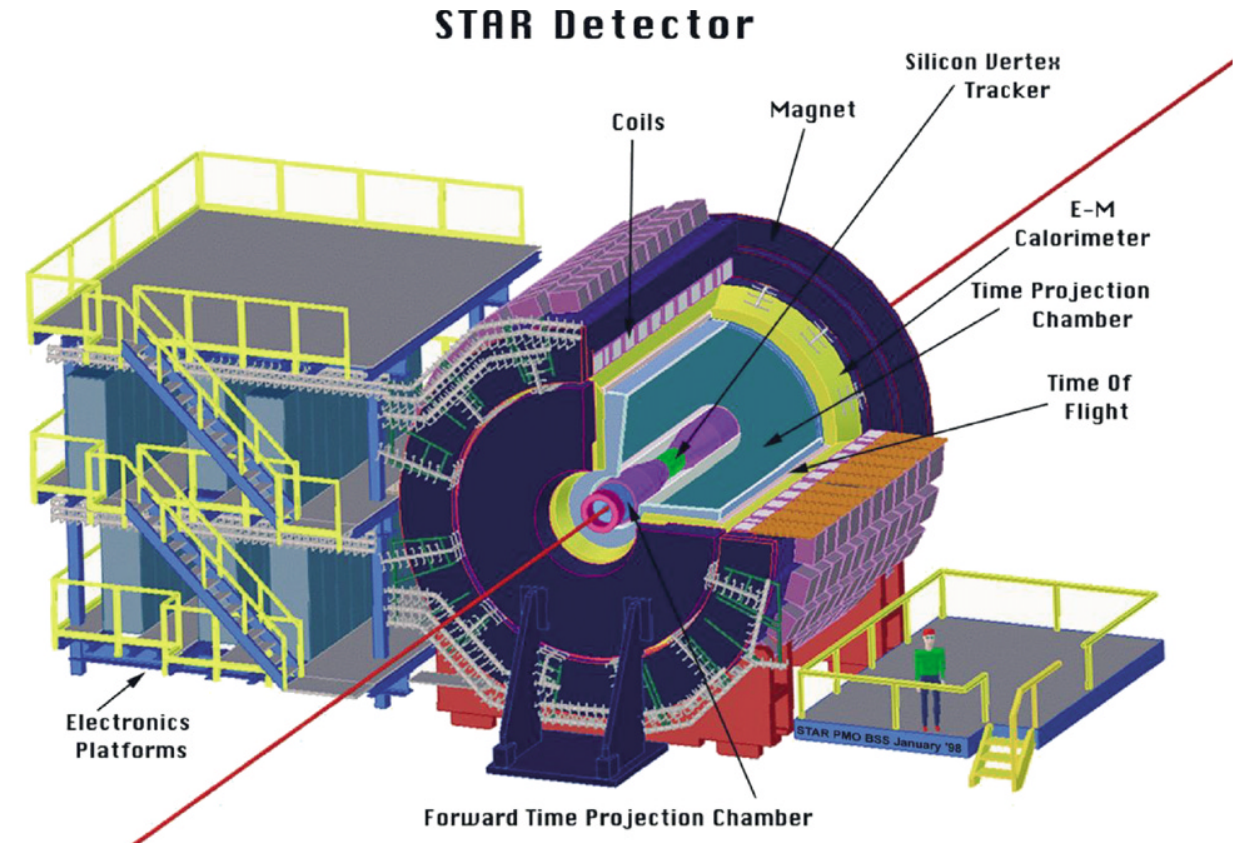}
    \includegraphics[width=0.50\textwidth]{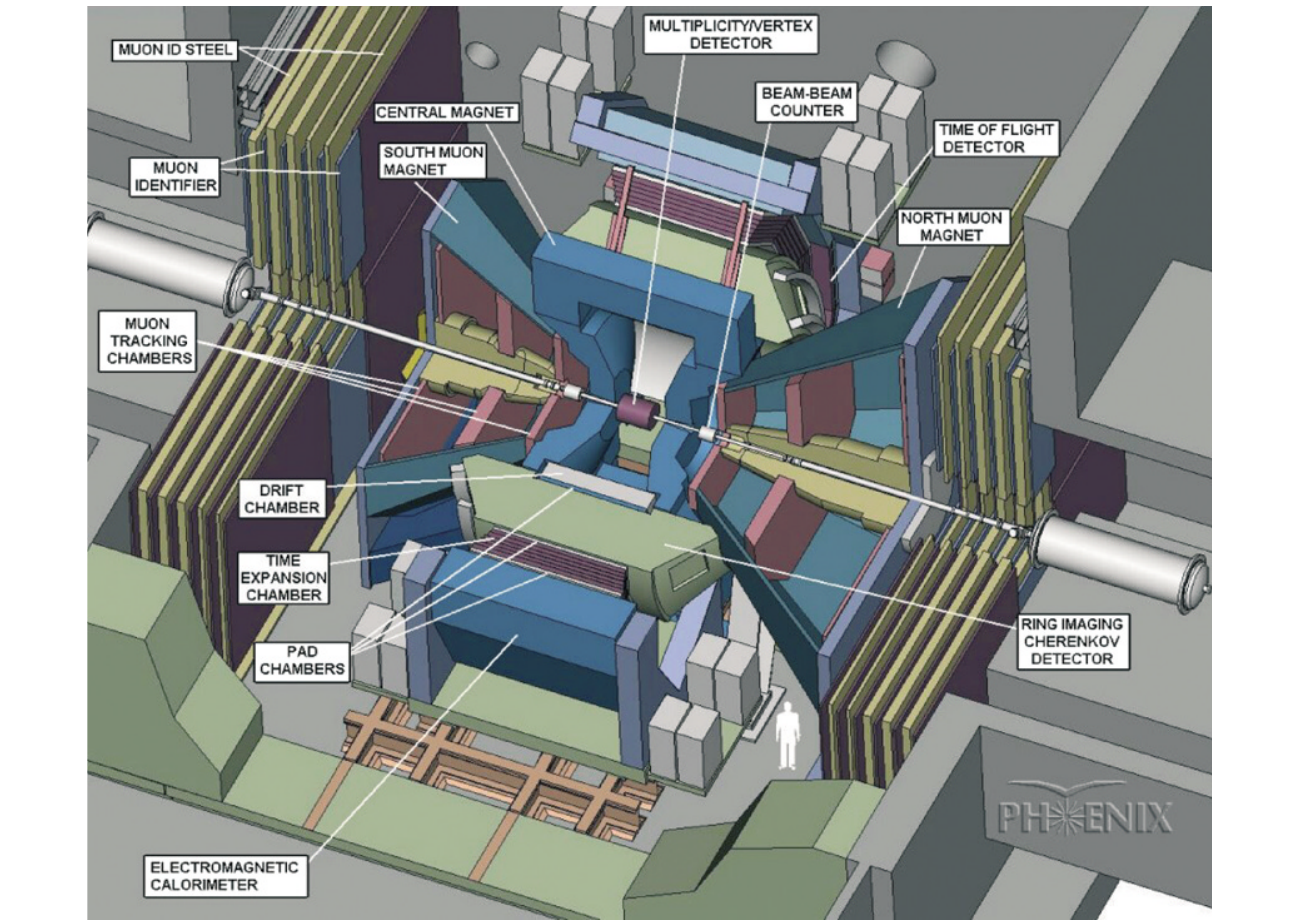}
    \caption{The two ``large'' detectors at the Relativistic Heavy Ion Collider: STAR (top) and PHENIX (bottom). Each performed early measurements of electromagnetic production of $e^+e^-$ in heavy-ion collisions~\cite{PhysRevC.70.031902,AFANASIEV2009321}.
    Reproduced from Ref.~\cite{LUDLAM2003428}.}
    \label{fig:rhic_detectors}
\end{figure}

In 2010, the Large Hadron Collider at the European Organization for Nuclear Research (CERN) began colliding heavy nuclei, and is to this day the highest energy collider of hadrons and heavy-ions. For the topics covered in this Report, three LHC detectors are of interest: The Compact Muon Solenoid (CMS~\cite{doi.10.1088/1748-0221/3/08/s08004}), A Toroidal LHC ApparatuS (ATLAS~\cite{doi.10.1088/1748-0221/3/08/s08003}) and the ALICE detector~\cite{doi.10.1088/1748-0221/3/08/s08002}

\begin{figure}
    \centering
    \includegraphics[width=0.51\textwidth]{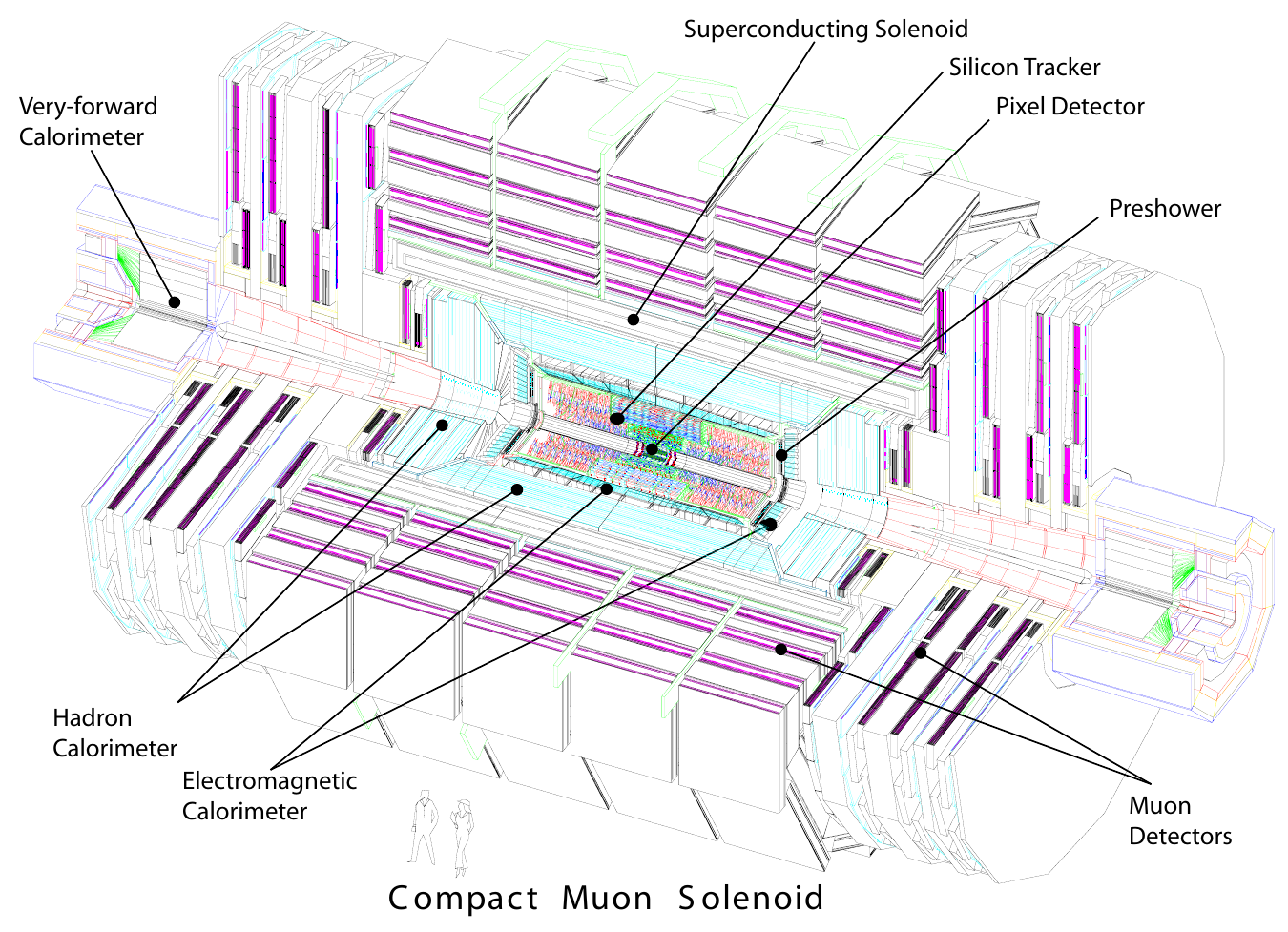}
    \includegraphics[width=0.51\textwidth]{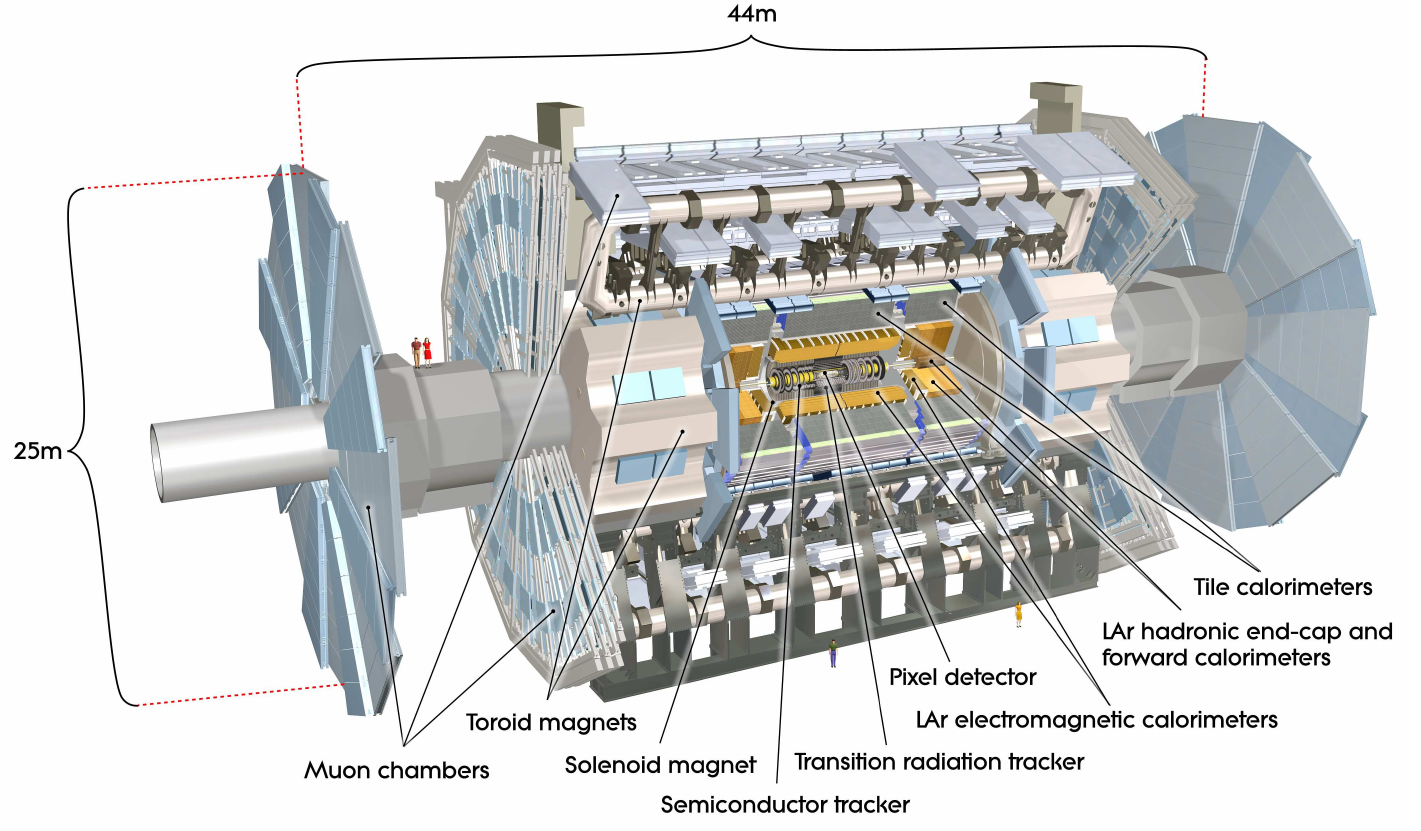}
    \includegraphics[width=0.51\textwidth]{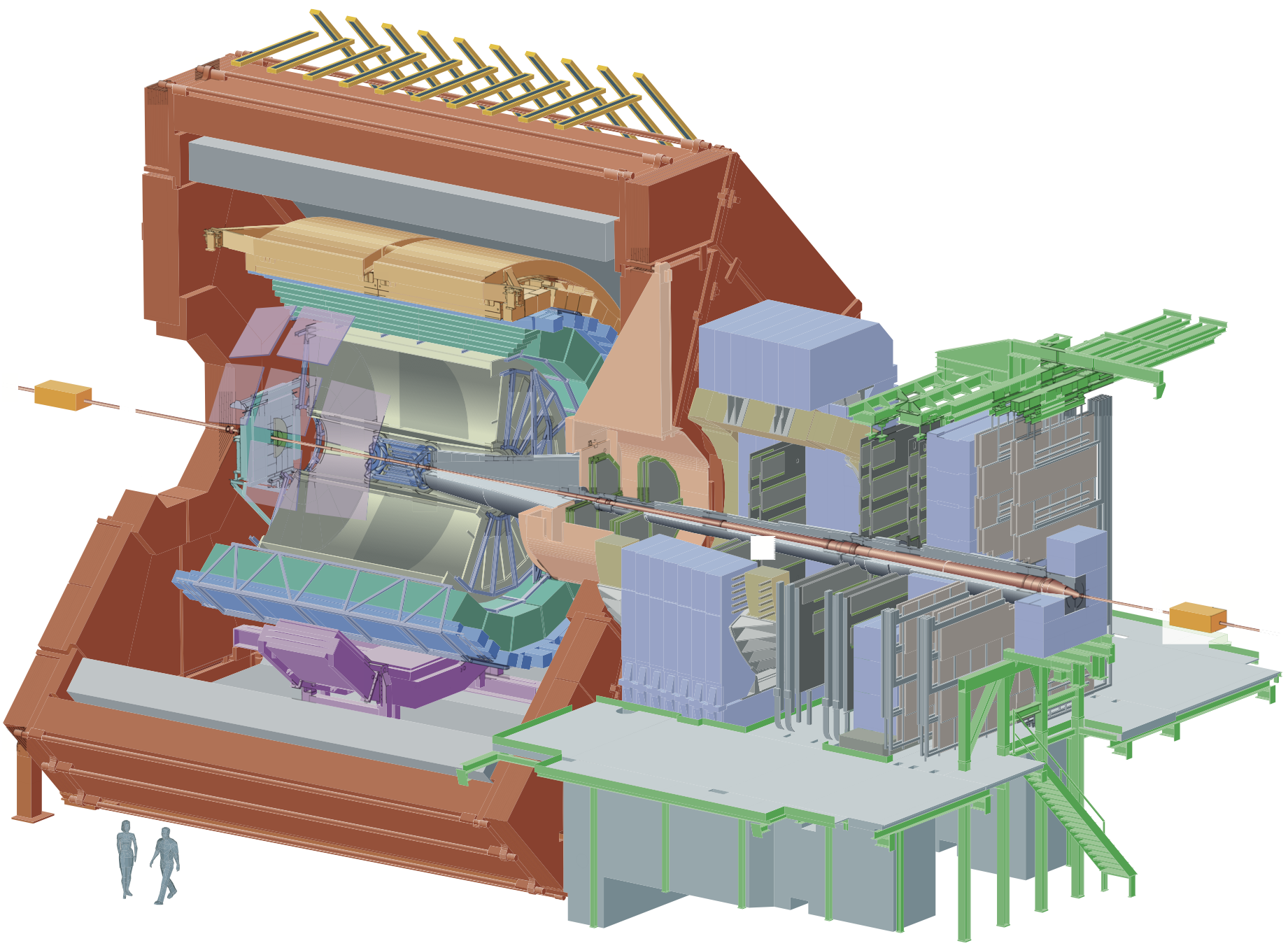}
    \caption{Relevant detectors at the LHC from top to bottom: CMS~\cite{doi.10.1088/1748-0221/3/08/s08004}, ATLAS~\cite{doi.10.1088/1748-0221/3/08/s08003}, and ALICE~\cite{doi.10.1088/1748-0221/3/08/s08002}.}
    \label{fig:lhc_detectors}
\end{figure}

\subsubsection{Detector Subsystems and Requirements}

\begin{figure*}
    \centering
    \includegraphics[width=0.89\textwidth]{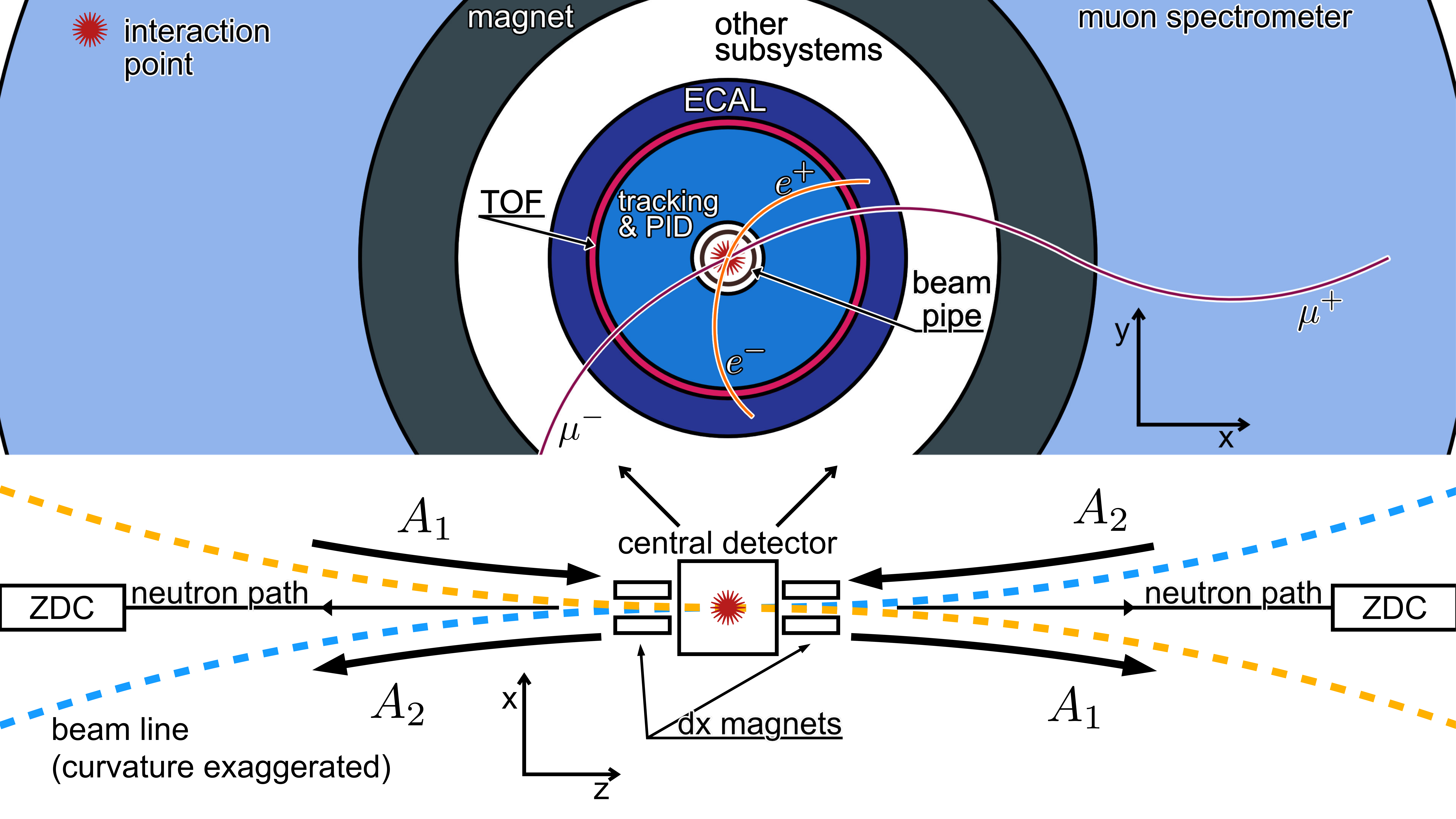}
    \caption{A schematic illustrating the generic layout of a high-energy collider experiment. Detailed schematics of the STAR and PHENIX experiments are shown in Fig.~\ref{fig:rhic_detectors} while the detailed schematics of the CMS, ATLAS, and ALICE detectors are shown in Fig.~\ref{fig:lhc_detectors}.  }
    \label{fig:generic_hep_experiment}
\end{figure*}

The STAR and PHENIX detectors at RHIC and the CMS, ATLAS, and ALICE detectors at the LHC are each unique general purpose experimental apparatuses for studying high-energy particle and heavy-ion collisions. 
Despite their differences, they share many similarities.  
Here, we focus on the aspects of these experimental apparatuses that are most relevant for the measurements that will be discussed in the next sections. 
These detector systems are structured like onions, with each layer serving a specific purpose. 
See Fig.~\ref{fig:generic_hep_experiment} for a schematic illustrating the typical layout of a general-purpose particle detector experiment. 
Note that the schematic does not directly correspond to any specific detector mentioned here but illustrates the relevant features common to those discussed here (STAR, PHENIX, CMS, ATLAS, and ALICE). \newline

\paragraph{ Tracker \newline} 
The tracking subsystem is often the central component in any high-energy collider experiment. 
The primary purpose of any tracking detector is to measure the trajectory of (charged) particles produced in an event by recording their interactions with material at precise locations. 
In general, they are embedded within a solenoid magnet (STAR, CMS, ALICE) that provides an inner volume with a constant and uniform magnetic field. 
However, powerful large volume solenoid magnets are heavy and expensive. 
Therefore, some designs (e.g., PHENIX and ATLAS) utilize a combination of other magnet configurations.  
The purpose of the magnetic field, regardless of its exact configuration, is to allow a particle's momentum to be determined from the measured radius of curvature $(r)$ of its trajectory as it traverses the magnetic field.

For a constant uniform magnetic field, the radius of curvature is related to the magnetic field strength $(B)$, the particle's charge $(q)$, and relativistic momentum $p=\gamma m |\vec{v}|$ as $r = p / (q B)$.
However, in reality, a particle's radius of curvature changes as it loses energy and/or scatters through interactions with material. 
For this and other cost-related reasons, tracking detectors are often placed closest to the interaction point, just outside the beam pipe.  
Finally, generally speaking, tracking detectors are only sensitive to charged particles and therefore do not measure the trajectories of neutrals, e.g., photons.

The STAR and ALICE experiments utilize a large volume gaseous tracker called a Time Projection Chamber (TPC) embedded within a $0.5$ Tesla magnetic solenoid. 
The low material gaseous TPC tracker within a relatively weak magnetic field provides optimal tracking for particles with momenta of a few hundred MeV/$c$. 
STAR and ALICE obtain optimal momentum measurement in the region $p_\perp < 100\ {\rm MeV}/c$ with a resolution of $dp_\perp / p_\perp \sim 1\%$.
In contrast, the CMS and ATLAS detectors utilize stronger magnetic fields of $4$ Tesla and $2$ Tesla, respectively. 
Within the magnetic field volume, they employ various technologies for tracking, including silicon-based detectors which provide precise measurement of particle interaction locations with a low material budget.
These detectors are optimized for measuring higher momentum particles, with a transverse momentum resolution of $200 - 300$ MeV$/c$ for particles with $p_\perp \lesssim 1.5$ GeV$/c$~\cite{aad_measurement_2015}.
The following references provide additional details about the specific tracking subsystems (and their performance) used in STAR\cite{ANDERSON2003659, doi.10.1016/j.nima.2010.12.006}, PHENIX\cite{Adcox2003,ADCOX2003469}, CMS\cite{doi.10.1088/1748-0221/9/10/p10009}, ATLAS\cite{Cornelissen2008,atlas2017technical,Cornelissendoi.10.1088/1742-6596/119/3/032013}, and ALICE\cite{doi.10.1142/S0217751X14300440}. \newline

\paragraph{ Identification of charged leptons and photons \newline} 

The energy density achieved in heavy-ion collisions provide a sufficient energy budget for producing a plethora of subatomic particles of various species. 
The STAR, PHENIX, CMS, ATLAS, and ALICE apparatuses built for high-momentum particle with extreme multiplicity and nuclear physics employ several specialized detector technologies for particle identification (PID) -- i.e., the determination of a long-lived particle's charge and mass. 
Here we focus on those most relevant for the identification of charged leptons (electrons, positrons, and muons) and photons. For particles observed in a tracking detector, their charge is directly determined from the sign of the measured curvature.

In general, a particle's mass may be identified either by measuring its Lorentz factor or by measuring its energy~\cite{doi.10.1016/S0168-9002(99)00323-X} lost as it traverses a medium of known material. 
For particles at low momentum, measurement of its Lorentz factor is optimal, while, for particles at higher momentum, measurement of the energy is generally optimal.

Various technologies have been developed for the identification of electrons, positrons, and muons at low momentum $(p<1 {\rm\ GeV})$. 
Ring Imaging Cherenkov (RICH) detectors, used e.g., in PHENIX, exploit the relationship between a particle's velocity $(\beta\equiv v/c)$ and the emission angle $(\theta)$ of Cherenkov light emitted as a particle passes through a medium with an index of refraction $(n)$ of:
\begin{equation}
    \cos{\theta} = 1/(n\beta).
\end{equation}

Time of Flight (TOF) detectors perform particle identification by measuring a particle's flight time $(\Delta t)$ over a known distance $(s)$, where $s$ is often provided by the tracking system. 
The STAR, PHENIX, and ALICE experiments employ specialized TOF detectors capable of measuring particle flight times to with a precision of $\sim$ a billionth of a second (ns, nanosecond). 
In conjunction with the momentum measured in a tracking system $(p)$, a TOF detector allows a particle's mass $(m)$ to be determined through the relation:
\begin{equation}
    m = p c^2 t^2 s^2 - 1,
\end{equation}
where powers of $c$, the speed of light in vacuum, are kept in this equation for clarity.
Both the STAR and ALICE experiments utilize large volume time-projection chamber tracking subsystems. In addition to their utility as tracking detectors, they also provide unique information for particle identification. 
As a charged particle travels through a TPC, it may ionize the gas within and lose energy as it travels. 
The mean ionization energy loss per unit length $\left(\langle dE/dx \rangle\right)$ can be correlated with the measured magnetic rigidity to provide particle identification~\cite{ANDERSON2003659}.

Identification of electrons and positrons at high momentum is generally carried out using electromagnetic calorimeters (ECAL). In high-energy physics experiments, any detector that measures the energy a particle loses as it passes through that detector is referred to as a calorimeter. 
They are designed in such a way as to force particles passing through them (within a certain energy range) to deposit all of their energy, therefore stopping within the calorimeter's volume.  
Depending on the material composition of a calorimeter, they can be designed to measure the energy from particles that interact via different fundamental forces/interactions. 
The two most common types of calorimeters utilize material compositions to measure energy deposited through the electromagnetic force (electromagnetic calorimeter) or through the strong force (hadronic calorimeters).
The STAR~\cite{doi.10.1016/S0168-9002(02)01970-8}, PHENIX~\cite{doi.10.1016/S0168-9002(02)01954-X}, and ALICE~\cite{doi.10.1016/j.nima.2009.10.010} experiments all employ Pb-plastic scintillator sampling electromagnetic calorimeters with some variations. 
The ATLAS experiment utilizes a Pb-liquid Argon calorimeter~\cite{doi.10.1088/1748-0221/3/08/s08003} while
the CMS experiment employs a Pb-tungstate scintillating crystal-based electromagnetic calorimeter~\cite{doi.10.1088/1748-0221/3/08/s08004}. Regardless of the specific material and approach utilized, the energy measured in the electromagnetic calorimeter $(E)$ along with the momentum measured by the tracker $(p)$ can be used to identify electrons (positrons) via the ratio:
\begin{equation}
    \frac{E}{p} = \left[ \frac{m_e^2 + p^2}{p^2}\right]^{1/2} \approx 1 {\rm,\ for\ } p \gg m_e.
\end{equation}
More information, like the shape of the electromagnetic shower, can be employed to further improve the identification of electrons with the information from electromagnetic calorimeters. 

Electromagnetic calorimeters are also essential for identifying photons. 
Since neutral particles undergo essentially no interaction with the materials in standard tracking detectors, their trajectory cannot be recorded. 
Even if a tracking detector were to be constructed capable of recording the trajectory of a neutral particle, the neutral particle will not be deflected by the magnetic field, limiting the usefulness of such a measurement. 
Therefore, neutral particles are primarily identified through isolated (i.e., without an associated charged track) deposits of energy observed within an ECAL. 
Further, identifying a neutral particle observed in an ECAL as a photon generally requires a veto on the amount of energy deposited in a hadronic calorimeter.

\begin{figure}
    \centering
    \includegraphics[width=0.60\textwidth]{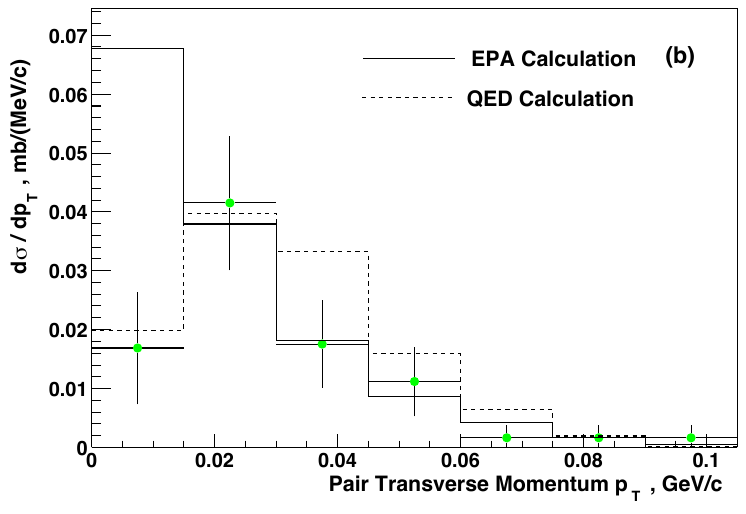}
    \caption{ STAR measurement of the transverse momentum distribution of photoproduced $e^+e^-$. The  data points are compared with predictions from the EPA (solid histogram) and lowest-order QED (dashed histogram) calculations. The error bars include both statistical and systematic errors. Reproduced from Ref.~\cite{PhysRevC.70.031902}}
    \label{fig:star_2004}
\end{figure}

\subsection{Early Measurements ($2000\sim2015$) }
Right from the start, the two ``large'' experiments at RHIC (STAR and PHENIX) studied the electromagnetic production of $e^+e^-$ from the fields of the colliding ultra-relativistic Au nuclei~\cite{PhysRevC.70.031902,AFANASIEV2009321}. The STAR Collaboration produced the first measurement of this kind from Au+Au collisions collected in the year 2001~\cite{PhysRevC.70.031902}. 
A minimum bias trigger was used to select events in which both gold nuclei broke up, by detecting events with one or more neutrons in zero degree calorimeters (ZDC). At that time STAR did not have a time of flight subsystem and could only identify electrons over a limited range in transverse momentum utilizing the  $\langle dE/dx\rangle$ measured in the TPC (the tracking detector). However, since this dataset was collected with a lower magnetic field ($0.25$ T) than STAR's nominal setup ($0.5$ T), electrons could still be well identified over a relatively large range of momentum.
Only 52 candidate events out of the 2001 data set, consisting of, $\sim$800\,000 minimum-bias events, survived the additional selection criteria imposed to identify electromagnetically produced $e^+e^-$. Figure~\ref{fig:star_2004} shows the observed differential cross section, in terms of the pair transverse momentum. The data exhibit reasonable agreement with the theoretical calculations based on the traditional EPA approach (solid line) and numerical lowest-order QED calculations (dotted line) shown for comparison.

While the two theoretical calculations were nearly identical in terms of the predicted total cross section, there was one crucial difference visible in the predicted pair transverse momentum distribution. The traditional EPA predicts that the cross section should continue to rise toward pair momentum of zero, while the QED calculation predicts a significant depletion of the differential cross section for pair momentum below 20 MeV$/c$.
Even given the large statistical uncertainties of the measurement due to the small sample of only 52 pairs, the shape of the transverse momentum distribution clearly favored the numerical lowest order QED calculation. It is at this point crucial to recognize that the theoretical developments discussed in Section~\ref{sec:theory_hic} came more than a decade after this early result from STAR. 
At the time, the discrepancy between the traditional EPA and QED calculations were interpreted as arising from the effect of photon virtuality~\cite{PhysRevC.70.031902}:
\begin{quote}
    The main difference between this calculation and the EPA approach is that the QED calculation includes photon virtuality.
\end{quote}
As discussed in Section~\ref{sec:theory_hic}, this is now understood to be an incorrect statement, with the difference resulting primarily from the impact parameter dependence of the process (included in the QED calculation but not in the traditional EPA) -- not from photon virtuality\cite{ZHA2020135089}. However, this interpretation was consistent with the representative consensus from the community at that point in time and expressed in a highly-cited review paper~\cite{doi:10.1146/annurev.nucl.55.090704.151526}. 
Over the next decade, this view became commonplace in the literature, with similar statements appearing in multiple papers over the next decade and a half~\cite{baltzTwophotonInteractionsNuclear2009,PhysRevLett.123.052001,aaboud_evidence_2017}. 
However, as we will discuss in the next sections, experimental progress began to challenge this interpretation. 

\begin{figure}
    \centering
    \includegraphics[width=0.60\linewidth]{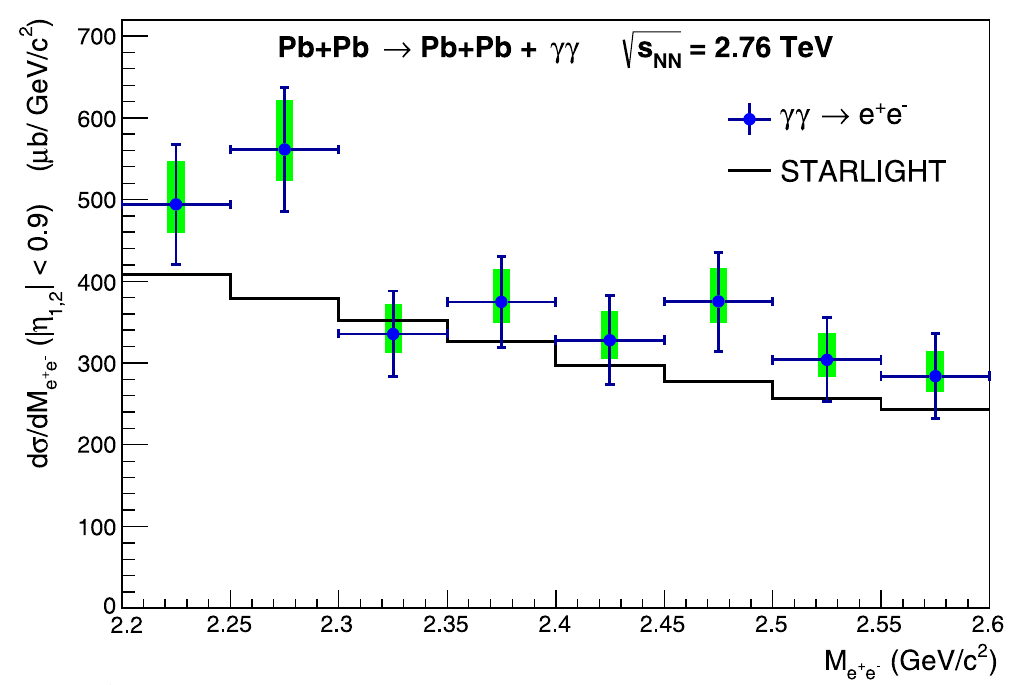}
    \caption{The invariant mass spectrum for photoproduced $e^+e^-$ in the range of $2.2 < M_{ee} < 2.6$ GeV$/c^2$. The data are compared to the predicted cross section based on traditional EPA calculations implemented by STARLight~\cite{doi.10.1016/j.cpc.2016.10.016}. Figures reproduced from Ref.~\cite{abbas_charmonium_2013}.}
    \label{fig:ALICE_low_mass}
\end{figure}

Progress measuring photoproduction processes continued over the next decade after STAR's initial measurement published in 2004. Due to the growing interest in utilizing heavy-ion collisions to study electromagnetic processes, multiple experiments performed measurements of the total cross section of electromagnetically produced $l^+l^-$ from ultra-peripheral heavy-ion collisions~\cite{AFANASIEV2009321,abbas_charmonium_2013,PhysRevLett.116.222301} or from exclusive p$+$p~\cite{doi.10.1016/j.physletb.2015.07.069} collisions.

In 2009, the PHENIX collaboration performed measurements of photoproduced $J/\psi$ and $e^+e^-$~\cite{AFANASIEV2009321}. However, the measurement was severely limited by statistical precision, with only 28 $e^+e^-$ reconstructed in the mass range of $2.0 < M_{ee} < 6.0 $ GeV$/c^2$, of which 14 were identified as continuum $e^+e^-$ with the remaining pairs resulting from $J/\psi$ decay. The observed production cross section for both the continuum $e^+e^-$ and $J/\psi$ were found to be consistent with traditional EPA predictions. Additionally, the transverse momentum spectrum for all observed $e^+e^-$ (continuum $e^+e^-$ and $J/\psi$ decay) was found to peak at low momentum ($\sim90$ MeV$/c$) as expected based on traditional EPA calculations. 

In 2013, ALICE made a similar but higher precision measurement of photoproduced $J/\psi$ and high mass $e^+e^-$~\cite{abbas_charmonium_2013}. The statistical power of their measurement allowed them to study the $e^+e^-$ invariant mass spectra over the wide range of $2.2 < M_{ee} < 6.0 $ GeV$/c^2$ with an order of magnitude more $e^+e^-$ compared to the PHENIX result~\cite{AFANASIEV2009321}. Of the $\sim 500$ $e^+e^-$ pairs observed within that mass range, $\sim300$ were found to be within the $J/\psi$ mass peak. The remaining $e^+e^-$ outside that mass range were identified as continuum $e^+e^-$. The production cross section of the continuum $e^+e^-$ were compared with traditional EPA prediction utilizing the Breit-Wheeler cross section in two different invariant mass ranges, one below and one above the $J/\psi$ mass peak.
The observed cross section in both regions were found to be $\sim20\%$ above the prediction, a disagreement of only about $1.5\sigma$ including all experimental uncertainties (and with no theoretical uncertainties accounted for). This measurement did not hint at any breakdown of the traditional EPA or the assumptions about photon virtuality. The observed cross section was interpreted to be evidence for the lack of any higher-order effects taking place in such photon-photon interactions, since higher-order effects are expected to reduce the observed cross section (with respect to the lowest-order process alone, see Sec.~\ref{sec:higher_order}) by as much as $\sim30\%$ for the ALICE experimental conditions~\cite{PhysRevC.80.034901,PhysRevLett.100.062302,PhysRevC.75.034903}. 

\begin{figure}
    \centering
    \includegraphics[width=0.60\linewidth]{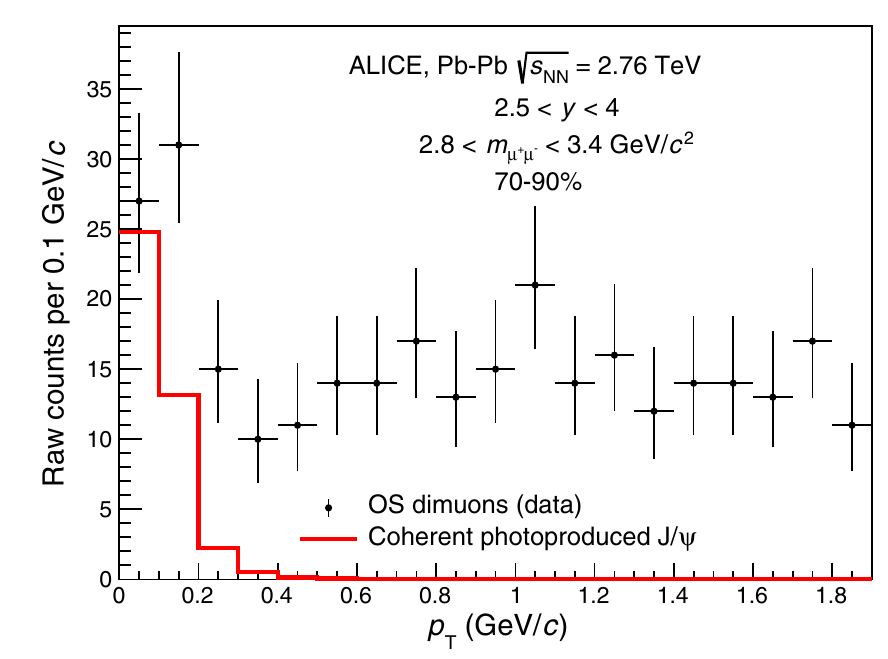}
    \caption{ Black data points show the raw transverse momentum distribution for $\mu^+\mu^-$ in the invariant mass range $2.8 < m_{\mu\mu} < 3.4$ GeV$/c^2$ from 70-90\% central collisions. The red curve indicates the transverse momentum distribution predicted by STARLight~\cite{doi.10.1016/j.cpc.2016.10.016} for coherent $J/\psi$ production 
    in Pb-Pb ultra-peripheral collisions and convoluted with the response function of the muon spectrometer. Reproduced from Ref.~\cite{PhysRevLett.116.222301}.}
    \label{fig:alice_ee}
\end{figure}

In both the PHENIX and ALICE measurements, continuum $e^+e^-$ and $e^+e^-$ from $J/\psi$ decay were observed together. These two contributions were primarily separated via an invariant mass selection window, whereby any $e^+e^-$ outside the $J/\psi$ mass peak were considered to be continuum $e^+e^-$ produced via photon-photon fusion. Both measurements found the $J/\psi$ photoprodction cross section to be consistent with EPA expectations for photonuclear interactions, where a WW photon from one nucleus fluctuates into a quark-antiquark pair capable of directly interacting with the target nucleus. The simultaneous observation of the continuum $e^+e^-$ along with $e^+e^-$ from $J/\psi$ decay obfuscate the nature of the interacting photons, since two real photons (with helicity +/- 1, zero forbidden) cannot form a vector meson (e.g., $J/\psi$). Therefore, these measurements can not experimentally demonstrate that the photons producing the observed $e^+e^-$ spectrum had only the allowed helicity states of real photons. Even if the experimental measurements had obtained significantly higher precision on the cross section, the range of theoretical predictions expected for the $\gamma+A\rightarrow J/\psi+X$ production mechanism~\cite{PhysRevC.60.014903,STRIKMAN200572,PhysRevLett.89.012301} would have made it difficult to definitively rule out contributions from virtual photon fusion e.g., $(\gamma^\prime+\gamma^\prime \rightarrow J/\psi + X)$.

\subsection{Observations of electromagnetic production in hadronic heavy-ion collisions ($2016 \sim 2018$)}

In 2016, The ALICE collaboration performed measurements of the $J/\psi$ particle in hadronic heavy-ion collisions over a large range of impact parameters $(0 < b < 2R)$ and over a large range of the decay $e^+e^-$ momentum. In peripheral collisions, they observed an anomalous excess of $J/\psi$ produced with very small transverse momentum~\cite{PhysRevLett.116.222301}.
As shown in Fig.~\ref{fig:alice_ee}, the yield of excess $J/\psi$ was found to be consistent with the expectation for the diffractive photonuclear process, even displaying the characteristic peak in cross section at very low transverse momentum for coherent photproduction processes.
However, the photon-photon and photonuclear interactions, had conventionally been considered only in ultra-peripheral collisions, where the hadronic (strong force) interaction does not take place. As discussed in Sec.~\ref{sec:theory_hic}, one ingredient in the theoretical description of coherent photoproduction is the treatment of the fields as external, with the nuclear charge distribution remaining undeflected throughout the interaction. Since the assumption of a straight-line trajectory before and after the collision is seemingly invalid in events with hadronic overlap, and the $J/\psi$ photoproduction requires photon diffraction coherently off the target nucleus as a whole, the observation of coherent photoproduction in such events was unexpected~\cite{brandenburgMappingElectromagneticFields2021a}. 

\begin{figure}
    \centering
    \includegraphics[width=0.60\linewidth]{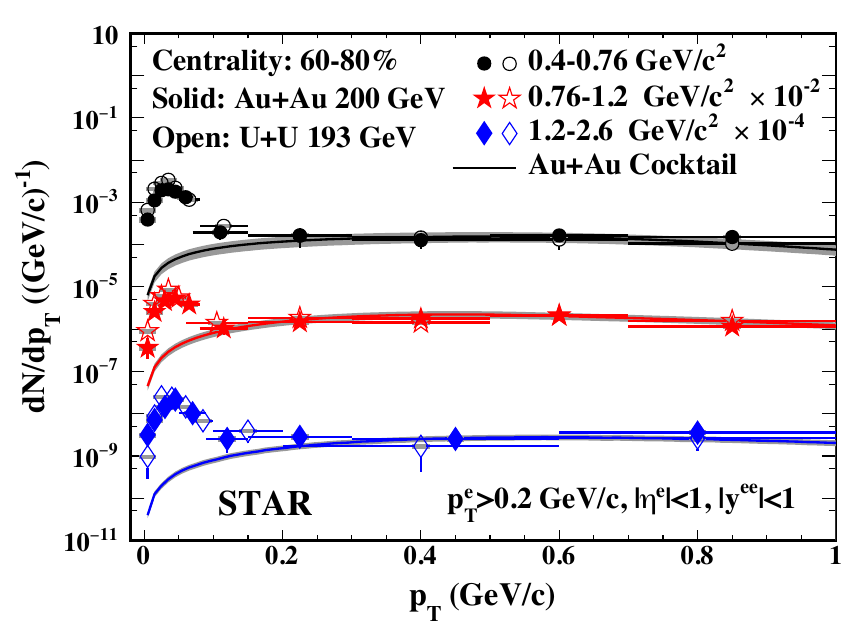}
    \caption{The transverse momentum distribution for $e^+e^-$ from coherent photon-photon interactions in 60-80\% central Au+Au and U+U collisions at $\sqrt{s_{NN}} = $ 200 GeV. The solid line represents the expected production from the cocktail of hadronic interactions and the gray bands depict the systematic uncertainties of the cocktails.. Reproduced from Ref.~\cite{PhysRevLett.121.132301}.}
    \label{fig:star_low_ptee}
\end{figure}

Shortly after, STAR pioneered the measurement of photon-photon production of $e^+e^-$ in the midst of hadronically interacting heavy-ion collisions~\cite{PhysRevLett.121.132301}, further confirming the ALICE finding that coherent photoproduction processes can occur even in events with hadronic overlap. Since high-energy hadronic interactions produce many electrons and positrons, isolating the electromagnetically produced pairs amid the numerous particles is a challenging task.  STAR utilized the excellent electron and positron identification capabilities provided by its various subsystems to remove background from other types of particles. Still, a single heavy-ion collision at RHIC energies can produce several $e^+e^-$ pairs per event~\cite{PhysRevLett.113.022301} making the identification of $e^+e^-$ from photoproduction difficult. However, coherent photoproduction shows a characteristic peak in cross section at very low transverse momentum, while the cross section for $e^+e^-$ from hadronic production tends to decrease for small values of transverse momentum (the ``Au+Au Cocktail'' in Fig.~\ref{fig:star_low_ptee}).
Therefore, STAR found that the $e^+e^-$ from photoproduction could be statistically isolated in peripheral collisions for pairs over a mass range of $0.4 < M_{ee} < 2.6$ GeV$/c^2$. However, increasingly central events (smaller $b$) tend to produce more $e^+e^-$ from hadronic interactions, swamping those from photoproduction and making the measurement untenable in mid-central to central collisions.

\begin{figure}
    \centering
    \includegraphics[width=0.60\linewidth]{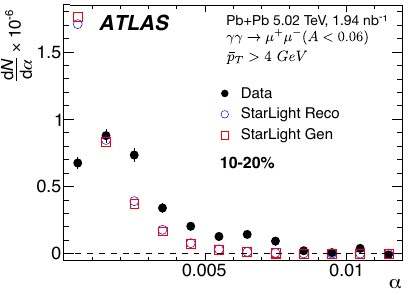}
    \caption{ ATLAS measurement of the acoplanarity $(\alpha)$ distribution from the $\gamma\gamma \rightarrow \mu^+\mu^-$ process in 10-20\% central Pb+Pb collisions at $\sqrt{s_{NN}} = 5.02$ TeV. Background from heavy-flavor and Drell-Yan processes are removed. Reproduced from Ref.~\cite{atlascollaborationMeasurementMuonPairs2022}.}
    \label{fig:atlas_broadening}
\end{figure}

Around the same time, the ATLAS experiment observed a significant centrality dependence in the electromagnetic production of $\mu^+\mu^-$~\cite{ATLAS:2018pfw}. Unlike the STAR result, their measurements spanned an impact parameter range from $b > 0$ outward, allowing investigation of the photoproduction cross section into the most central events. The observation of $\mu^+\mu^-$ instead of $e^+e^-$ was the primary difference allowing ATLAS to measure photoproduction even in central collisions. ATLAS's superior muon identification and the focus on muons with $p_T \gtrsim 4$ GeV$/c$ helped reduce the background from other types of particles to very low levels, allowing photoproduced $\mu^+\mu^-$ to be identified in head-on collisions with $b\approx0$.  Figure~\ref{fig:atlas_broadening} shows an ATLAS measurement of the $\alpha$ distribution for $\mu^+\mu^-$ pairs from coherent photon-photon production in 10-20\% central Pb+Pb collisions. The total production cross section measured in central collisions is in good agreement with traditional EPA calculations (STARLight). However, the $\alpha$ distribution was found to be significantly broader than the traditional EPA prediction. At that time the broadening was interpreted as arising due to final state interaction with the produced hadronic medium~\cite{ATLAS:2018pfw,PhysRevLett.122.132301}. 

The ALICE~\cite{PhysRevLett.116.222301} measurement demonstrated that coherent diffractive photonuclear interactions can occur in hadronic interactions. Then the STAR~\cite{PhysRevLett.113.022301} and ATLAS~\cite{ATLAS:2018pfw} measurements conclusively demonstrated that coherent photon-photon interactions occurs not only in ultra-peripheral collisions, but also in hadronic interactions. Since traditional EPA calculations have included the impact parameter dependence of the photon flux (see Eq.~1 of Ref.~\cite{PhysRevC.97.054903}) they have been able to provide predictions for the photoproduction cross section in hadronic events. However, for impact parameter ranges extending below $b_{\rm min} \approx 2R$, a realistic nuclear charge distribution was found to make a significant impact on the expected photon flux and the resulting photoproduction cross section~\cite{ZHA2018182}.  With these considerations taken into account, the measured cross sections from ALICE, STAR, and ATLAS were found to be consistent with traditional EPA predictions extended for hadronic events~\cite{PhysRevC.97.054903}.The ATLAS measurement even demonstrates that coherent photoproduction occurs in head-on collisions where the nuclei are broken almost completely apart. 

While the total cross section measured by STAR and ATLAS were consistent with coherent photoproduction, both measurements showed a puzzling deviation from the traditional EPA predictions. Both STAR~\cite{PhysRevLett.121.132301} and ATLAS~\cite{PhysRevLett.121.212301} found a significant broadening of the lepton pair's transverse momentum in hadronic heavy-ion collisions in comparison to those in UPCs and to traditional EPA calculations. 
The STAR Collaboration characterized the broadening by measuring the $P_{\bot}^{2}$ and the invariant mass spectra of lepton pairs in Au+Au and U+U collisions with respect to traditional EPA calculations. 
STAR utilized theoretical and phenomenological models to qualitatively describe the broadening by introducing the effect of a magnetic field trapped in an electrically conducting QGP~\cite{PhysRevLett.121.132301}).
The ATLAS Collaboration quantified the broadening effect via the acoplanarity ($\alpha$) of lepton pairs in different centrality events, in contrast to the same measurements in UPCs. 
Alternatively, the ATLAS collaboration proposed that the broadening effect may be due to the electromagnetic (EM) scattering of leptons in the hot and dense medium~\cite{PhysRevLett.121.212301}. 
However, in each case, the broadening was measured with respect to a ``baseline'' determined by measurements in UPCs and/or from traditional EPA calculations. These approaches assumed that there is no impact-parameter dependence of the transverse momentum distribution for the lepton pair from the initial photon-photon collision since such assumption was almost held as self-evident by the community up to this point of time.

\begin{figure}
    \centering
    \includegraphics[width=.60\linewidth]{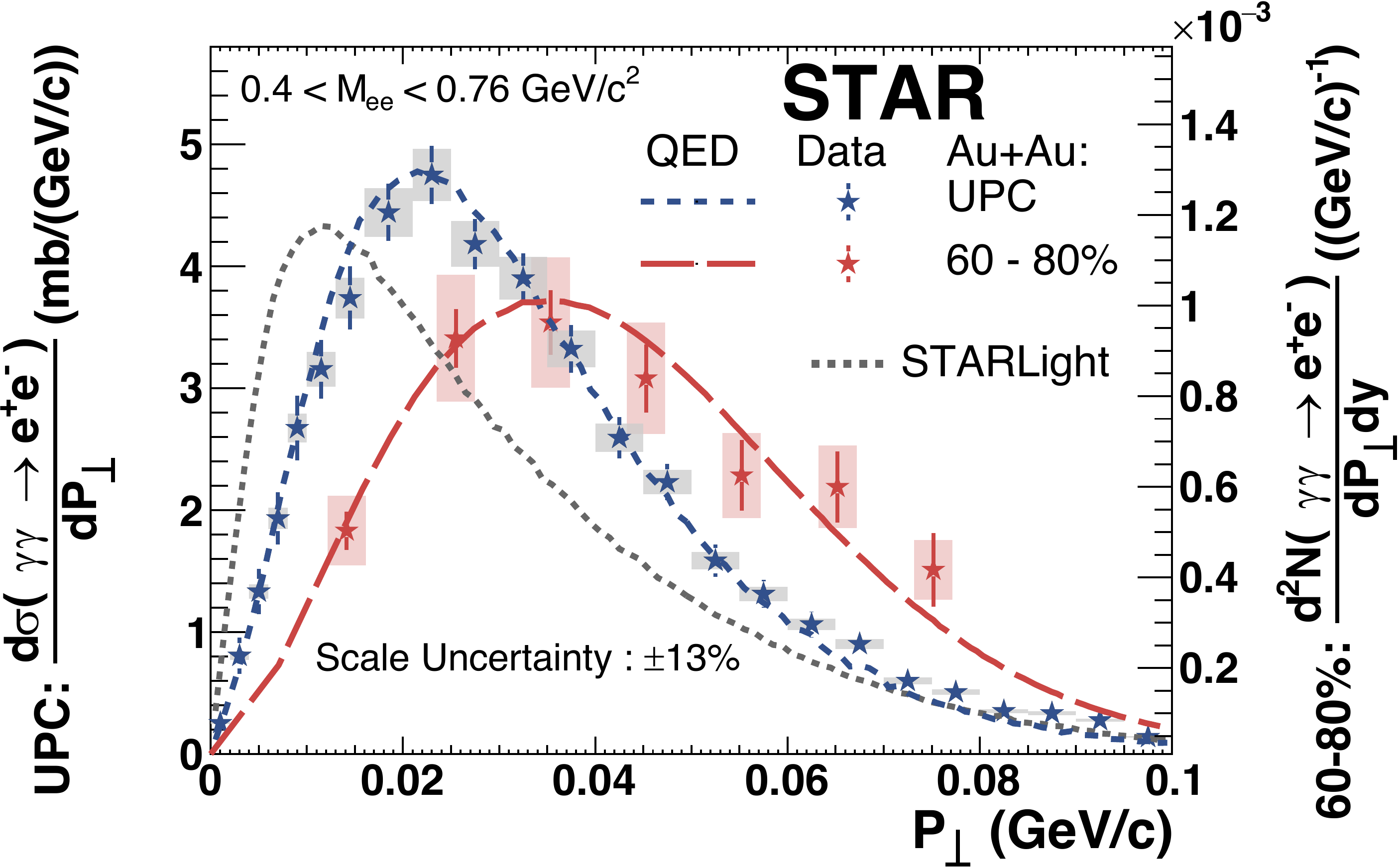}
    \caption{ (color online) STAR measurement of the $P_\perp$ distribution for the $\gamma\gamma \rightarrow e^+e^-$ process in UPC and $60-80\%$ central hadronic interactions. The measured $P_\perp$ distributions are incompatible with the $k_\perp$-factorization result. Reproduced from Ref.~\cite{PhysRevLett.127.052302}
    }
    \label{fig:exp_impact1}
\end{figure}

\begin{figure}
    \centering
    \includegraphics[width=.60\linewidth]{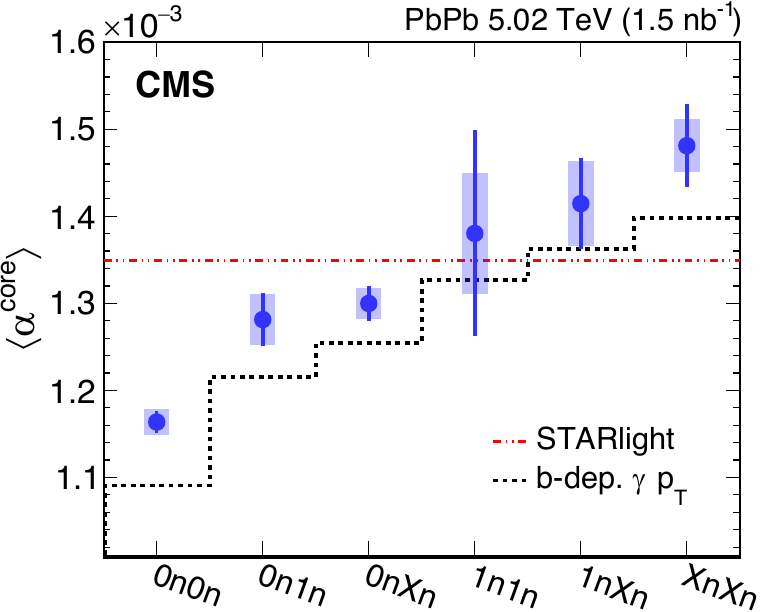} 
    \caption{ (color online) CMS measurement of $\langle \alpha_{\rm core} \rangle$ from the $\gamma\gamma \rightarrow \mu^+\mu^-$ process for various neutron multiplicities compared to the STARlight model and to lowest order QED calculations (b-dep. $\gamma\ p_T$). A strong neutron multiplicity dependence is observed. Reproduced from Ref.~\cite{PhysRevLett.127.122001}.
    }
    \label{fig:exp_impact2}
\end{figure}

\subsection{Observation of impact parameter dependence on photon kinematics}

According to the traditional EPA and the $k_\perp-$factorization approach, the photon $k_\perp$ distribution is primarily a result of the uncertainty principle with no dependence on $b$~\cite{doi.10.1016/j.cpc.2016.10.016} (see specifically Eq.~19 of Ref.~\cite{baltzTwophotonInteractionsNuclear2009}). These two concepts go hand-in-hand, since one can not generally describe a set of conjugate variables as a function of one-another. Up to this point in time ($\sim 2018$), there had been no direct evidence that the photon kinematics predicted by the traditional EPA were incorrect. In fact, the traditional EPA had successfully described experimental measurements over nearly two decades and specifically succeeded in describing the $P_T$ and $\alpha$ distributions observed in UPCs by experiments at RHIC~\cite{PhysRevC.70.031902,AFANASIEV2009321} and the LHC~\cite{PhysRevLett.121.212301}. 

While the aforementioned measurements from STAR~\cite{PhysRevLett.121.132301} and ATLAS~\cite{PhysRevLett.121.212301} showed transverse momentum broadening that is inconsistent with traditional EPA predictions (see Fig.~\ref{fig:atlas_broadening} and Fig.~\ref{fig:exp_impact1}, they were otherwise consistent in terms of total cross section. 
For this reason, the discrepancy was initially interpreted as evidence for final state effects driven by interaction with the hadronic matter (or even a QGP)~\cite{PhysRevLett.122.132301,PhysRevD.102.094013} -- and not evidence that the initial photon kinematics were incorrectly described by the traditional EPA.

In order to determine the primary source of the broadening, it was necessary to determine if the observed broadening was a result of the initial photon flux or a result of final state interactions. 
While STAR had already measured the $e^+e^-$ transverse momentum spectrum from UPCs in their 2004 paper~\cite{PhysRevC.70.031902}, the statistical precision was insufficient to perform additional differential studies capable of investigating the impact parameter dependence. 
Over the next several years, various experimental techniques were developed to test the impact parameter dependence of the photon flux and kinematics of the produced dileptons.

For highly charged nuclei, $Z\alpha_{em} \approx 0.6$ ($\alpha_{em} \approx 1/137$ and $Z_{\rm{Au}} = 79$, $Z_{\rm{Pb}} = 82$), meaning that the density of photons is appreciable, and therefore, the nuclei may exchange multiple photons in a single passing.
In UPCs, the quasi-exclusive $\gamma\gamma \rightarrow l^+l^-$ process may be selected in collisions where additional exchanged photons lead to the excitation and subsequent dissociation of the nuclei. Mutual Coloumb excitation (MCE) is the process by which at least two photons (in addition to those mediating the semi-exclusive process of interest) cause one or both nuclei to become excited~\cite{stelsonCoulombExcitation1963}. 
The cross section is dominated by the Giant Dipole Resonance (GDR)~\cite{newtonObservationGiantDipole1981} which peaks at low energy ($E_\gamma\approx 14$ MeV/$c$ for gold and lead nuclei). 
The GDR excitation is responsible for several final states with one or two neutrons emitted and has been measured with high precision by various experiments~\cite{veyssierePhotoneutronCrossSections1970}. All the experiments discussed herein are well-equipped to detect beam energy neutrons via Zero Degree Calorimeters.

The STAR and CMS collaborations employed a neutron tagging approach to experimentally test the impact parameter dependence of the $\gamma\gamma \rightarrow l^+l^-$ process and to specifically investigate the photon $k_\perp$ distributions in events without hadronic overlap. 
The STAR collaboration first demonstrated that the lepton pair momentum depends strongly on the impact parameter range of the colliding nuclei~\cite{PhysRevLett.127.052302} -- in stark contrast to the long-accepted behavior predicted by the $k_\perp-$factorization approach used in the traditional EPA models. At the same time, theoretical progress (summarized in Sec.~\ref{sec:theory_hic}) via lowest order QED calculations for the $\gamma\gamma \rightarrow l^+l^-$ process indicated that the kinematic distribution of the initial photon flux contains strong impact parameter dependence~\cite{ZHA2020135089,PhysRevC.47.2308}. Figure~\ref{fig:exp_impact1} shows measurements of the $P_\perp$ distribution from the $\gamma\gamma\rightarrow e^+e^-$ process in ultra-peripheral and peripheral collisions. The precision measurement from UPCs (with two orders of magnitude more statistics than the 2004 STAR measurement) shows a significantly broader $P_\perp$ spectrum compared to the traditional EPA (STARLight) prediction. Unlike the STAR 2004 measurement, the measurement shown in Fig.~\ref{fig:exp_impact1} utilized events with predominately one neutron in each of the east and west STAR ZDCs. The presence of neutrons biases the nucleus-nucleus impact parameter distribution to smaller values, compared to events without neutrons in the final state.  
Furthermore, the transverse momentum spectra from both ultra-peripheral and peripheral collisions can be well described by the same lowest-order QED calculations, suggesting that the previously observed broadening in hadronic heavy-ion collisions mainly results from the initial photon flux, not from final state interactions. 

Figure~\ref{fig:exp_impact2} shows additional measurements by the CMS collaboration of the $\alpha$ distribution ($\pi\alpha \simeq P_\perp / W$) of dimuons in events with various neutron emission scenarios.
A significant dependence of the $\langle\alpha_{\rm core}\rangle$ distribution with neutron multiplicity is observed, where $\alpha_{\rm core}$ is the statistically isolated $\alpha$ distribution from coherent $\gamma\gamma \rightarrow \mu^+\mu^-$ interactions~\cite{PhysRevLett.127.122001}. 
In the CMS measurement, the narrow signal ($\alpha_{\rm core}$) and broad background distributions were isolated via empirical fit functions. 
The observation by the CMS collaboration confirmed that made by the STAR collaboration, demonstrating that significant broadening of the lepton pair momentum results from the initial photon flux in the absence of any hadronic medium that may modify the final distributions. While these measurements do not rule out potential medium interactions proposed previously by the STAR and ATLAS collaborations, they demonstrate that the primary source of the observed broadening is the initial photon flux.

\begin{figure}
    \centering
    \includegraphics[width=0.60\textwidth]{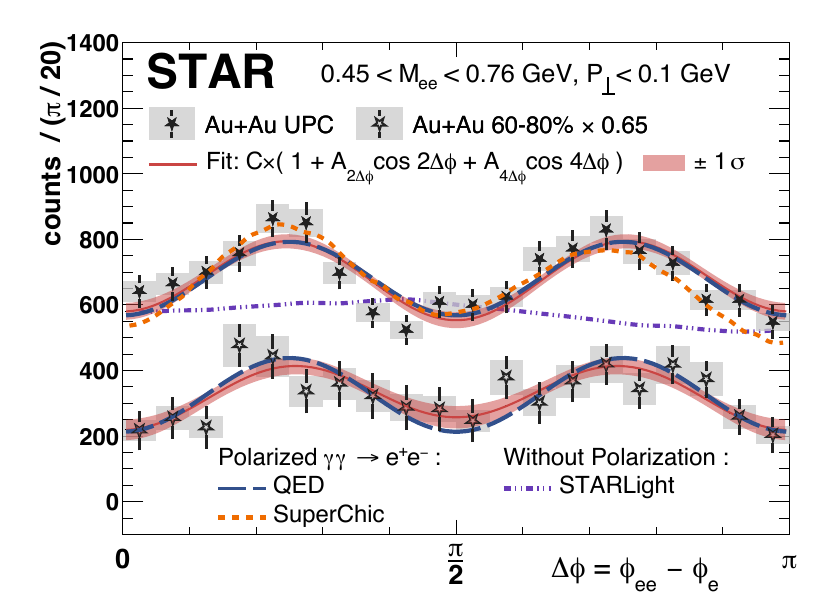}
    \caption{The $\Delta\phi = \phi_{ee} - \phi_{e}$ distribution from UPCs and 60 -- 80\% central collisions for $0.45 < M_{ee} < 0.76$ GeV compared with calculations from lowest-order QED~\cite{doi.10.1016/j.physletb.2019.07.005,PhysRevD.101.034015}, STARLight~\cite{doi.10.1016/j.cpc.2016.10.016} and the publicly available SuperChic3 code~\cite{SuperChic3}.}
    
    \label{fig:star_phi}
\end{figure}

\subsection{Observation of Photon Polarization Effects}

Ultra-relativistic nuclei produce a highly Lorentz contracted radial electric field emanating from the nucleus, with a magnetic field circling the nucleus. Both the electric and magnetic field are almost entirely Lorentz-contracted into the plane perpendicular to the direction of motion. Therefore, at any given point, the fields appear as a nearly transverse linearly polarized electromagnetic wave. It was therefore expected that the corresponding photons manifest from these fields are linearly polarized in the transverse plane with respect to the beam~\cite{1952PhDT21T}. 
Despite this expectation, until recently there was no proposed technique for accessing the photon polarization information, since the primary effects are asymmetries in the azimuthal (plane perpendicular to the beam axis) distribution of the produced $e^+e^-$. 

Since each heavy-ion collision occurs with a random impact parameter orientation, such azimuthal asymmetries are generally washed out over multiple events.   
For the kinematic ranges applicable for $e^+e^-$ measurements in ultra-peripheral collisions with STAR, lowest-order QED calculations predict a $\cos4\phi$ modulation of approximately $-17\%$. Figure~\ref{fig:star_phi} shows the measurement of $\phi$ ($\Delta\phi$ in  the figure) from the Breit-Wheeler process in ultra-peripheral and peripheral collisions~\cite{PhysRevLett.127.052302}. The data are compared with three model calculations: lowest-order QED~\cite{PhysRevD.101.034015} and SuperChic3~\cite{SuperChic3} which both include the photon polarization effect, and STARLight~\cite{doi.10.1016/j.cpc.2016.10.016}, which does not -- predicting a roughly isotropic angular distribution. The data are in good agreement with both the lowest-order QED and SuperChic3 predictions, demonstrating sensitivity to the photon polarization effect. The implications of this measurement in unison with the other aspects of the STAR measurement of the Breit-Wheeler process will be discussed in Sec.~\ref{sec:discussion}.
 
We examine whether a precision angular distribution in the transverse plane could have been observed and measured by other experimental collaborations from previous data sets and detectors as presented in past publications. The data from CERES~\cite{BAUER1994471}, PHENIX~\cite{AFANASIEV2009321}, STAR~\cite{PhysRevC.70.031902}, CDF~\cite{PhysRevLett.98.112001,PhysRevLett.102.242001} and ALICE~\cite{abbas_charmonium_2013} do not have enough statistics to perform a $\cos{(4\Delta\phi)}$ measurement while those measurements presented from ATLAS~\cite{aad_measurement_2015,aaboud_evidence_2017} and CMS~\cite{chatrchyan_search_2012,sirunyan_observation_2018,doi.10.1016/j.physletb.2019.134826} do not have sufficient pair momentum resolution ($200-300$ MeV$/c$) to perform this measurement.

\subsection{Light-by-Light Scattering}
Light-by-light scattering is a purely quantum mechanical process in electrodynamics.  Within the standard model, it proceeds at lowest order via virtual one-loop box diagrams that involve charged fermions or W bosons. Light-by-light scattering is challenging to observe experimentally because it occurs at order $\alpha_{EM}^4 ~ 3 \times 10^{-9}$.  However, the large photon fluxes produced in ultra-peripheral heavy ion collisions at the LHC finally made this observation possible.  In 2016, the ATLAS collaboration presented the first evidence 
~\cite{aaboud_evidence_2017}, 
for the direct observation of this phenomenon in Pb+Pb collisions at $\sqrt{s_{NN}} = 5.02$ TeV, $Pb + Pb \rightarrow Pb^{(*)} + Pb^{(*)} + \gamma +\gamma$.  The two lead nuclei continue down the beam pipe, although they may be left in an excited state (represented by the *) and subsequently emit neutrons. A Feynman diagram for the LbyL processes is shown in Fig.~\ref{fig:vb_pvlas_diagrams}(a).

The observed signal is two exclusive photons in the central detector, with possibly some neutrons deposited in the ZDCs from the decay of the excited lead nuclei.  The primary background is from electrons misidentified as photons.  This can happen if, for example, the electron track is not reconstructed and the electron is only detected through energy deposited in the electromagnetic calorimeters. The dominant source of electrons is the process $\gamma \gamma \rightarrow e^+ e^-$.  The exclusive 2-photon final state can also be produced via the strong interaction through the central exclusive process, $g g \rightarrow \gamma \gamma$. The backgrounds are effectively suppressed by selecting events with low di-photon transverse momentum, where the acoplanarity of the two photons is very small $(\alpha < 0.01)$. The ATLAS dataset collected in 2015, with a luminosity of 0.480 nb$^{-1}$, gave a signal with a significance of 4.4 standard deviations. The observation was confirmed by the CMS collaboration ~\cite{doi.10.1016/j.physletb.2019.134826} - a signal with a significance of 4.1 standard deviation was observed in the 2015 CMS dataset with a luminosity of 0.39 nb$^{-1}$.  A subsequent measurement by the ATLAS collaboration with data collected in 2018, with a luminosity of 1.73 nb$^{-1}$, brought the significance of the observation to 8.2 standard deviations~\cite{PhysRevLett.123.052001}. The measured two-photon production cross sections are consistent with standard model predictions for light-by-light scattering.  

\begin{figure}
    \centering
    \includegraphics[width=0.60\textwidth]{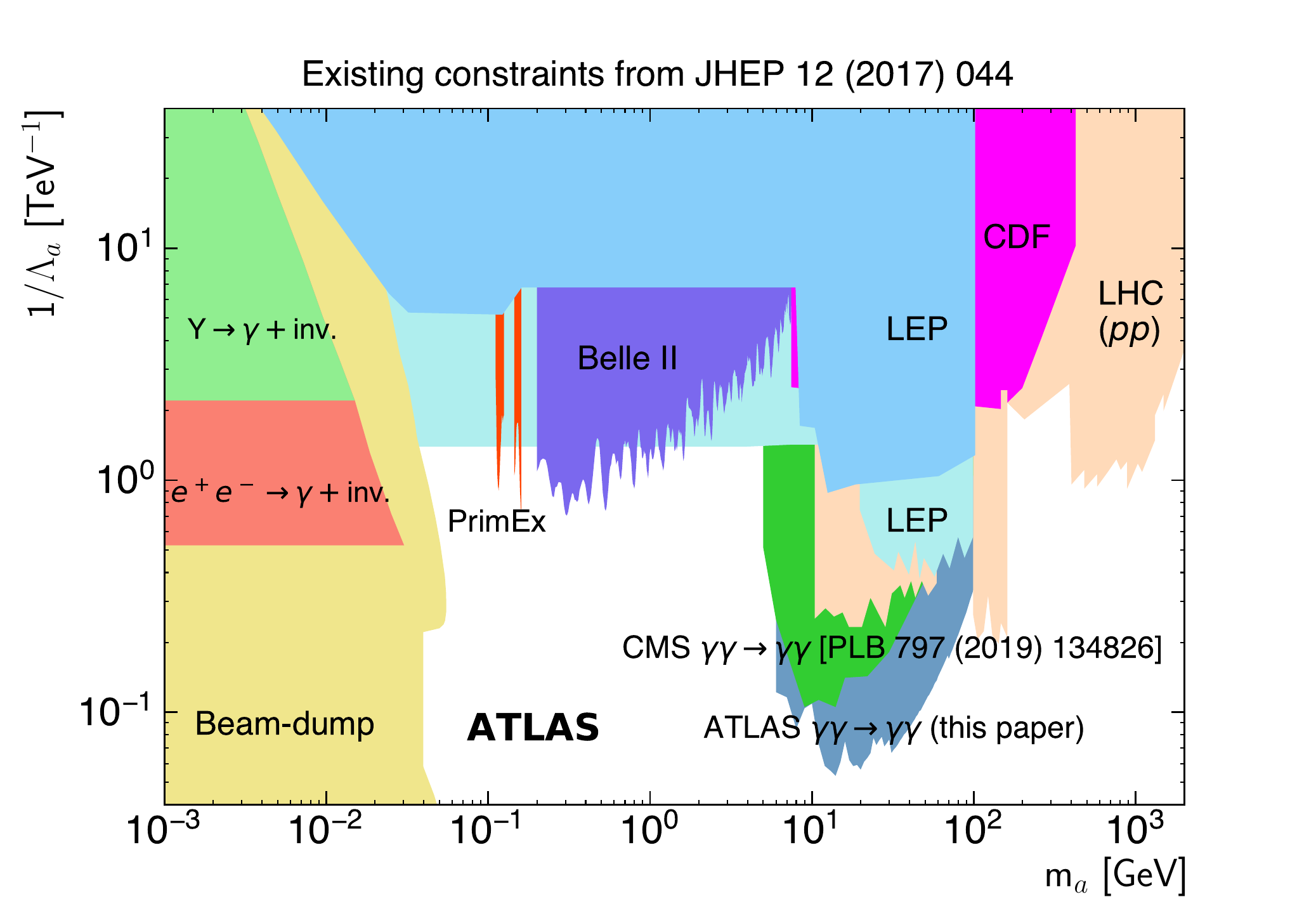}
    \caption{Compilation of exclusion limits at 95\% CL in the ALP-photon coupling ($1/\Lambda_{\alpha}$) versus
ALP mass ($m_{\alpha}$) plane obtained by different experiments. All measurements assume a 100\% ALP decay branching fraction into photons. Reproduced from ~\cite{JHEP.243}.}
    \label{fig:ATLASAxionLimit}
\end{figure}

Both ATLAS and CMS used these same datasets to search for evidence of axion-like particles that couple directly to photons ~\cite{PhysRevLett.118.171801}.  These could be produced in an s-channel process in $\gamma \gamma$ collisions ($\gamma \gamma \rightarrow a \rightarrow \gamma \gamma$) and would be observed as a narrow resonance structure in the invariant mass spectrum of the 2-photon final state.  In this case, light-by-light scattering would be yet another background process.  Neither ATLAS nor CMS has observed evidence of such a resonant structure to date, and the absence has allowed them to place new exclusion limits on the production of axion-like particles.  The ATLAS limit, based on the combined 2015 and 2018 datasets, is shown in Figure~\ref{fig:ATLASAxionLimit}. 
This limit is the most stringent to date in the mass range of 6-100 GeV ~\cite{JHEP.243}.


\section{Discussion}
\label{sec:discussion}
\subsection{The Breit-Wheeler Process}

In Section~\ref{sec:key_issues_bw} we listed a few objections, challenges, or questions to the interpretation and observation of the linear Breit-Wheeler process in heavy-ion collisions.
In the following discussion, we now revisit these issues in the context of the tremendous experimental and theoretical progress made in the last few decades, as covered in Sec.~\ref{sec:exp_elec_pos}-\ref{sec:hic_exp}. 

\subsubsection{Do the highly-Lorentz contracted fields produced in heavy-ion collisions provide a valid source of photons for the Breit-Wheeler process?}
Literature from within the high-energy particle and nuclear physics community over the last three decades has reported conflicting and contradictory statements regarding the viability of WW photons for achieving the Breit-Wheeler process~\cite{PhysRevC.47.2308,PhysRevA.51.1874,PhysRevA.55.396,baltzTwophotonInteractionsNuclear2009,doi.10.1016/j.cpc.2016.10.016}. 
The immediate source of such statements is somewhat unclear, since WW photons from highly Lorentz contracted EM fields is precisely the photon source proposed by Breit and Wheeler in their seminal work~\cite{PhysRev.46.1087}:
\begin{quote}
    In the considerations of Williams, however, the large nuclear electric fields lead to large densities of quanta in moving frames of reference. This, together with the large number of nuclei available in unit volume of ordinary materials, increases the effect to observable amounts. 
\end{quote}
From this quotation it is clear that Breit and Wheeler certainly expected the Lorentz-boosted fields of ultra-relativistic nuclei, understood theoretically through the Weiszäcker-Williams approximation, to provide a viable source for achieving the Breit-Wheeler process in laboratory.
To determine if this claim, made in 1934, is still merited, one must ask if any theoretical or experimental advancement has been made that altered these conclusions? Even after Quantum Electrodynamics was firmly established at the end of the 1940s, the Weiszäcker-Williams method of equivalent photons was extensively used by Wheeler's student to study theoretical aspects of the Breit-Wheeler process in greater detail~\cite{1952PhDT21T,PhysRev.113.1649}.

Why then, in recent decades, have WW photons from Lorentz-boosted nuclear EM fields been disregarded as a viable source of photon for the Breit-Wheeler process?
We believe that one source of confusion may be due to claims about the photon virtuality resulting in the ``special case'' believed to be important for the production of $e^+e^-$ compared to other heavier leptons~\cite{doi:10.1146/annurev.nucl.55.090704.151526}:
\begin{quote}
    These photons are almost real, with virtuality $-q^2 < (\hbar/R_A)^2$. Except for the production of $e^+e^-$ pairs, the photons can usually be treated as real photons.
\end{quote}
This distinction between $e^+e^-$ production (the original concept of the Breit-Wheeler process) and the production of heavier lepton-pairs is crucial. If the above claim were correct, one would expect to observe a qualitative difference in the kinematic distribution for photoproduction of $e^+e^-$ and $\mu^+\mu^-$, since the former would be characterized by virtual photon collisions, while the latter would be characterized by real photon collisions (due to the much heavier mass of the muon). 
However, as discussed in Sec.~\ref{sec:hic_exp}, recent measurements have verified that the kinematics of both the $\gamma\gamma \rightarrow \mu^+\mu^-$ process and the $\gamma\gamma \rightarrow e^+e^-$ exhibit exactly the same features, features predicted for real photons. Specifically, the broadening of the pair transverse momentum, 
has been proven to be a result of the electromagnetic field distribution -- not a result of virtuality via the uncertainty principle~\cite{brandenburgMappingElectromagneticFields2021a}.

Even with the above clarification, WW photons from highly-Lorentz contracted EM fields do not provide, carte blanche, real photons capable of interacting via the Breit-Wheeler process. As discussed in Sec.~\ref{sec:theory_hic}, only a small fraction of the $e^+e^-$ cross section from photoproduction may be considered a result of the Breit-Wheeler process, with the remaining overwhelming fraction resulting from processes which include virtual photons (e.g., the Bethe-Heitler and Landou-Lifshitz processes~\cite{PhysRevC.47.2308,PhysRevSTAB.17.051003,PhysRevA.51.1874,Wang:2022ihj}).

\subsubsection{Can the Breit-Wheeler process be isolated from background?}
Even with a viable source of photons for achieving the Breit-Wheeler process, definitive observation requires a procedure for clearly identifying the process and separating it from  background processes. This experimental challenge is not unique to high-energy nuclear physics experiments. For instance, Laser-based experiments like the E-144 experiment at SLAC had to carefully estimate the number of positrons observed in the ``Laser off'' configuration in order to identify the distribution of positrons from the multi-photon Breit-Wheeler process.

In heavy-ion experiments, the photons capable of achieving the Breit-Wheeler process are only found in a small region of phase-space, resulting in $e^+e^-$ pairs in which the daughters are produced nearly back-to-back. This constraint results in the characteristic peak in cross section at very  low pair transverse momentum. As mentioned previously, this feature is crucial for the separation of the Breit-Wheeler process from other sources of $e^+e^-$, since all other processes that produce an $l^+l^-$ pair have a cross section that decreases at low pair transverse momentum. This is most plainly illustrated in Fig.~\ref{fig:star_low_ptee}, where one can see that the yield of the ``Au+Au Cocktail'' (all non Breit-Wheeler sources of $e^+e^-$ pairs) decreases precisely in the region where the Breit-Wheeler cross section peaks. 

The separation of processes involving real vs. virtual photons via a selection in phase space has also been carried out in the well recognized observations of Light-by-Light scattering by the ATLAS and CMS Collaborations~\cite{doi.10.1016/j.physletb.2019.134826}. 
In these cases, WW photons from highly Lorentz contracted fields are used as a source for studying Light-by-Light scattering. However, in order to separate the LbyL process from those involving virtual photons, only high-mass, mid-rapidity di-photon production is considered~\cite{PhysRevLett.111.080405}. In the same way, the Breit-Wheeler process can be isolated from background processes by considering mid-rapidity, high-mass, low transverse momentum $e^+e^-$ pairs. 

\subsubsection{Are higher order effects present, and if so, are they separable or inseparable from the lowest-order Breit-Wheeler process?}
The discovery of higher order QED effects in high energy particle and nuclear collisions would itself be an important finding. Potential signatures of Coulomb corrections in high-energy electromagnetic scattering processes have been thoroughly studied over the last few decades~\cite{2001NuPhA695395B,PhysRevC.80.034901,Sun:2020ygb,Zha2021.10.1007/JHEP08(2021)083}.  While significant theoretical differences and uncertainties exist, several authors have predicted that Coulomb corrections modify the total pair photoproduction cross section by as much as $30\%$. However, significant experimental uncertainties prevent measurement of the photoproduction cross section to better than $\sim8\%$~\cite{Baltz:1998ex,baltzTwophotonInteractionsNuclear2009,ZDC130xs2002}. With such a level of uncertainty, cross section measurements alone cannot unambiguous rule out the presence of Coulomb corrections. 

In the specific kinematic region of the STAR Measurement for which the Breit-Wheeler process is expected to be applicable, QED calculations find that Coulomb corrections exactly cancel. In this regime, the pair production process is accurately described by the lowest-order process alone. 
Experimentally, the unique $\phi$ modulation measured by the STAR Collaboration shows that the process results from the spin carried by two real photons at lowest order~\cite{PhysRevD.101.034015}. The exploration of Coulomb correction at the LHC and future EIC are points of significant interest (see Sec.~\ref{sec:future}).

\subsubsection{Has the Breit-Wheeler process already been observed in heavy-ion collisions}
The production of $e^+e^-$ pairs has been measured by different experiments in hadron and ultra-peripheral heavy-ion collisions at the SPS~\cite{PhysRevLett.69.1911,BAUER1994471}, RHIC~\cite{PhysRevC.70.031902,AFANASIEV2009321}, the Tevatron~\cite{PhysRevLett.98.112001,PhysRevLett.102.242001} and the LHC~\cite{abbas_charmonium_2013,chatrchyan_search_2012,sirunyan_observation_2018,doi.10.1016/j.physletb.2019.134826,aad_measurement_2015,aaboud_evidence_2017} facilities over the past three decades. All experiments reported cross sections for $e^+e^-$ pair production within their kinematic acceptance. All experiments compared their results with models implementing two-photon collisions. The experimental uncertainty of these measurements range from about 15\% to 30\% with the exception that the WA93 experiment detected only the positrons at forward angle in a fixed-target configuration.
However, unambiguous identification of the Breit-Wheeler process has not been achieved until recently since it requires experimental demonstration that the colliding photons have the energy spectrum and quantum spin states of real photons, and that any approximations do not alter the physics result of real photon collisions.

It should be noted that the photoproduction of vector mesons (e.g. $J/\psi$, $\Upsilon$) and intermediate bosons ($Z^0$) from photon-hadron interactions dominates at their respective mass over production from the two-photon process. All the previous measurements ($J/\psi$ from PHENIX~\cite{AFANASIEV2009321}, CDF~\cite{PhysRevLett.102.242001} and ALICE~\cite{abbas_charmonium_2013}, $\Upsilon$ and $Z^0$ from CDF~\cite{PhysRevLett.98.112001}, ATLAS~\cite{aad_measurement_2015,aaboud_evidence_2017} and CMS~\cite{chatrchyan_search_2012,sirunyan_observation_2018,doi.10.1016/j.physletb.2019.134826}) had to exclude those mass ranges from the measurement of the two-photon cross sections. The WA93~\cite{PhysRevLett.69.1911}, CERES~\cite{BAUER1994471} and previous STAR~\cite{PhysRevC.70.031902} studies did not have the reach in invariant mass or statistics for the vector meson exclusion measurements. 
None of the previous measurements had the capability of identifying the unique smooth feature in the mass range where vector mesons are known to be present, but absent from the observed production only in the case of an exclusive Breit-Wheeler process.

Prior to Ref.~\cite{PhysRevLett.127.052302} from STAR, no measurement made in ultra-relativistic heavy-ion collisions claimed to be an observation of the Breit-Wheeler process.
We also emphasize that the previous studies that went beyond cross section measurements all reached conclusions that the data indicated significant photon virtuality (and therefore were identified as a Landau-Lifshitz process)~\cite{PhysRevC.70.031902} and/or that there may be significant final-state interactions~\cite{PhysRevLett.121.132301,PhysRevLett.121.212301} in hadronic events. Note, for instance, recent papers by the STAR~\cite{PhysRevLett.121.132301} and ATLAS~\cite{PhysRevLett.121.212301} collaborations with measurements of $P_{\perp}$ broadening (STAR) and increased acoplanarity, $\alpha$ (ATLAS) that were both interpreted as evidence for final state effects. With the new measurements in UPCs from the LHC and RHIC in the last few years, it has been proven that these conclusions were influenced by the assumptions in the specific EPA model used~\cite{ZHA2020135089}. On one hand, many of the publications in the literature avoided rigorous identification of the Breit-Wheeler process and the approximation necessary for the models and experiments; on the other hand, photon-photon collisions with the Breit-Wheeler cross section have been assumed in many model calculations and any discrepancies with experimental data would be attributed as a disproof of the Breit-Wheeler process. 
In contrast, STAR performed the first measurements providing unambiguous observation of the linear Breit-Wheeler process by demonstrating that: 1) the colliding photon energy spectrum is consistent with the collision of real photons (i.e., not determined by photon virtuality) and 2) that the quantum spin states of the colliding quanta are those of real photons. Furthermore, the observation of the $\cos4\phi$ spin interference effect indicates that the process is governed by the lowest order interaction, with no significant effect from higher-order interactions. The availability of high luminosity from ultra-relativistic heavy-ion colliders and the advances in detector technology in the last decade have provided unique opportunity with multiple tools and kinematic variables for the discovery.

\subsection{Vacuum magnetic birefringence}

\subsubsection{What is novel about the STAR measurement, i.e., how is it unique compared to the SLAC E-144 measurement, with respect to polarization effects?}
The SLAC E-144 measurement of the multi-photon Breit-Wheeler process utilized a positron calorimeter to identify events with excess positron production compared to background processes present in the ``laser off'' case. The excess positron yield and momentum spectrum was found to be in good agreement with the expectation for the multi-photon Breit-Wheeler process, for a mean number of photons $\langle n\rangle\simeq5$ colliding in each interaction. 
However, since a copious amount of electrons were scattered and produced in each event, the E-144 measurement could not identify the electron partner of each positron, and therefore could not uniquely identify the entire $e^+e^-$ pair. With limited numbers of positrons observed and the lack of full pair information, no differential measurements could be made that might be sensitive to photon polarization effects. 

Unlike the multi-photon Breit-Wheeler measurement, the measurement performed by the STAR collaboration allowed full identification of the electron and positron produced in the linear Breit-Wheeler process. The STAR experimental setup, consisting of charged particle tracking and particle identification, allowed events to be selected with exactly one $e^+e^-$ pair per event. With information from both the electron and positron, reconstruction of the full kinematic information for the Breit-Wheeler process is possible. As discussed in Sec.~\ref{sec:discussion}A, measurement of the $e^+e^-$ momentum and energy spectrum is sufficient to uniquely identify the process as the  Breit-Wheeler process. The energy and momentum spectra also provides constraints on the charge distribution producing the Lorentz contracted electromagnetic fields~\cite{brandenburgMappingElectromagneticFields2021a,Wang:2022ihj}. 

With the statistical precision and precise trajectory measurement capabilities of the STAR tracker, additional differential measurements sensitive to the photon polarization states are possible. 
The key experimental challenge for measuring polarization effects in heavy-ion collisions results from the random orientation of the photon polarization vectors in each event, due to the random nucleus-nucleus impact parameter. However, as described in Sec.~\ref{sec:theory}, WW photons in the regime of validity for the Breit-Wheeler process have polarization vectors in the projected transverse directions that correspond with their transverse momentum vectors. This relationship allows photon polarization effects to be studied through quantum spin-momentum correlations in the final state $e^+e^-$ pair~\cite{doi.10.1016/j.physletb.2019.07.005}, which leads to the $\cos4\phi$ modulation observed by the STAR collaboration. Even though one must generally integrate over the possible combinations of photon-photon polarization to calculate the observable cross section, a net effect due to photon polarization persists into the integrated cross section because of a mismatch in the number of photons (luminosity) colliding with parallel vs. perpendicular polarization vectors. 

\subsubsection{How is the azimuthal angle $(\phi)$ modulation observed by STAR related to vacuum birefringence and/or dichroism?}
The $\cos4\phi$ modulation observed by the STAR collaboration demonstrates that the colliding photons are linearly polarized. In terms of the angle $\phi$, the differential cross section for the Breit-Wheeler process may be expressed as~\cite{PhysRevD.101.034015,BUDNEV1975181}:
\begin{equation}
    d\sigma/d\phi \propto \left[ \mathcal{A} \left( \frac{\sigma_\parallel + \sigma_\perp}{2} \right) + \tau_{\gamma\gamma} \cos 2\phi \left( \mathcal{B} + \mathcal{C} \cos 2\phi \right) \right],
\end{equation}
and $\mathcal{A}$, $\mathcal{B}$, and $\mathcal{C}$ are numerical constants. In the above, we make explicit the dependence on $\tau_{\gamma\gamma} \equiv \sigma_\parallel - \sigma_\perp$, to mirror the derivation by Toll~\cite{1952PhDT21T,BUDNEV1971470}.
Since $\mathcal{B} \propto m^2/p_\perp^2$, it vanishes for e.g., $e^+e^-$ production with $p_\perp > \sim 200$ MeV, as required by the experimental conditions at the STAR and ALICE experiments. However, $\mathcal{C}$, not being suppressed by powers of the lepton mass, is visible within the kinematic ranges accessed by existing experiments.
With the above factorization, the amplitude of the $\cos 4\phi$ modulation measured in an ensemble of events can be expressed as:
\begin{equation}
    A_{4\phi} \propto \frac{ \tau_{\gamma\gamma}^{\star} \mathcal{C}} { \mathcal{A} },
\end{equation}
where $\tau_{\gamma\gamma}^\star = I_\parallel - I_\perp$ is the difference in intensity for photon-photon collisions with parallel vs. perpendicular polarization vectors. The STAR kinematic acceptance results in a measurement of $e^+e^-$ well above the pair mass threshold. Just above threshold $\tau_{\gamma\gamma}$ is large, but approaches zero for pairs produced at higher invariant mass. 
Therefore, STAR is not directly sensitive to a difference in the absorption cross section and vacuum dichroism.

The observed $\cos4\phi$ modulation for the Breit-Wheeler process results from the behavior of the helicity amplitudes relevant for polarized photons~\cite{SuperChic3,KLUSEKGAWENDA2021136114}. While this particular observation is not sensitive to $\tau_{\gamma\gamma}$, it demonstrates the physical splitting of the wavefunction governing the momentum distribution of the $e^+e^-$ for parallel vs. perpendicular photon polarizations.
For two real photons, the absorption process (the Breit-Wheeler process) and the forward scattering process (Light-by-Light scattering) are related by the Optical Theorem~\cite{PhysRev.86.1}, with exact analytical expressions for the dispersive and absorptive parts of the amplitude. 
Additional theoretical work is needed to relate the observed modulation directly to birefringence in the  forward scattering process.
As an example, according to the SuperChic3~\cite{SuperChic3} code, the light-by-light scattering process exhibits a $\cos 2\phi$ modulation when restricted to approximately the same kinematics as the STAR Breit-Wheeler measurement (mid-rapidity with large invariant mass).
These effects result from the structure of the helicity amplitudes for the Breit-Wheeler process and Light-by-Light scattering, taking into account the photon polarization determined by the electric field lines of the source from the Lorentz contracted Coulomb field. The angular modulation effect is similar to proposed laser experiments for probing vacuum birefringence  and dichroism~\cite{PhysRevLett.119.250403}. As a final note, we recognize that these observed effects operate in a purely perturbative regime of QED as depicted by the tree-level diagrams shown in Fig.~\ref{fig:vb_pvlas_diagrams} and do not address the strong field regime of QED~\cite{PhysRevD.104.116013}, at least for heavy-ion collisions at the ultra-relativistic energies with the specific kinematics discussed currently.

\section{Future Opportunities and Facilities}
\label{sec:future}

Significant progress through multiple discoveries has been made in the physics of photon interactions. Experimental measurements and theoretical descriptions have been progressing from the initial discoveries toward quantitative and precise comparisons. At this point in time, polarized photons have been used and proposed as a tool to test and define the photon Wigner function~\cite{ZHA2020135089,PhysRevD.104.056011,PhysRevD.102.094013,KLUSEKGAWENDA2021136114,PhysRevLett.127.122001,Sun:2020ygb,Zha2021.10.1007/JHEP08(2021)083}, to probe the properties of the Quark-Gluon Plasma~\cite{brandenburgMappingElectromagneticFields2021a,PhysRevLett.121.132301,PhysRevLett.121.212301,PhysRevLett.122.132301,Wang:2021oqq,An:2021wof,Klusek-Gawenda:2018zfz}, to measure nuclear charge and mass radii~\cite{Wang:2022ihj,STAR:2022wfe,Budker:2021fts,brandenburgMappingElectromagneticFields2021a}, to study gluon structure inside nuclei~\cite{Hatta:2021jcd,Xing:2020hwh,Bor:2022fga} and to investigate new quantum effects~\cite{STAR:2022wfe,Zha:2018jin,Zha:2020cst,Xing:2020hwh,Dyndal:2020yen}. In this section, we provide some examples of these future opportunities.

\begin{figure}
    \centering
    \includegraphics[width=.60\linewidth]{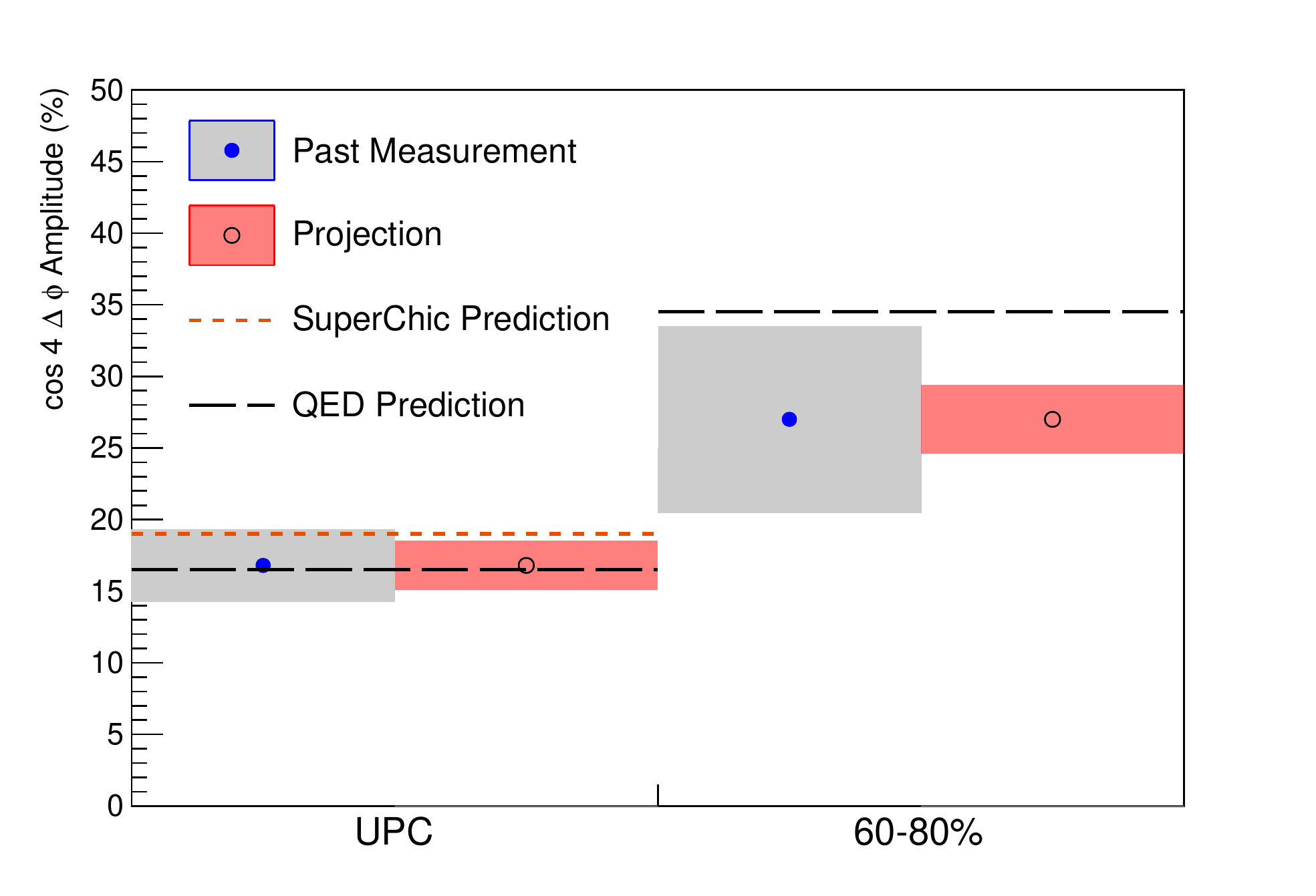}
    \caption{ STAR projection of sensitivity expected from future measurements of the $\cos{4\Delta\phi}$ asymmetry for both peripheral $(60-80\%)$ and ultra-peripheral collisions. Reproduced from Ref.~\cite{SN0755STARBeam}. }
    \label{fig:future2}
\end{figure}

\subsection{Opportunities as electromagnetic probes of Quark-Gluon Plasma} 
As discussed in Ref.~\cite{brandenburgMappingElectromagneticFields2021a}, the observation of these $\gamma\gamma$ processes in events with hadronic overlap potentially allow the pure QED processes to be used as a probe of the produced nuclear medium. Both RHIC and the LHC plan future data collection that will allow high precision multi-differential analysis of these $\gamma\gamma$ processes~\cite{SN0755STARBeam,ProspectsMeasurementsPhotonInduced}. Future measurements at STAR are expected to provide significantly higher precision measurements of the $e^+e^-$ transverse momentum spectra and the $\cos4\phi$ modulation. Additionally, multi-differential measurements, such as the $\cos4\phi$ modulation strength versus pair $p_T$, will be possible. The increased precision on the pair $p_T$ will provide additional constraining power to investigate the proposed final-state broadening effects. In addition to their effect on the $p_T$ spectra, final state interactions would wash out the $\cos4\phi$ modulation strength that results from the initial colliding photon polarization. Figure~\ref{fig:future2} shows the predicted future precision that will be achieved for the $p_T$ (a) and $\cos4\phi$ modulation measurements (b) in future STAR analyses. The added precision in the $cos4\phi$ modulation measurement is expected to allow experimental verification of impact parameter dependence predicted by the lowest order QED calculations (and therefore further exclude the $k_\perp$-factorization plus TMD treatment of the photon polarization~\cite{doi.10.1016/j.physletb.2019.07.005,SuperChic3}). The future data taking campaigns planned for the LHC experiments will also allow improved measurements from ALICE of the $\gamma\gamma \rightarrow e^+e^-$ process (Fig.~\ref{fig:ALICE_low_mass}) in a similar region of phase space as measured by STAR, but in collisions with a much larger Lorentz-boost factor. Such measurements will provide further constraints on the treatment of the photon kinematic distributions over a range of photon energies. Similarly, future data taking and analyses by CMS and ATLAS~\cite{atlascollaborationMeasurementMuonPairs2022,ProspectsMeasurementsPhotonInduced} will allow additional precision measurements of the $\gamma\gamma \rightarrow \mu^+\mu^-$ process in events with hadronic overlap, possibly shedding light on the presence (or lack) of medium induced modifications via differential measurements of the produced dilepton kinematics.

\subsection{Probe the geometry of nuclear charge and gluon distributions}

As discussed in the previous sections, the Breit-Wheeler process is sensitive to the nuclear charge distribution. 
Unlike in the case for an $e^++e^-$ collider, the photon flux does not diverge in UPCs because the low-energy photon flux is regulated by the finite Lorentz factor of the ions and the high-energy photon flux is naturally cut off by the finite field strength due to the finite size of the ion’s charge distribution in the form factor. 
However, there is an additional important factor which makes the Breit-Wheeler process sensitive to the nuclear geometry. 
The WW photons are linearly polarized, and the two Feynman diagrams~\cite{PhysRevA.51.1874} in Eq.~\ref{eq:cross_section_pair} cancel at low $k_{\perp}$. The phase modulation is of the form $\exp{(-i\vec{b}\cdot\vec{k_{\perp}})}$, and depends on the impact parameter, which is related to the nuclear geometry. 
This is also what results in an impact-parameter dependence of the Breit-Wheeler process. Therefore, in high-energy ultra-peripheral heavy-ion collisions, the low-$k_{\perp}$ is modulated by $\exp{(-i\vec{b}\cdot\vec{k_{\perp}})}$
and high-$k_{\perp}$ by the form factor. Both of these factors are a function of the nuclear geometry~\cite{Wang:2022ihj}. The first attempt of extracting the nuclear charge radius was performed using the available Au+Au collisions at 200GeV, and shows comparable result with that from low-energy electron scattering~\cite{Wang:2022ihj}. Future theoretical and experimental progress could yield quite precise charge radius measurements. 

When two relativistic heavy nuclei pass one another at a distance of a few nuclear radii, the photon from one nucleus may interact through a virtual quark-antiquark pair with gluons from the other nucleus forming a short-lived vector meson (e.g. $\rho^0$) as described in Section~\ref{sec:photoprod}. The polarization was utilized in diffractive photoproduction to observe a unique spin interference pattern in the angular distribution of $\rho^0\rightarrow\pi^+\pi^-$ decays~\cite{STAR:2022wfe}. The observed interference is a result of an overlap of two  wave functions at a distance an order of magnitude larger than the $\rho^0$ travel distance within its lifetime. The observable is an example of the quantum interference of non-identical particles. Independent of theoretical models, the interference patterns observed in Au+Au and U+U measurements can be quantified by studying the polarization dependence of the $|t|$ distribution in two dimensions. In this way, the effects of photon transverse momentum and two-source interference can be removed to extract the nuclear radius of gold and uranium. The strong-interaction nuclear radii were extracted from these diffractive interactions, and found to be $6.53\pm 0.06$ ${\rm fm}$ ($ {^{197} {\rm Au }}$) and $7.29\pm 0.08$ ${\rm fm}$ ($ {^{238} {\rm U}}$), larger than the nuclear charge radii~\cite{STAR:2022wfe}. These could be used to extract neutron skins~\cite{STAR:2022wfe}. The new measurement of neutron skin ($S$) for $^{197}Au$ of $0.17 \pm 0.03 {\rm(stat.)} \pm 0.08 {\rm(syst.)}$ fm seems to follow the trend of world measurements at low energies~\cite{CentellesPhysRevLett.102.122502} while that of $^{238}U$ of $0.44\pm0.05 {\rm(stat.)} \pm 0.08{\rm(syst.)}$ fm significantly non-zero and indicates a value larger than that expected for neutron skins of similar nuclei~\cite{PREXPbSkinPhysRevLett.126.172502} [in terms of the fraction of neutron excess (N-Z)/A] ~\cite{CentellesPhysRevLett.102.122502}. It further demonstrates that this spin-induced orbital angular momentum interferometry offers a new avenue for studying nuclear geometry and gluon distribution within large nuclei at a quantitative level.

The photo-nuclear interactions can also be viewed as photon-Pomeron fusion process, and are therefore sensitive to the gluon distribution in nuclei, especially for heavy quarnkonia photoprdocution, in which the process can be treated with perturbative QCD. The momentum transfer of this process is determined by the local density, which enables probing the spatial gluon distribution within nuclei. Furthermore, the linear polarization feature of photoproduction in relativistic heavy-ion collision makes it possible to align the initial collision geometry, which allows 2D tomography of the gluon distribution~\cite{STAR:2022wfe,Xing:2020hwh,Zha:2020cst}. Currently, various measurements of $J/\psi$ photoproduciton have been made in ultra-peripheral collisions~\cite{STAR:2008llz,AFANASIEV2009321,doi.10.1007/JHEP09(2015)095,doi.10.1016/j.physletb.2021.136481,PhysRevC.96.054904,STAR:2019yox}, however most of them are only performed in one dimension (magnitude of transverse momentum). The photoproduction processes could be accompanied by violent hadronic interaction when achhieved in events with nuclear overlap~\cite{PhysRevLett.116.222301,STAR:2019yox}. In these events, the photon produced products could serve as a novel probe to detect the properties of the hot medium created in the overlap region. However, the existing experimental measurements are lack of statistics to infer that. Both RHIC and LHC plan future data collection that will allow high precision multi-differential analysis of these photoproduction processes~\cite{SN0755STARBeam,ATLAS:2018xqt} and provide significantly higher precision of the two-dimensional transverse momentum spectra of photon produced vector meson. Future measurements at STAR with iTPC upgrade are expected to make the first observation $\phi$ photoproduction in heavy-ion collisions. Such future data taking campaigns planned at RHIC and LHC  will significantly improve our understanding of the little known gluon distribution in nuclei and provide powerful tools to quantitatively extract the properties of matter governed by strong interaction.     

\subsection{The Breit-Wheeler process for Measuring tau $g-2$ and Axion Searches}

The Breit-Wheeler process, realized as the photoproduction of a pair of tau leptons $(\gamma\gamma \rightarrow  \tau^+\tau^-)$, provides a unique opportunity to measure the anomalous magnetic moment of the tau lepton: $a_\tau = (g_\tau - 2)/2$~\cite{Aguila1991}. Due to the heavy mass of the tau, measurement of the anomalous magnetic moment provides stringent tests of fundamental predictions from QED and potentially probes physics beyond the standard model~\cite{PhysRevD.102.113008} with precision $m^2_\tau/m^2_\mu\sim280$ times more than for measurements of the muon. Currently, the highest precision experimental measurement of $a_\tau$ results from a 2004 measurement by the DELPHI collaboration~\cite{DELPHI2004} yielding $a_\tau=-0.018(17)$. The central value of this measurement is surprisingly an order of magnitude larger than the QED prediction $a_\tau^{\rm QED} = 0.00117721(5)$, allowing significant room for physics beyond the standard model. 

Past measurements of the electron and muon have yielded $a_e$ to a precision of $\sim0.28$ ppt~\cite{PhysRevLett.100.120801} and $a_\mu$ to a precision of $\sim0.7$ ppb~\cite{PhysRevLett.89.101804}. Considering that the value of $a_\mu$ shows a $3-4\sigma$ tension with the standard model predictions~\cite{PhysRevLett.109.111808,Davier2017,Jegerlehner2018,PhysRevD.97.114025}, measurement of $a_\tau$ may prove an extremely important tool for uncovering new physics beyond the standard model. Measurement of $a_\tau$ via the Breit-Wheeler process in ultra-peripheral heavy-ion collisions provides multiple benefits: the $Z^2$ enhancement of the coherent photon field leads to large cross sections, clean events allow individual tracks from tau decays to be identified, and their unique event characteristics that allow efficient triggering and selection. By employing a recently proposed analysis, experiments at the LHC may be able to improve upon the DELPHI measurement, increasing the precision on $a_\tau$ by a factor of $\sim3$ with existing datasets~\cite{PhysRevD.102.113008}. Future datasets to be collected at the High Luminosity LHC will further allow constraints an order of magnitude more stringent than the DELPHI result. 

Photon-photon fusion processes can also be used to search for the existence of Axion-like particles (ALP, $a$) with lepton-flavor-violating couplings that could potentially lead to a anomalous magnetic moment of the leptons. In the Standard Model, the leading order Light-by-Light process proceeds through box diagrams of virtual charged particles. However, measurement of the process has also been proposed as an avenue for testing physics beyond the standard model, since the process may proceed through predicted ALPs ($\gamma\gamma \rightarrow a \rightarrow \gamma\gamma$)~\cite{bauerColliderProbesAxionlike2017}. Existing measurements from the ATLAS collaboration have utilized the $\gamma\gamma\rightarrow \mu^+\mu^-$ to set the most stringent limits on the existence of ALPs with an invariant mass between $10-100$ GeV~\cite{PhysRevLett.118.171801}. Event more stringent limits are expected with the planned 10 nb$^{-1}$ data to be collected in Run 3 and Run 4. Finally, the strong electromagnetic fields of ultra-relativistic heavy nuclei may provide sufficient energy density to manifest as dark photons$(A^\prime)$~\cite{Filippi2020}, a massive gauge boson that couples to the Standard Model through kinetic mixing. The existence of dark photons may lead to an anomalous signal in the yield of $l^+l^-$ within the kinematic region dominated by the Breit-Wheeler process due to production via one or two dark photons ($\gamma A^\prime\rightarrow l^+l^-$ or $A^\prime A^\prime\rightarrow l^+l^-$)~\cite{brandenburgMappingElectromagneticFields2021a}.  

\begin{figure}
    \centering
    \includegraphics[width=0.99\linewidth]{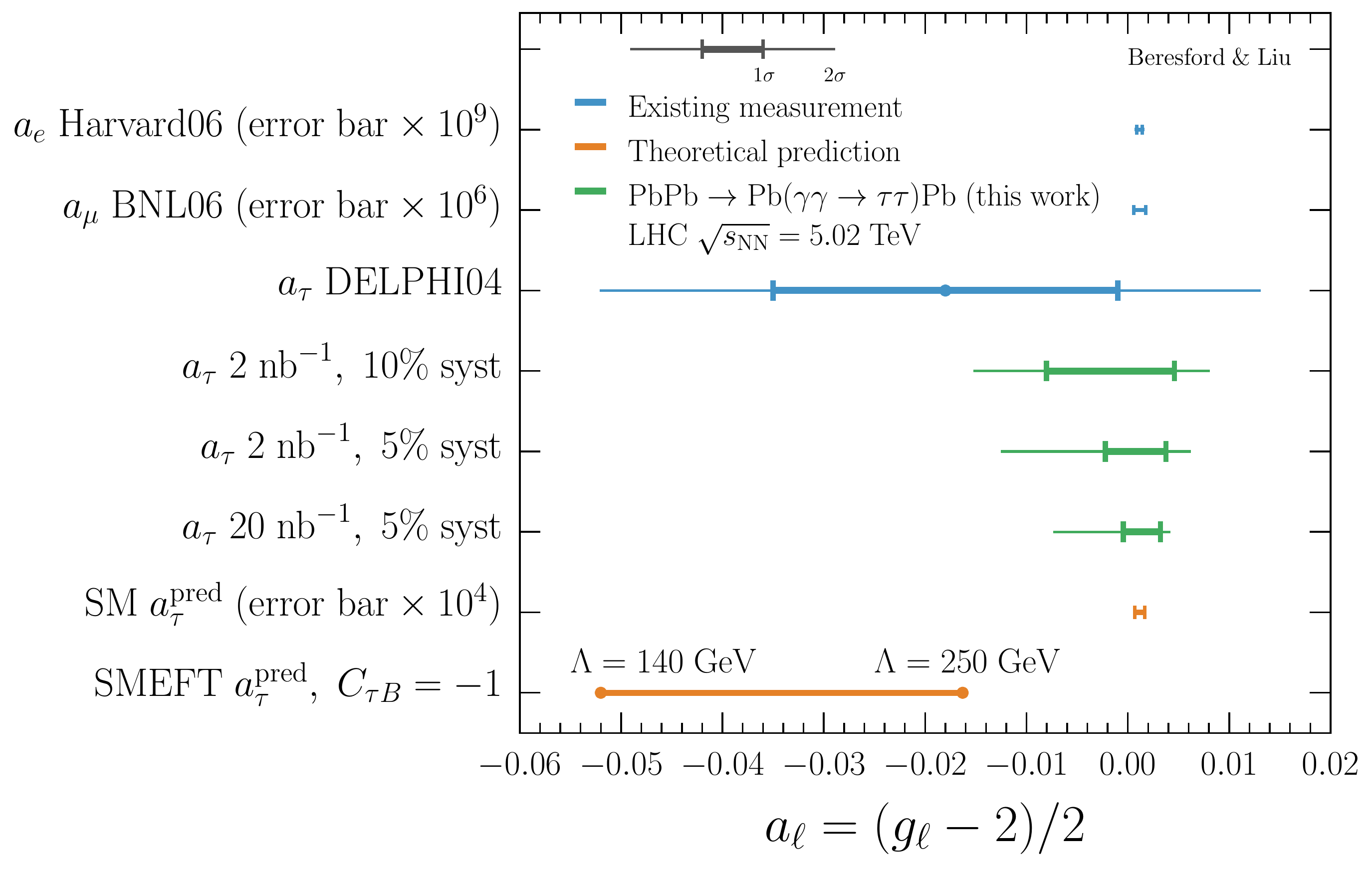}
    \caption{Summary of lepton anomalous magnetic moments measurements for $a_l$ with theoretical predictions for $\alpha_\tau$ and experimental projections based on existing LHC data. For visual clarity, the $1\sigma$ error bars on $a_e$ $(a_\mu)$ are inflated by $10^9$ $(10^6)$, and $10^4$ for the standard model $a_\tau$ prediction in shown in orange. Reproduced from Ref.~\cite{PhysRevD.102.113008}.}
    \label{fig:dipoles}
\end{figure}

\subsection{Future Laser Facilities}
In 1996 the NOVA Petawatt laser~\cite{Perry:99} at Lawrence Livermore National Laboratory achieved the first petawatt (PW) laser pulse via chirped pulse amplification~\cite{stricklandCompressionAmplifiedChirped1985}. Since then the high-field laser community has been making rapid progress~\cite{PhysRevLett.127.114801} with multiple petawatt grade facilities such as the Extreme Light Infrastructure~\cite{Gales_2018} in Europe, the Shanghai Superintense Ultrafast Laser Facility~\cite{Gan2021} in China, and the Laboratory for Laser Energetics (LLE) at the University of Rochester, in New York, USA~\cite{boehlyInitialPerformanceResults1997,maywarOMEGAEPHighenergy2008}.
Even higher powered laser facilities are expected to come online in the next decade, such as the 100-PW laser at the Station of Extreme Light (SEL) in China, the proposed 180-PW laser at the Exawatt Center for Extreme Light Studies (XCELS) in Russia, or the Optical Parametric Amplifier Line (OPAL), a 75-PW laser upgrade proposed at the LLE~\cite{PhysicistsArePlanning}. Though, up til now, the highest intensities achieved by laser are still sub-critical $(2\times10^{22}$ W cm$^{-2} )$. While many of the aforementioned facilities are meant primarily for laser-driven fusion research, this rapid progress in power toward that needed to reach the critical field strength, make it feasible to test effects of strong-field QED as well. In light of the recent advances in laser power, experiments have even been proposed for the express purpose of testing QED in the strong field regime, such as the LUXE experiment at the European X-Ray Free-Electron Laser Facility~\cite{borysovStrongfieldQEDExperiment2021}. 

\section{Conclusion}
\label{sec:conclusion}
Motivated by the recent discoveries in polarized photon collisions by the STAR Collaboration~\cite{PhysRevLett.127.052302}, we have reviewed the endeavor of photon-photon collisions from the time of Breit and Wheeler's proposal through efforts in particle accelerators from $e^+e^-$ to heavy-ion collisions at RHIC and the LHC. The experimental measurements of the cross section, transverse momentum, acoplanarity, angular distributions, and other observables have been summarized and comparisons to various theoretical calculations have been summarized. We have sought to clarify confusions about the role of photon virtuality and to define a criterion for the specific domain of validity for the Breit-Wheeler process in relativistic heavy-ion collisions. The unique features of polarized photon interactions have been discussed with connection to the pioneering work by Toll. The importance of experimental demonstration of polarized photons from highly Lorentz-contracted Coulomb fields has been highlighted and some examples of future opportunities of their utilization have been given. It is evident from the tremendous progress made in the last half decade with multiple discoveries and measurements related to photon induced processes at both RHIC and the LHC that indeed relativistic heavy-ion collisions can be used to test QED and beyond.  

\section{Acknowledgment}
\label{sec:ack}
The authors would like to thank Dr. Spencer Klein, Prof. Gorden Baym, Prof. Jian Zhou, Prof. Jinfeng Liao, Prof. Shi Pu, Prof. Qun Wang, Prof. Uli Heinz, Prof. Yuri Kovchegov, Prof. Bowen Xiao, Dr. Feng Yuan and many members of STAR, ATLAS, ALICE, CMS, and PHENIX Collaborations.
This work was funded in part by the U.S. DOE Office of Science under contract No. DE-sc0012704, DE-FG02-10ER41666, DE-FG02-96ER40991,
and DE-AC02-98CH10886, the National Natural Science Foundation of China (12175223) and MOST(2018YFE0104900). 
\bibliographystyle{apsrev4-1}
\bibliography{clean_ropp}

\end{document}